\pdfoutput=1
\documentclass[apj,twocolumn,twocolappendix,numberedappendix]{openjournal}
\usepackage[dvipsnames]{xcolor}
\definecolor{red}{rgb}{0.9, 0,0}
\definecolor{navy}{rgb}{0.05, 0.05,0.8}

\usepackage[colorlinks,hypertexnames=false]{hyperref}
\hypersetup{
    colorlinks = true,
    citecolor  = red,
	linkcolor  = navy
}

\usepackage{amsmath}
\usepackage{amssymb}
\usepackage{amsfonts}
\usepackage{bm}
\usepackage{multirow}
\usepackage{mathtools}
\usepackage{booktabs}
\usepackage{makecell}

\DeclareMathOperator*{\argmin}{argmin}

\def\vec#1{\mathbf{#1}}

\begin{document}

\title{Detectable subhalo impacts in Milky Way streams}

\author{Junyang Lu$^{1 *}$ }
\author{Elias Bernreuther$^1$ }
\author{Tongyan Lin$^1$ }
\author{Vincent S. H. Lee$^{1,2}$ }
\author{Ana Bonaca$^3$}
\author{Ethan O.~Nadler$^4$}
\thanks{$^*$E-mail: jul143@ucsd.edu}
\affiliation{$^1$ Department of Physics, University of California, San Diego, La Jolla, CA 92093, USA \\
  $^2$ Department of Physics, University of California, Berkeley, Berkeley, CA 94720, USA\\
  $^3$  The Observatories of the Carnegie Institution for Science, Pasadena, CA 91101, USA\\
  $^4$ Department of Astronomy \& Astrophysics, University of California, San Diego, La Jolla, CA 92093, USA \\
}

\begin{abstract}
Dark matter subhalos leave gravitational imprints in the stellar streams of the Milky Way. Observing individual strong impacts of subhalos offers a compelling way to constrain and discover potentially dark subhalos down to $10^6 M_\odot$, allowing for new tests of the particle physics properties of dark matter. We develop a pipeline and statistical framework to forecast the expected number of detectable subhalo impacts on stellar streams, based on morphological and kinematic data from surveys such as LSST and Via. Starting from a catalog of confirmed stellar streams, we focus our efforts on 14 promising streams that are relatively well-modeled with a particle spray algorithm.
Our criteria for a detectable impact is a deviation at 95\% CL from the best-fit polynomial proxy model for the stream, which accounts for stream modeling uncertainties and regulates the effect of distant impacts that are degenerate with these uncertainties.  Among the 14 streams studied, we find that 5 streams have an expected number of detectable impacts greater than 0.2. With LSST and Via data, the stream Jet has $5.15^{+1.10}_{-0.95}$ expected detectable impacts, followed by Orphan-Chenab ($1.40^{+0.62}_{-0.47}$), ATLAS-Aliqa Uma ($1.25^{+0.60}_{-0.44}$), GD-1 ($0.55^{+0.43}_{-0.28}$), and Palomar 5 ($0.40^{+0.39}_{-0.23}$), where error bars are the 95\% containment on the Poisson mean.
These values rely on the assumed subhalo population, which can give a factor of few systematic uncertainty in the predictions. We also consider effects of different particle dark matter models on the number of impacts, finding a suppression by a factor of $\sim 4$ for warm dark matter and fuzzy dark matter models at their current mass bounds and an $O(1)$ enhancement for a toy model of self-interacting dark matter.
\end{abstract}

\maketitle

\section{Introduction}
\label{sec:intro}

The predictions of the $\Lambda$CDM paradigm match well with observations of matter clustering, from the largest scales of the universe down to galactic scales~\citep{chabanier2019}. Dark matter (DM) substructure below the $10^8\, M_\odot$ scale is much less constrained~\citep{2017ARA&A..55..343B}, however. DM halos have been constrained with strong lensing for mass down to $ 2 \times 10^8 M_\odot$ (peak mass prior to the effects of tidal stripping)~\citep{2020MNRAS.491.6077G,2026arXiv260405237N} and with Milky Way satellites for peak mass down to $3 \times 10^8 M_\odot$~\citep{2018MNRAS.473.2060J,2010MNRAS.404L..16M,2020ApJ...893...48N}.
Many of the halos with peak mass below $10^7 - 10^8\, M_\odot$ are below the threshold for galaxy formation~\citep{2020MNRAS.498.4887B,2023MNRAS.524.2290N,Nadler2025ApJ...983L..23N} and thus are completely dark.
Measurements of these low mass subhalos would allow for new tests of DM theories beyond CDM, such as fuzzy DM (FDM)~\citep{Hu_2000}, warm DM (WDM)~\citep{Colombi_1996,Bode_2001}, or self-interacting DM (SIDM)~\citep{Spergel_2000}, since these models alter the abundance and/or internal structure of low-mass halos and subhalos (see \citealt{bechtol2023snowmass2021} for an overview).

Stellar streams, tidally disrupted globular clusters and dwarf galaxies~\citep{NewbergCarlin2016}, are a promising probe of such low mass subhalos. With their cold kinematics and elongated structures, they are sensitive to gravitational perturbations induced from a close encounter with a subhalo~\citep{Johnston_2002,Ibata_2002,Yoon2011ApJ...731...58Y,Carlberg2012}. It has been shown that stream observations could probe subhalos down to $10^5-10^6\, M_\odot$ in present-day halo mass~\citep{drlicawagner2019probing,Bovy2017,lu2025detectabilitydarkmattersubhalo}.
The cumulative effect of many  impacts can leave signatures such as increased density fluctuations and stream heating~\citep{Bovy2017,2022MNRAS.513.3682D,Carlberg_2023arXiv230108991C,Nibauer_2025ApJ...983...68N,2026arXiv260609629A}, which have been studied in GD-1~\citep{Grillmair_2006} through the density power spectrum~\citep{Banik2021a,Banik:2019smi} and velocity dispersion~\citep{DESI_2026arXiv260420958J,Nibauer:2025ezn,carlberg2026jointmodelinggd1c19}.

One compelling signature of subhalos on streams arises from single strong impacts localized along the stream. Such notable features have been observed in GD-1 and attributed to a dark perturber~\citep{Bonaca2019,Bonaca2020}, although other explanations have also been proposed~\citep{deBoer2020,Ibata2020}. Previous work has estimated the rate for subhalo encounters, typically finding O(1) strong impacts over the lifetime of the stream~\citep{Erkal_2016,menker2024,Adams:2024zhi}. These works have focused on a few well-studied streams such as GD-1 and Palomar 5 and gap formation, or density fluctuations, as the primary observable of a localized strong impact. Adding in kinematic and morphological data (angular deflection of stars) would aid in identifying such impacts, as well as potential reconstruction of the perturber properties such as mass and impact geometry ~\citep{Erkal_2015_2,hilmi2024inferringdarkmattersubhalo,lu2025detectabilitydarkmattersubhalo,2025arXiv251207960N}.

In recent years, the number of observed streams has exploded, motivating an assessment of how many such strong impacts could be present, as well as the most promising streams for detecting impacts. To date, over 100 stellar streams of the Milky Way have now been identified \citep{BONACA2025101713,Mateu_2023,Ibata_2023_STREAMFINDER}, with many more candidate streams in \cite{Shih:2023jfv} and \cite{chen2025starstreamgaiastreamdiscovery} as well as those expected with LSST~\citep{pearson2024forecastingpopulationglobularcluster}. However, the amount of information that can be gained from each stream has not been systematically studied. Furthermore, while significant improvements in stream observations are expected in the near future from photometric surveys such as LSST~\citep{lsstsciencebook2009} and spectroscopic surveys such as Via~\citep{theviacollaboration2026projectoverviewscienceinstrument}, analyzing streams for subhalo impacts requires dedicated resources, including observing time and simulation of individual streams.

In previous work by some of the same authors~\citep{lu2025detectabilitydarkmattersubhalo}, we studied the minimum detectable subhalo mass over 50 Milky Way streams, allowing us to provide a first assessment of the promising streams for subhalo detection. Our analysis led to a fitting formula for the minimum detectable subhalo mass as a function of stream distance, width, and density. However, that work did not address the question of subhalo encounter rate, and also relied on a simplifying analytic model of subhalo impacts.

In this work, we perform a detailed statistical analysis of the expected number of detectable subhalo impacts in a selection of 14 relatively well-modeled Milky Way (MW) streams. We start with a systematic study of a catalog of known streams~\citep{BONACA2025101713}, performing detailed fits of streams to determine stream age $T_{\rm form}$ and applying selection criteria to identify those that are relatively well-modeled by a particle spray algorithm. We account for multiple observables including stream density, transverse angle, and radial velocities under both present-day (Gaia+DESI) and near-future (LSST+Via) observational scenarios. Our work also improves upon the analytic modeling of \cite{lu2025detectabilitydarkmattersubhalo} with particle spray simulations of subhalo impacts.

To leading order, streams can be promising subhalo detectors if they have high geometric encounter volume (proportional to length times time) and encounter high subhalo density. From \cite{Erkal_2016}, for a stream on a circular orbit, the number of encounters within an impact parameter $b_{\rm max}$ can be estimated as:
\begin{align}
    N_{\rm enc}^{\rm circ} = \sqrt{\frac{\pi}{2}}  n_\mathrm{sub} \sigma b_{\rm max} l T_{\rm form}
    \label{eq:Nenc}
\end{align}
for physical stream length $l$, stream age $T_{\rm form}$, halo velocity dispersion for one component $\sigma$, and subhalo density $n_\mathrm{sub}$. While this gives an overall scaling, not all encounters are strong enough to leave a statistically significant feature. This simple scaling also does not account for the expected dependence of detectable impacts on stream width and density.
Converting this rate into a number of {\emph{detectable}} impacts requires a detection threshold, which we apply to simulated impacts.

To handle the growth of encounters with impact parameter, $N_{\rm enc} \propto b_{\rm max}$, we define detectable impacts relative to a polynomial proxy model, rather than a particular smooth-stream simulation.
Distant encounters cause large-angle variations in the stream, which are in part degenerate with uncertainties in smooth-stream modeling from the underlying potential and mass-loss history.
A polynomial proxy model for the stream can automatically absorb the large-angle perturbations in the stream, allowing us to handle both uncertainties in modeling and convergence of impacts  with $b_{\rm max}$.
We then define detectable impacts as statistically significant deviations from the best-fit third-order polynomial for the stream.
We demonstrate that without this subtraction, many detectable impacts may lie outside the commonly used cut of $b < 5 r_s$ where $r_s$ is the scale-radius of a subhalo. With the subtraction, detectable impacts are concentrated within  $b < 5 r_s$.

Our main finding is that with data from LSST and Via, 5 out of the 14 streams we study have an expected number of  detectable impacts above 0.2 (see Fig.~\ref{fig:vs_dm_models} for our main results).
Jet~\citep{2018MNRAS.480.5342J} is the standout with $5.15^{+1.10}_{-0.95}$ detectable impacts, followed by Orphan-Chenab ($1.40^{+0.62}_{-0.47}$), ATLAS-Aliqa Uma ($1.25^{+0.60}_{-0.44}$), GD-1 ($0.55^{+0.43}_{-0.28}$), and Palomar 5 ($0.40^{+0.39}_{-0.23}$). These have already been identified as promising for subhalo detection~\citep{Erkal_2016,menker2024,Barry:2023ksd,Adams:2024zhi,hilmi2024inferringdarkmattersubhalo,
Ferguson_2021,lu2025detectabilitydarkmattersubhalo,2026arXiv260413374D}. Jet benefits from a number of effects, including having the largest $T_{\rm form}$ of 4.8 Gyr among the 14 streams we study, as well as large physical length $l$.

These values are idealized forecasts under the assumption of a static Milky Way potential and an analytic CDM subhalo mass function, and should be read as best-case estimates. The subhalo population is a large source of uncertainty in our predictions, giving a factor of few effect.
We also consider effects of different DM models on the number of impacts, finding a suppression for warm dark matter and fuzzy dark matter models and an enhancement for an SIDM-like scenario.

The remainder of the paper is organized as follows. Sec.~\ref{sec:stream_catalog} describes the stream catalog, our selection criteria, and the determination of stream ages. Sec.~\ref{sec:modeling} details our simulated mock observations, and introduces the polynomial proxy model used for smooth streams. Sec.~\ref{sec:impacts} describes our subhalo impact pipeline, involving a three-layer selection starting from sampling all impacts with a loose $b_{\rm max}$ and subhalo mass cut, and ending with a high-statistics simulation of impacts for accurate assessment of detectability. We present detailed results in Sec.~\ref{sec:results}, where we show the distribution of encounters as well as the impact of the polynomial subtraction method. We also assess uncertainties in our predictions under different assumptions for CDM subhalos as well as alternate DM models.  We conclude in Sec.~\ref{sec:conclude}.

\section{Stream catalog and selection}
\label{sec:stream_catalog}

In this section, we introduce the catalog of stellar streams on which we base our subhalo-impact analysis. We summarize the relevant observational data contained in the catalog and describe how we obtain values for the progenitor mass and age of each stream. These serve as inputs for simulating the time-evolution of a stream and calculating the number of subhalo impacts over its lifetime. To ensure that the streams in our analysis are modeled with sufficient accuracy, we specify a set of stream selection criteria, which are summarized at the end of this section. These define the set of 14 streams studied in this paper and are listed in Table~\ref{tab:selected_streams}.

Our starting point is the catalog of 131 stellar streams assembled in
\cite{BONACA2025101713}, which is based in part on the \texttt{galstreams} package \citep{Mateu_2023}. These streams have been identified through methods based on matched-filtering in color-magnitude space and incorporating proper motions with the STREAMFINDER algorithm \citep{Ibata_2023_STREAMFINDER}, and through other Gaia-based studies~\cite{Awad_2024,Bonaca2020,Chandra_2022,Ferguson_2021,Grillmair_2019,Grillmair_2022,Huang_2019,10.1093/mnras/stad551,S5:2020tao,Shipp_2019,Shipp_2020,Yang_2022,Yang_2023,10.1093/mnras/staa3673}.
The catalog contains information about the distances and sky coordinates as well as radial and angular velocities of each stream. Moreover, if known, it provides the total present-day stellar mass $M_\mathrm{stellar}$ of the stream and the type of its progenitor, which may be either a globular cluster (GC) or a dwarf galaxy.

In the following we will not consider streams with unknown $M_\mathrm{stellar}$ or angular length of less than 20 degrees. We also skip the following three streams. Omega Centauri has an unusual morphology and overlapping arms (see {\emph e.g.,} \citealt{BONACA2025101713}), which render it unsuitable for our analysis. We also skip C-19 due to its relatively uncertain origin and characterization: its length and stellar mass as given in \cite{BONACA2025101713} are significantly different from other estimates in the literature such as in \cite{2025A&A...698A..82Y}, and furthermore, it has properties consistent with both a GC origin as well as a dwarf galaxy \citep {2022MNRAS.514.3532E,2025ApJ...988...96C}. Nevertheless, we emphasize that its potential high stellar mass and stream length \citep{2025A&A...698A..82Y} could lead to a very old stream with high subhalo encounter rate, making C-19 a promising stream for DM subhalo detection.
We also skip New-25 since its best-fit orbit is unbound (total orbital energy $E_{\rm tot} > 0$ in \cite{BONACA2025101713}).

\subsection{Progenitor mass and stream age}
\label{sec:stream_age}

\begin{figure*}[t]
\centering
\includegraphics[width=0.75\textwidth]{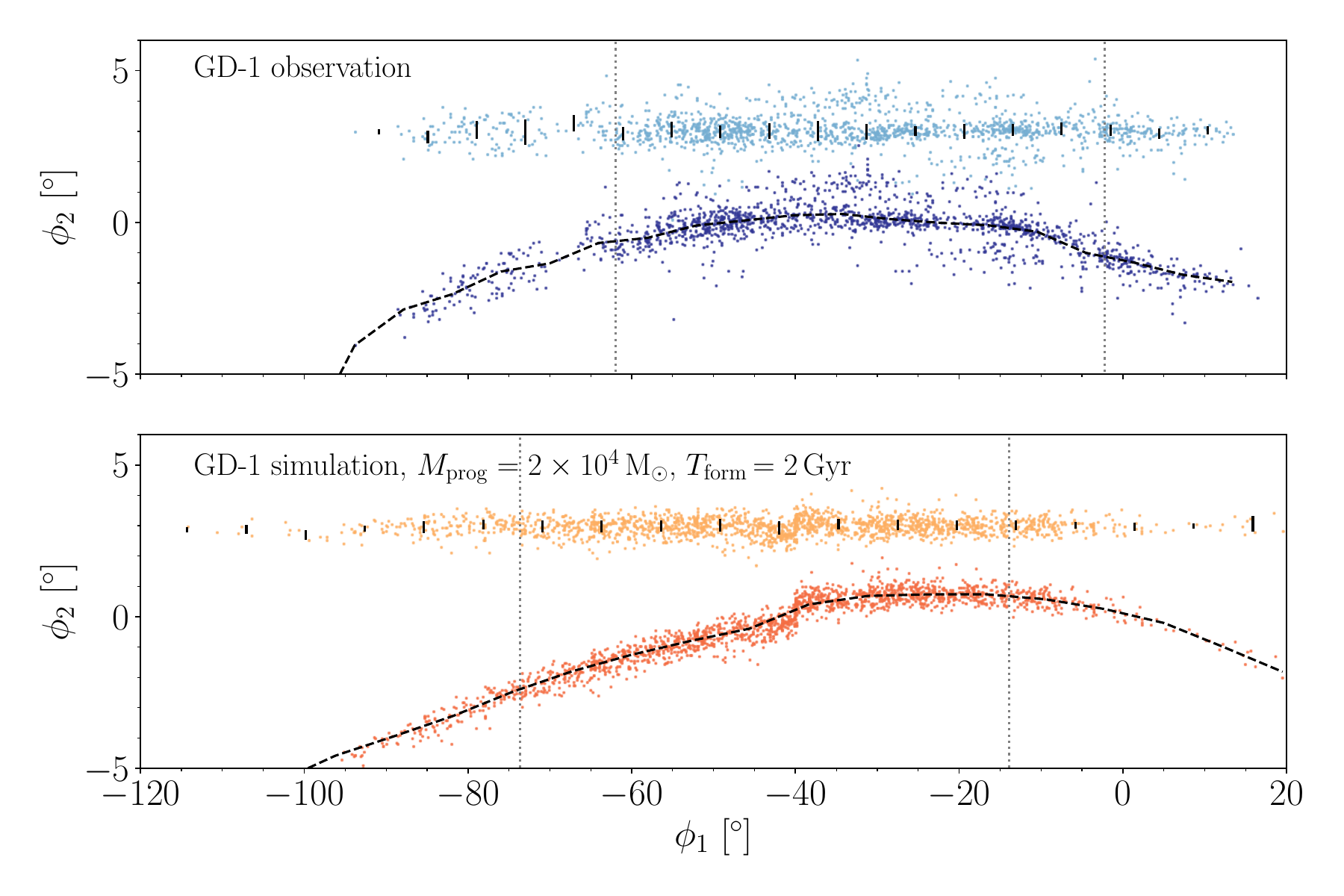}
\caption{Illustration of the stream length and width measurements used for fitting the stream age $T_\mathrm{form}$ and checking the accuracy of our simulation (see Sec.~\ref{sec:stream_catalog} for details). As an example, we show the observed GD-1 stream (top panel, dark blue) and its simulation with the best-fit value of $T_\mathrm{form}=2$~Gyr (bottom panel, dark orange) in the stream-aligned coordinate system. No subhalo impacts are included in the simulation at this point. Vertical gray dotted lines mark the 10$^\text{th}$ and 90$^\text{th}$ percentile of star positions in $\phi_1$, defining the 90$^\text{th}$-to-10$^\text{th}$-percentile length $\ell^{90-10}$. At an offset, we show the segment-wise rotated streams (light blue and light orange) used for measuring the stream widths. The MAD width in each bin is indicated by vertical black lines.}
\label{fig:stream_fitting}
\end{figure*}

In addition to the observational data in the catalog, modeling a stream requires knowledge of its age $T_\mathrm{form}$ and the original mass of its progenitor $M_\mathrm{prog}$. Here, we set $M_\mathrm{prog}$ based on previous estimates from the literature and stellar masses in the catalog, while we determine $T_\mathrm{form}$ by fitting simulated mock streams to data.

For setting $M_\mathrm{prog}$, we apply different procedures depending on the type of progenitor: For globular cluster progenitors which have not fully dissolved, we set the original progenitor mass to be $M_\mathrm{prog} = M_{\mathrm{prog},0} + M_\mathrm{stellar}$, where $M_{\mathrm{prog},0}$ is the present-day mass of the progenitor from \cite{Baumgardt_2023}. For fully dissolved globular cluster progenitors or dwarf galaxy progenitors, we use $M_\mathrm{prog}$ values from previous studies. The streams in this category (with the respective reference for their progenitor mass) are 300S~\citep{cohen2025siftingstreammorphology300s}, ATLAS-Aliqa Uma~\citep{S5:2020tao}, GD-1~\citep{Nibauer:2025ezn}, Indus~\citep{DES:2018imd}, Leiptr~\citep{Atzberger2025Chemical}, Orphan-Chenab~\citep{Mendelsohn:2022jjm}, and Wukong~\citep{10.1093/mnras/stae969}. For 300S, ATLAS-Aliqa Uma and GD-1, these values are similar to the catalog values of $M_\mathrm{stellar}$ but have been updated in some instances with more recent studies.

If the progenitor type is unknown, we set $M_\mathrm{prog} = M_\mathrm{stellar}$, implicitly assuming that the stream originated from a fully dissolved globular cluster. For Jet, whose stellar mass is not included in \cite{BONACA2025101713}, we use the value from \cite{Ferguson_2021}. The resulting progenitor masses for all streams considered here are listed in Table~\ref{tab:fit_all_streams}. Note that Wukong's  progenitor is estimated to be extremely heavy, with $M_\mathrm{prog} \sim 10^{10}~M_\odot$, which makes it challenging to reproduce the observed stream in simulation. Hence, we drop Wukong from our analysis at this point.

Having fixed the progenitor masses as described, we determine the age $T_\mathrm{form}$ of each stream by simulating mock streams of varying ages with \textsc{Gala}~\citep{gala,adrian_price_whelan_2020_4159870} and fitting their length to observations.
To simulate a stream with a given $T_\mathrm{form}$, we first integrate the progenitor orbit backwards in the Milky Way potential \texttt{MilkyWayPotential2022} from the time of observation $t=0$ to $t=T_\mathrm{form}$.  We then use the particle spray method of \cite{Fardal_2015} as implemented in \textsc{Gala} to evolve the stream forward in time to $t=0$. Note that throughout the work, we use lookback time $t$.

We determine the total number of stars to be released in the simulation from the stellar mass $M_\mathrm{stellar}$ of the stream and the initial mass function (IMF)~\citep{Chabrier:2001dc}. We implement this using the package \texttt{imf}\footnote{\url{https://github.com/keflavich/imf}} for the initial mass function and package \texttt{minimint}~\citep{minimint} for the MIST isochrone. Moreover, in determining the number of stars, we assume a metallicity of $-2$ and an age of 10~Gyr for each stream. While these are rough approximations, we find that their effect on our fit results is negligible.
To specify the number of particles released at each step, we adopt a mass-loss model following~\cite{2010MNRAS.409..305L}, where%
\begin{equation}
\label{eq:mass-loss}
\frac{dM(t)}{dt} =\alpha M(t)^{1-\gamma} \, ,
\end{equation}
with $M(t=0)=M_{\mathrm{prog},0}$. We set $\gamma=1$, corresponding to a constant mass-loss rate, and fix $\alpha$ by the condition $M(t=T_\mathrm{form})=M_\mathrm{prog}$. (For fit results assuming a range of different values of $\gamma$ see Fig.~\ref{fig:stream_age_gamma}.)

We then adjust our simulation to approximately reproduce Gaia~\citep{GaiaEDR3} observations by excluding simulated stream stars which are too faint to be observed. To this end, we use the same IMF and MIST isochrones~\citep{Dotter_2016, Choi_2016} to assign an absolute magnitude to each star in the stream, here using metallicity from the latest literature~\citep{Koposov_2026, Ibata_2023_STREAMFINDER, Li_2022_S5}, as shown in Table~\ref{tab:selected_streams}. Combined with the distance of each star, we convert these absolute magnitudes to apparent magnitudes.
We then eliminate stars with apparent magnitude above the Gaia detection limit of 20.7.

To fit the stream length, we transform the positions of stars in the simulated stream into the coordinate system defined by the stream-aligned longitude $\phi_1$ and the corresponding latitude $\phi_2$, where we use the coordinate system defined in \cite{BONACA2025101713}. We measure the 90$^\text{th}$-to-10$^\text{th}$ percentile stream length as $\ell^{90-10} \equiv \phi_1^{90}-\phi_1^{10}$. Here $\phi_1^{90}$ and $\phi_1^{10}$ denote the 90th and 10th percentile of the distribution of stars in $\phi_1$, respectively. We then find the best-fit value of the stream age $T_\mathrm{form}$ by requiring
\begin{equation}
    \ell^{90-10}_\mathrm{sim}(T_\mathrm{form}) = \ell^{90-10}_\mathrm{data} \, ,
\label{eq:length_fit}
\end{equation}
where $\ell^{90-10}_\mathrm{sim}$ is the length of the simulated stream and $\ell^{90-10}_\mathrm{data}$ is the analogously defined length of the observed stream, computed using the stream members compiled in \cite{BONACA2025101713}. An illustration of this stream length is shown in Fig.~\ref{fig:stream_fitting}, where the gray dotted lines indicate the locations of  $\phi_1^{90}$ and $\phi_1^{10}$.
In case of multiple solutions, we choose the smallest $T_\mathrm{form}$ that solves Eq.~\ref{eq:length_fit}. The thus obtained best-fit values of $T_\mathrm{form}$ for all streams considered in the fit are shown in Table~\ref{tab:fit_all_streams}. The stream ages we find are in reasonable agreement with results from previous studies, see e.g.\ \cite{Erkal_2016, Chen_2025}. Note, however, that the ages we infer for GD-1 and Palomar 5 are somewhat lower than previously assumed in \cite{Erkal_2016, Kuepper_2015}.

\begin{figure*}[t]
\centering
\includegraphics[width=\textwidth]{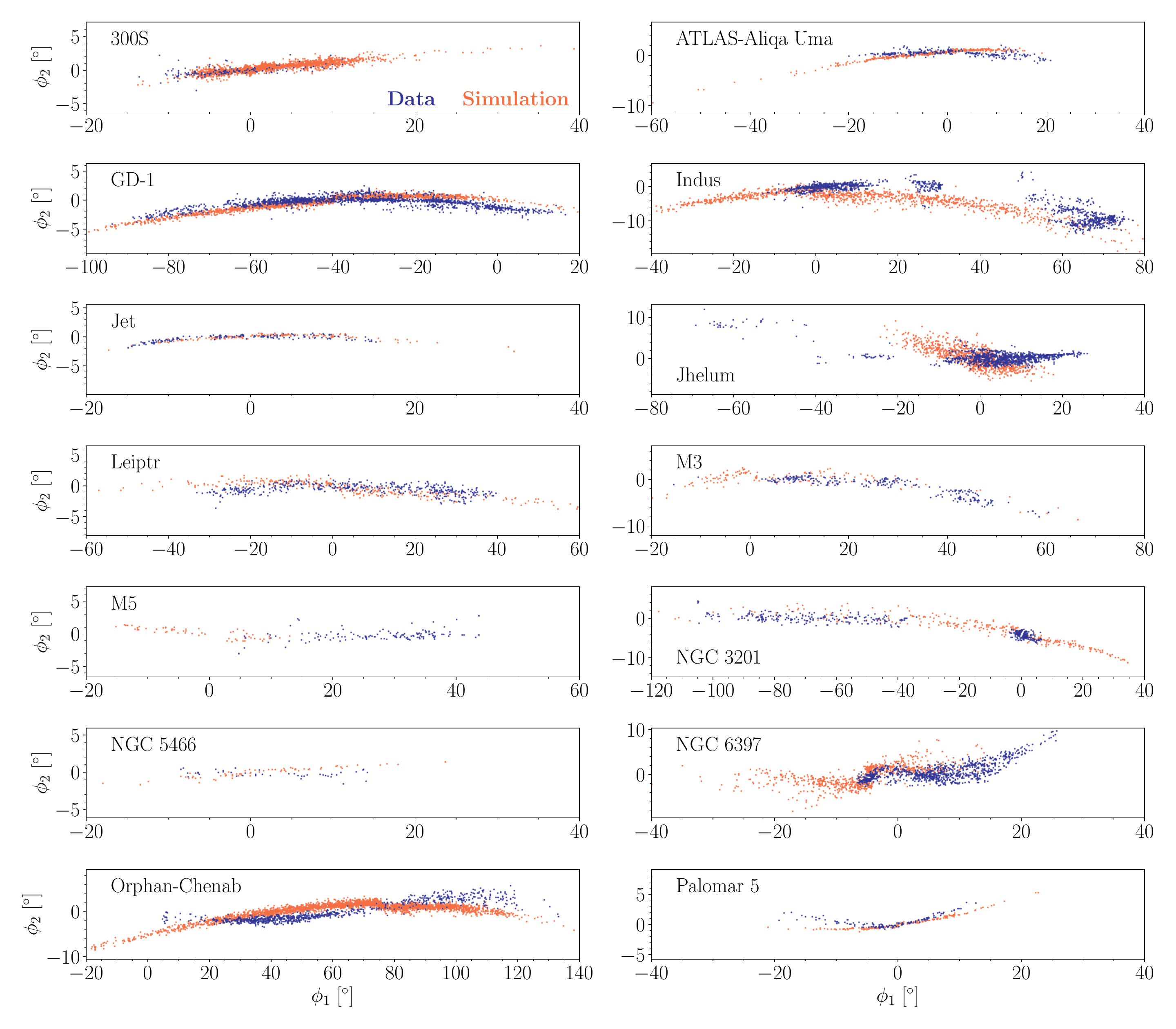}
\caption{Sky plots of the streams included in our analysis based on the selection criteria described in Sec.~\ref{sec:stream_catalog}, shown in their associated coordinate systems defined by the stream-aligned longitude $\phi_1$ and latitude $\phi_2$. Observed streams are shown in blue, simulated streams in orange. The properties of the shown streams are summarized in Table~\ref{tab:selected_streams}. Here, a magnitude cutoff of $G < 20.7$ (corresponding to Gaia sensitivity) is applied to the simulated streams. For expected LSST10 sky plots see Fig.~\ref{fig:sky_plots_lsst10} in Appendix~\ref{app:extras}.
}
\label{fig:sky_plots}
\end{figure*}

\begin{table*}[t]
	\begin{center}
		\begin{tabular}{cccccccccc}
            \toprule
            Name & Progenitor Type & $M_\mathrm{prog}$ & $M_{\mathrm{prog},0}$ & $M_\mathrm{stellar}$ & [Fe/H] & $\ell^{90-10}$ & $T_\mathrm{form}$ & $w^\mathrm{sim}$ & $w^\mathrm{data}$\\
            \addlinespace[3pt]
             &  & [$M_\odot$] & [$M_\odot$] & [$M_\odot$] &  & [$^\circ$] & [Gyr] & [$^\circ$] & [$^\circ$]\\
            \midrule
            300S & GC & $5.0 \times 10^4$ & $0$ & $5.0 \times 10^4$ & $-1.4$ & 15 & 1.1 & 0.19 & 0.24 \\
            ATLAS-Aliqa Uma & GC & $2.0 \times 10^4$ & $0$ & $2.0 \times 10^4$ & $-2.2$ & 25 & 3 & 0.11 & 0.18 \\
            GD-1 & GC & $2.0 \times 10^4$ & $0$ & $2.0 \times 10^4$ & $-2.2$ & 60 & 2 & 0.18 & 0.23 \\
            Indus & dwarf & $6.5 \times 10^6$ & $0$ & $3.4 \times 10^4$ & $-2.1$ & 72 & 0.64 & 0.69 & 0.75 \\
            Jet & unknown & $2.5 \times 10^4$ & $0$ & $2.5 \times 10^4$ & $-2.4$ & 22 & 4.8 & 0.11 & 0.15 \\
            Jhelum & dwarf & $1.3 \times 10^7$ & $0$ & $1.7 \times 10^4$ & $-2.1$ & 23 & 0.14 & 0.95 & 0.58 \\
            Leiptr & dwarf & $1.0 \times 10^5$ & $0$ & $3.0 \times 10^3$ & $-2.2$ & 54 & 0.89 & 0.32 & 0.36 \\
            M3 & GC & $4.1 \times 10^5$ & $4.1 \times 10^5$ & $2.0 \times 10^3$ & $-1.7$ & 39 & 0.66 & 0.5 & 0.43 \\
            M5 & GC & $3.9 \times 10^5$ & $3.9 \times 10^5$ & $7.1 \times 10^2$ & $-1.4$ & 22 & 0.19 & 0.24 & 0.28 \\
            NGC 3201 & GC & $2.0 \times 10^5$ & $1.9 \times 10^5$ & $2.1 \times 10^3$ & $-1.6$ & 81 & 0.6 & 0.44 & 0.39 \\
            NGC 5466 & GC & $5.8 \times 10^4$ & $5.6 \times 10^4$ & $1.9 \times 10^3$ & $-2.1$ & 19 & 0.97 & 0.087 & 0.12 \\
            NGC 6397 & GC & $8.5 \times 10^4$ & $8.2 \times 10^4$ & $2.5 \times 10^3$ & $-2$ & 20 & 0.13 & 0.74 & 0.57 \\
            Orphan-Chenab & dwarf & $1.1 \times 10^6$ & $0$ & $1.3 \times 10^5$ & $-2$ & 76 & 1.9 & 0.32 & 0.39 \\
            Palomar 5 & GC & $3.0 \times 10^4$ & $1.3 \times 10^4$ & $1.7 \times 10^4$ & $-1.2$ & 18 & 2 & 0.07 & 0.14 \\
            \bottomrule
		\end{tabular}
	\end{center}
	\caption{Streams passing the selection described in Sec.~\ref{sec:stream_catalog}. Shown are the progenitor type, the estimated original progenitor mass $M_\mathrm{prog}$, the mass of the progenitor today $M_{\mathrm{prog}, 0}$,  the stellar mass $M_\mathrm{stellar}$, the metallicity [Fe/H], the 90$^\mathrm{th}$-to-10$^\mathrm{th}$-percentile length $\ell^{90-10}$, the best-fit stream age $T_\mathrm{form}$, the width of the resulting simulated stream $w_\mathrm{sim}$, and the width of the observed stream $w_\mathrm{data}$ (see main text for details). \label{tab:selected_streams}}
\end{table*}

\subsection{Stream selection}

After simulating each stream with the resulting best-fit value of $T_\mathrm{form}$, we again compare the simulated stream to observation to make sure that all streams included in our later analysis are well-modeled. Since the simulated and observed stream agree in length by definition in our fit, we make use of the stream width as an independent check. If the widths of the simulated and observed stream differ greatly, we consider the stream as not well-characterized and do not include it in our study.

For the purpose of this check, we need a measure to characterize the stream width. The catalog widths in \cite{BONACA2025101713} are defined to be either the scaled median absolute deviation (MAD) of the $\phi_2$ coordinates of all stars in the stream (given by $1.4826 \times \mathrm{median}(|\phi_2 - \mathrm{median}(\phi_2)|)$) or their standard deviation. However, these definitions do not account for the stream curvature.
To take this curvature into consideration, we divide each stream into 20 segments along $\phi_1$ (or fewer to ensure a minimum segment length of 2$^\circ$). We then rotate the star coordinates in each individual segment such that the local stream direction is aligned with $\phi_1$. Subsequently, we measure the local stream width as the scaled MAD of the $\phi_2$ coordinates of the stars in the segment.  We carry out this segmentation and width measurement separately for the simulated and the observed stream, as illustrated for GD-1 in Fig.~\ref{fig:stream_fitting}. We then define the stream width as the weighted average
\begin{equation}
    w \equiv \frac{1}{N_\mathrm{star}} \sum_{i=1}^{N_\mathrm{seg}} \, n_{\mathrm{star}, i} \, w_i \, ,
    \label{eq:weighted_width}
\end{equation}
with $N_\mathrm{seg}$ denoting the number of segments, $n_{\mathrm{star}, i}$ the number of stars in segment $i$, $w_i$ the scaled MAD width in segment $i$, and $N_\mathrm{star}$ the total number of stars in the stream. We only keep streams whose simulated and observed widths, $w_\mathrm{sim}$ and $w_\mathrm{data}$, differ by less than a factor of 2, \emph{i.e.},\ streams with $\frac{1}{2} w_\mathrm{data} < w_\mathrm{sim} < 2 w_\mathrm{data}$. This criterion eliminates almost all streams with unknown progenitor type, except Jet.

In summary, in this section we have narrowed down the catalog of 131 streams from \cite{BONACA2025101713} in the following steps:

\begin{itemize}
    \item Eliminate 58 streams with catalog length $< 20^\circ$.
    \item Eliminate 20 streams for which no stellar mass is given (except Jet, for which we use a value from \citealt{Ferguson_2021}).
    \item Eliminate Omega Centauri due to its unusual morphology, C-19 due to the large uncertainty in its stellar mass and stream length, New-25 due to its unbound orbit, and Wukong due to its large progenitor mass (see above for details).
    \item Eliminate 35 streams for which $w_\mathrm{sim} < w_\mathrm{data}/2$ or $w_\mathrm{sim} > 2 w_\mathrm{data}$, where $w_\mathrm{sim}$ and $w_\mathrm{data}$ denote the widths (as measured with Eq.~\ref{eq:weighted_width}) of the simulated and observed streams, respectively.
\end{itemize}

At the end of this process, we are left with 14 streams, which are listed in Table~\ref{tab:selected_streams}.  Our selection retains all confirmed GC streams longer than 20 degrees (except M2, M68, NGC 288, and Omega Centauri) in addition to several dwarf galaxy and unknown progenitor type streams.

In Table~\ref{tab:fit_all_streams} in App.~\ref{app:extras} we show the estimated number of subhalo encounters $N_\mathrm{enc}^\mathrm{circ}$, given by Eq.~\ref{eq:Nenc}, for the 14 selected streams as well as the 35 streams that do not meet the final selection criterion. The physical length is estimated by $l=r_h \ell^\mathrm{90-10}$, and the subhalo density $n_\mathrm{sub}$ is evaluated at the time-averaged distance to Galactic Center $\left<r_\mathrm{gc}\right>=\frac{r_\mathrm{apo}+r_\mathrm{peri}}{2}(1+\frac{ecc^2}{2})$ for a Keplerian elliptical orbit. While at least a few of these 35 streams have $N_\mathrm{enc}^\mathrm{circ}$ comparable to the selected 14, they generally have lower stellar mass, which makes them less promising targets for a subhalo-impact analysis. A possible exception is Kwando; however, the simulated width is a factor of 6 smaller than the width of the observed stars, indicating the need for improved modeling.

The sky plots of observed and simulated streams of the 14 streams are shown in Fig.~\ref{fig:sky_plots}. The simulated streams have been generated with the best-fit $T_{\rm form}$ and with a magnitude cut corresponding to the Gaia detection limit, as described in Sec.~\ref{sec:stream_age}. Our simulated streams largely have similar number density and length as the catalog of observed stream members. However, some streams clearly have a more complex morphology not captured by our setup (notably Indus and Jhelum). In addition, our simulation of the stream 300S gives a much higher number density, with roughly an order of magnitude more stars. This can be traced to the updated stellar mass we use, which is based on completeness of stream members in the $S^5$ survey \citep{2024MNRAS.529.2413U,cohen2025siftingstreammorphology300s}; the difference here may reflect an incomplete catalog of stream members with respect to Gaia.

\section{Smooth-stream modeling}
\label{sec:modeling}

We now summarize our method for generating mock observational data (Sec.~\ref{sec:mock_data}) and introduce polynomial proxy models for the smooth streams (Sec.~\ref{sec:proxy_model}). In principle, we could define a detectable subhalo impact as a statistically significant perturbation away from a simulated smooth stream. However, there are two main challenges to this approach: first, %
the number of subhalo encounters increases with impact parameter, Eq.~\ref{eq:Nenc}, so naively summing over arbitrarily distant subhalos causes the total number of detectable impacts to diverge, and second, comparing to a fiducial simulated smooth stream does not account for uncertainties in smooth-stream modeling itself, which could be significant. These two effects are related: distant subhalo encounters produce large angular scale perturbations in the stream, which are indistinguishable from uncertainties in the smooth-stream model itself. Properly accounting for these modeling uncertainties thus renders the number of detectable impacts finite. Here we introduce simple polynomial proxy models for smooth streams, which can absorb large-angle perturbations from these effects. We can then define a detectable subhalo impact as a deviation from the best-fit proxy model.

By defining a detectable subhalo impact relative to the proxy models, our results are more robust to the details of stream modeling and arbitrary cutoffs in impact parameter for subhalo encounters. This approach also mimics a more data-driven approach. For instance, previous studies of perturbations in GD-1 also subtracted the best-fit polynomial model~\citep{Banik:2019smi}. Since not all streams can be described by a polynomial over the full length, we will identify angular regions of the smooth stream that are well-described by the proxy model. These are the regions that will be used for subhalo-impact analysis. This naturally has the effect of removing regions near the progenitor where epicyclic density perturbations~\citep{2008MNRAS.387.1248K,2012MNRAS.420.2700K} are expected to be present. For the streams passing our selection criteria in Sec.~\ref{sec:stream_catalog} and listed in Tab.~\ref{tab:selected_streams}, we show in Sec.~\ref{sec:proxy_model} that, on average, they are well-described by the proxy model over an angular region that is $\sim80\%$ of the original angular length.

\subsection{Mock data for smooth streams}
\label{sec:mock_data}

For our mock observations for unperturbed streams, we use the particle spray model from~\cite{Fardal_2015} to simulate smooth streams, using $M_\mathrm{prog}$, $M_{\mathrm{prog},0}$, $T_\mathrm{form}$, and $M_\mathrm{stellar}$ as shown in Tab.~\ref{tab:selected_streams}. The stellar mass $M_\mathrm{stellar}$ is used to determine the total number of stars in the stream through the initial mass function (IMF)~\citep{Chabrier:2001dc}. The particle spray model outputs the phase space information for each stream star at the time of observation, including the angular position $\phi_1$, $\phi_2$ in the stream coordinates, as well as the radial velocity $v_r$.

Based on the results of  \cite{lu2025detectabilitydarkmattersubhalo}, the three most important observables for detecting a subhalo impact are: the density along the stream $\rho$, the transverse angle $\phi_2$ and the radial velocity $v_r$. We will only consider these three observables in this work. For most of our analysis, we assume the best-case observational scenario, where the angular positions are from LSST~\citep{lsstsciencebook2009} 10-year (LSST10) data with magnitude cutoff $r<27$, and the radial velocities are from Via~\citep{theviacollaboration2026projectoverviewscienceinstrument} spectroscopic survey with magnitude cutoff $G<24$. As discussed in \cite{lu2025detectabilitydarkmattersubhalo}, this neglects effects such as contamination from galaxies. As a comparison, we will also perform our analysis assuming a present-day observational scenario, where the angular positions are from Gaia~\citep{GaiaEDR3} with $G<20.7$ and the radial velocities are from DESI~\citep{Cooper_2023} with $G<20$. %

For each simulated stream we carry out the following steps:
\begin{enumerate}
    \item
    For each star, we keep $\phi_1$, $\phi_2$ only if the apparent magnitude is below the cutoff for position data, and keep $v_r$ only if the apparent magnitude is below the cutoff for the radial velocity for the observational scenario being considered.
    \item For each visible star, we add a random Gaussian noise on the radial velocity to mimic the observational error:
    \begin{equation}
        v_r\rightarrow v_r + V_r \, ,
    \end{equation}
    where $V_r$ is a Gaussian random variable with zero mean and standard deviation $\sigma_{v_r, \mathrm{obs}}$.
    The level of error $\sigma_{v_r, \mathrm{obs}}$ as a function of apparent magnitude is based on the spectroscopic survey being considered. For Via~\citep{theviacollaboration2026projectoverviewscienceinstrument}, we use the combined statistical and systematic errors for the red giant branch (RGB) assuming metallicity [Fe/H]=-2 and 10 hours of observation. For DESI~\citep{2024MNRAS.533.1012K}, we download the dataset from DESI Data website\footnote{\url{https://data.desi.lbl.gov/doc/releases/edr/vac/mws/}}, and extract the mean errors from stars with metallicity [Fe/H] between -2.5 and -1.5. The observational errors for the position data $\phi_1$, $\phi_2$ ($\sim 10^{-5} \deg$) are negligible compared to the stream dispersion ($\sim 0.1 \deg$)~\citep{lsstsciencebook2009}.
\end{enumerate}

This process generates mock data for individual stars with observational errors. We further aggregate the data by binning the stars into 1-degree $\phi_1$ bins.
We label the bins by $i$ and denote the bin center by $\phi_{1,i}$. For each bin, we compute the average $\phi_{2,i}$ and $v_{r,i}$, as well as the density $\rho_i$, which is effectively the number of observed stars in bin $i$.
For $v_{r,i}$, we use the inverse-variance weighted average
\begin{equation}
    v_{r,i} = \frac{\sum_j v_{r,ij}/(\sigma^2_{v_{r,i,\mathrm{disp}}}+\sigma^2_{v_{r,ij,\mathrm{obs}}})}{\sum_j 1/(\sigma^2_{v_{r,i,\mathrm{disp}}}+\sigma^2_{v_{r,ij,\mathrm{obs}}})}
\end{equation}
to give higher weight to stars with smaller observational error. Here the index $j$ runs over the stars within bin $i$, and $\sigma^2_{v_{r,i,\mathrm{disp}}}$ and $\sigma^2_{v_{r,ij,\mathrm{obs}}}$ denote the intrinsic stream dispersion and the observational error, respectively.

Going forward, we will use $X_i$ to denote the stream observables, $\phi_{2,i}$, $v_{r,i}$ or $\rho_i$, and $X^{\rm smooth}_i$ to indicate observables from a simulated smooth stream.

\subsection{Proxy models for smooth streams}
\label{sec:proxy_model}

\begin{figure}[t]
\centering
\includegraphics[width=\columnwidth]{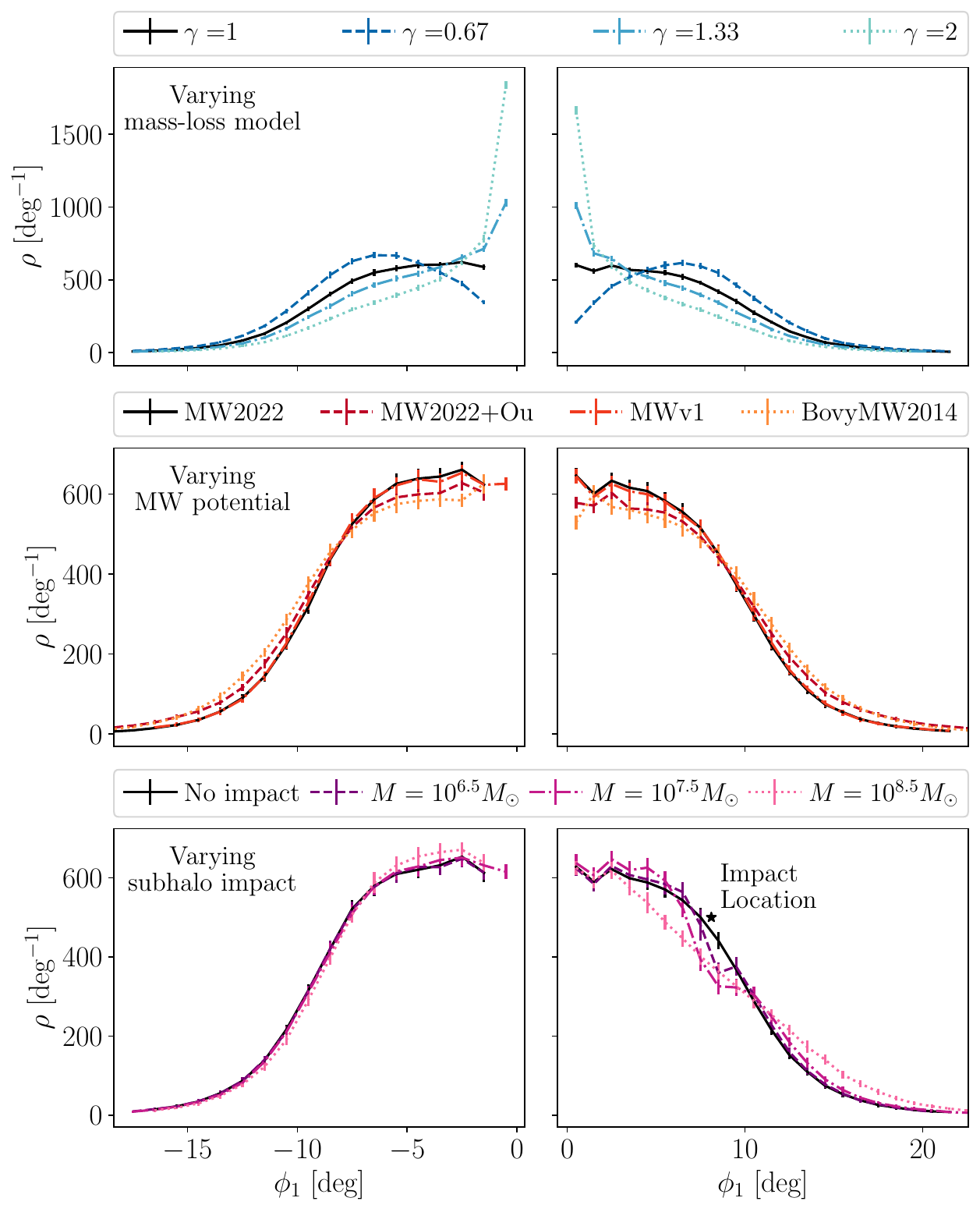}
\caption{ Comparison of stream modeling uncertainties to a single subhalo impact. \textbf{Top}: Unperturbed Jet stream with varying mass-loss model, parameterized by $\gamma$ as in Eq.~\ref{eq:mass-loss}. Our fiducial value, $\gamma=1$, is shown in black.
\textbf{Middle}: Unperturbed Jet stream with varying Milky Way potential, including \texttt{gala.MilkyWayPotential2022} (default in black)~\citep{Eilers_2019,darraghford2023textttescargotmappingverticalphase},
\texttt{gala.MilkyWayPotential2022} with the dark halo replaced according to ~\cite{ou2025decodinggalactictwirldownfall},
\texttt{gala.MilkyWayPotential("v1")}, whose disk model is based on~\cite{2015ApJS..216...29B},
and \texttt{gala.BovyMWPotential2014}~\citep{2015ApJS..216...29B}.
\textbf{Bottom}: Smooth Jet stream (black) compared to perturbations by different DM subhalo impacts.  The impact is located at the center of the right arm (marked by a black star). The impact parameter is chosen to be $b=1.05 \times (M/10^8 M_\odot)^{0.5}$ kpc. In all cases, the average density over 50 stream realizations $\langle \rho_i \rangle$ is shown, with error bars corresponding to the per-bin uncertainty $\sigma_{\rho,i}$ for LSST 10-year sensitivity.}
\label{fig:mass-loss-large-scale-sim}
\end{figure}

We now illustrate the two issues with defining subhalo impacts relative to the mock-generated smooth streams -- subhalo encounters at large impact parameter $b$ and uncertainties in smooth-stream modeling -- as well as the polynomial proxy model for handling these.

In Fig.~\ref{fig:mass-loss-large-scale-sim}, the top row shows how the smooth-stream model for Jet is affected by variations in the mass-loss model, which mostly impacts the density along the stream. We vary the parameter $\gamma$ in the mass-loss model (see Eq.~\ref{eq:mass-loss}), where $\gamma=1$ corresponds to a constant mass-loss rate, $\gamma<1$ corresponds to a decreasing mass loss typical for clusters without black holes, and $\gamma>1$ corresponds to an increasing loss rate, appropriate for clusters with black holes~\citep{Gieles_2023}.
Varying $\gamma$ leads to significant differences in the stream on large angular scales.
Beyond the smooth mass-loss model of Eq.~\ref{eq:mass-loss}, there may also be bursts of tidal stripping near pericentric passages~\citep{Palau2025,chen2024improvedparticlesprayalgorithm} and dependence on the stream evaporation time $t_\mathrm{ev}$~\citep{roberts2025stellarstreamsblackholerich}, which further modulates the shape of possible density profiles along the stream.

The middle row of Fig.~\ref{fig:mass-loss-large-scale-sim} is similar to the top row, but we vary the Milky Way potential instead. Again, the difference caused by different Milky Way potentials is greater than the 1$\sigma$ statistical errors assuming LSST 10-year sensitivity. Allowing for time-dependent potentials or other distant Milky Way perturbers introduces additional uncertainties beyond this.
One could attempt to account for all of these effects by profiling (or marginalizing) over them when fitting a stream, but this is computationally very expensive.

The bottom row of Fig.~\ref{fig:mass-loss-large-scale-sim} shows how such large-scale variations in the stream can be partially degenerate with distant subhalo impacts (see Sec.~\ref{sec:impacts} for a description of our subhalo impact simulation). We show different scales of impacts with $(M,~b) = (10^{6.5}\,M_{\odot},~0.19\,\mathrm{kpc})$,
$(10^{7.5}\,M_{\odot},~0.59\,\mathrm{kpc})$, and
$(10^{8.5}\,M_{\odot},~1.87\,\mathrm{kpc})$. The scaling $b\propto \sqrt{M}$ ensures similar gap depths according to the analytic estimation from~\cite{Erkal_2015}. Low-mass nearby subhalos tend to leave localized perturbations, as can be seen for the impacts from $10^{6.5} M_\sun$ and $10^{7.5} M_\sun$ subhalos. However, more distant and massive subhalos $\sim 10^{8.5} M_\sun$ tend to perturb the stream over large angular scales, similar to changes in the mass-loss history or MW potential. %
It is expected that as subhalo impacts become more distant, they become more degenerate with changes in the smooth MW potential.  Furthermore, the number of subhalo impacts grows with impact parameter $b$ for the $\phi_2$ and $v_r$ observables. As we will show quantitatively in Sec.~\ref{sec:convergence_with_b}, the signal strength does not drop sufficiently rapidly to offset the growth of the encounter rate.

We address the considerations above by introducing a polynomial proxy model for smooth streams. By defining subhalo impacts relative to the best-fit polynomial model, we automatically subtract out variations over large angular scales. This enables convergence of detectable subhalo impacts with impact parameter $b$. At the same time, profiling over the proxy models gives an inexpensive way for us to account for the smooth-stream modeling uncertainties, without performing expensive profiling over mass-loss models or MW potentials.

We denote the proxy models $X^{\mathrm{model}}(\phi_1;\bm \theta)$, with $X=\rho, \phi_2, v_r$. These are smooth functions in $\phi_1$ with some free parameters collectively denoted as $\bm \theta$. When comparing to binned data, we evaluate them at the bin centers, $X^{\mathrm{model}}_i(\bm \theta)\equiv X^{\mathrm{model}}(\phi_{1,i};\bm \theta)$.
In general, we find that all three observables $\phi_{2,i}$, $v_{r,i}$ or $\rho_i$ are well-fit by a cubic polynomial
\begin{equation}
    X^{\mathrm{model}}(\phi_1;{\bm \theta}) = a_0 + a_1\phi_1 + \frac{1}{2}a_2\phi_1^2 +  \frac{1}{6}a_3\phi_1^3,
    \label{eq:proxy1}
\end{equation}
where ${\bm \theta} = (a_0,a_1,a_2,a_3)$. The fits are performed on each observable and each arm of the stream independently. We have chosen a third-order polynomial in order to balance the goals of adequately fitting the stream over much of its length, while minimizing the number of free parameters. %

\begin{figure}[t]
\centering
\includegraphics[width=\columnwidth]{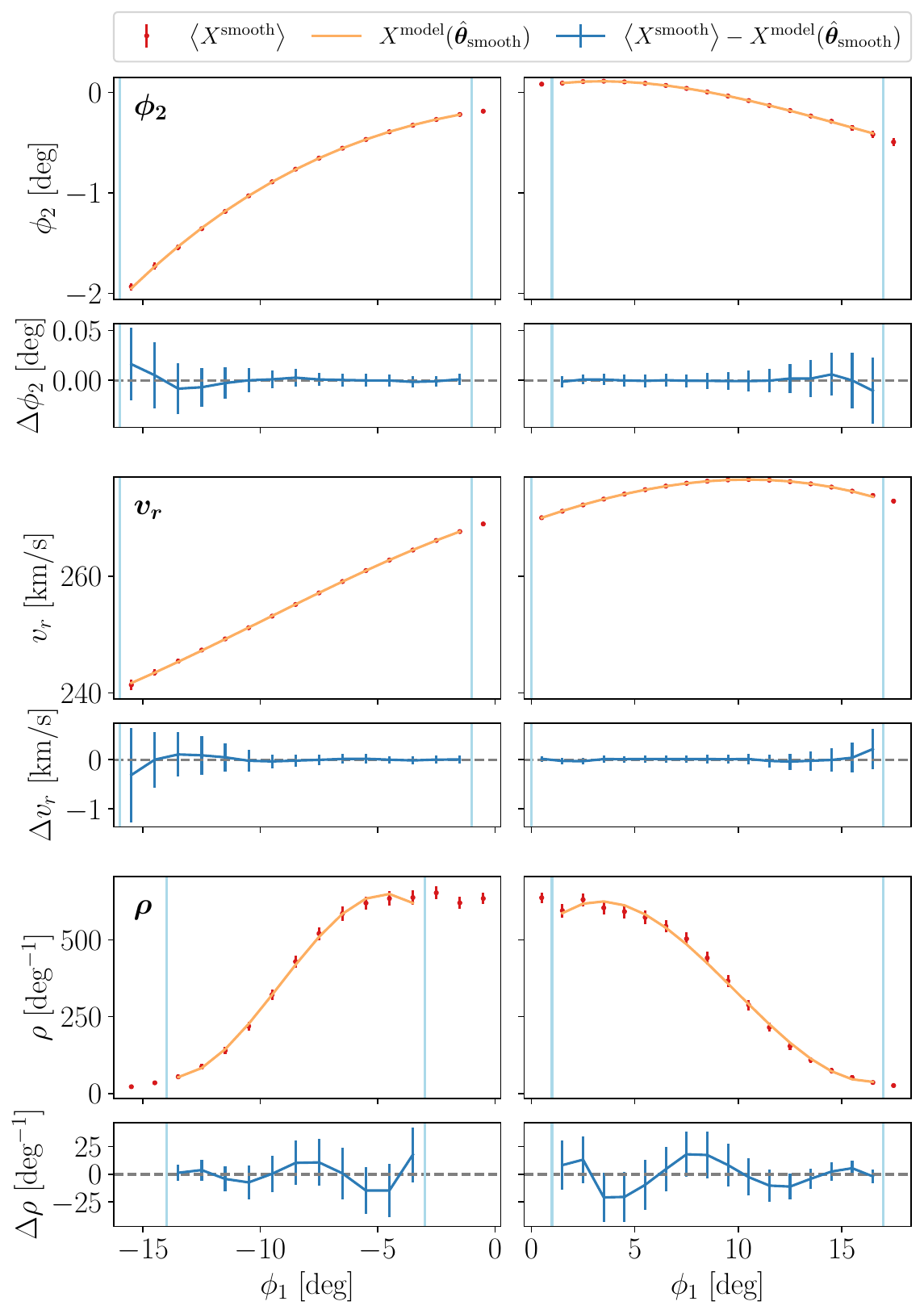}
\caption{Fitting the smooth stream observables $\phi_2$, $v_r$ and $\rho$ for Jet with the proxy model defined in Eq.~\ref{eq:proxy1}. Shown are the average smooth stream over 250 realizations (red points), the best-fit polynomial model (orange lines), and the residual after subtracting the best-fit polynomial model (dark blue lines). The error bars correspond to the per-bin uncertainties $\sigma_{X,i}$. Light blue vertical lines indicate the boundaries of the valid $\phi_1$ regions for later impact analysis, {\it i.e.,} those where the proxy model is a good fit to the smooth stream (satisfying Eq.~\ref{eq:cut}).}
\label{fig:fit_proxy}
\end{figure}

In order to determine the baseline $X^{\mathrm{model}}$ for each stream, we generate 250 realizations of the observables $X_i^\mathrm{smooth}$ from the smooth-stream simulations described in Sec.~\ref{sec:mock_data}. For each observable and each stream arm, we calculate the mean binned observable $\langle X_i^\mathrm{smooth}\rangle$ and the covariance matrix
\begin{equation}
    (C_X)_{ij} \equiv \big\langle (X_i^\mathrm{smooth}-\langle X_i^\mathrm{smooth}\rangle)(X_j^\mathrm{smooth}-\langle X_j^\mathrm{smooth}\rangle) \big\rangle \, ,
    \label{eq:covariance_on_sims}
\end{equation}
where $\langle\cdot\rangle$ denotes the average over 250 smooth stream realizations. Note that we only include the full covariance matrix for $\phi_2, v_r$ and $\rho$ separately, and do not include cross-correlations between observables; we have verified that this is a small effect on our downstream analysis. The best-fit value of the model parameters $\hat{\bm \theta}_\mathrm{smooth}$ for the mean smooth stream $\langle X_i^\mathrm{smooth}\rangle$ is then obtained as
\begin{align}
    \hat {\bm \theta}_\mathrm{smooth}=\argmin_{\bm \theta} \!\! & \sum_{i,j~\mathrm{bins}} \! \left(\langle X_i^\mathrm{smooth}\rangle - X^{\mathrm{model}}_i(\bm \theta)\right)\left(C_X^{-1}\right)_{ij} \nonumber \\ & \qquad\;\;\times \left(\langle X_j^\mathrm{smooth}\rangle-X^{\mathrm{model}}_j(\bm \theta)\right) \,
    \label{eq:theta_smooth}
\end{align}
for each stream arm and observable.

The aim of using the proxy model is to model the large-scale variations, but the polynomial does not necessarily model the entirety of the smooth stream, including the epicyclic density fluctuations near the progenitor.
For each observable, we will only include $\phi_1$ regions where the difference between the smooth stream and the proxy model is within 1$\sigma$ variation of the smooth stream:
\begin{equation}
\label{eq:cut}
|\langle X_i^\mathrm{smooth}\rangle-X^{\mathrm{model}}_i(\hat{\bm \theta}_\mathrm{smooth})| < \sigma_{X,i}.
\end{equation}
Here $\sigma_{X,i}$ is obtained from the diagonal entries of the covariance matrix, $\sigma_{X,i} \equiv \sqrt{(C_X)_{ii}}$. For a given observable of a given stream arm, we use the longest continuous $\phi_1$ region satisfying Eq.~\ref{eq:cut} as the valid region for subhalo impact detection.
Specifically, to find the longest valid region, we start from the original angular length of the arm and gradually reduce the length being considered. For each possible angular length, we apply a sliding window to the original arm and redo the polynomial fit until we find a region satisfying Eq.~\ref{eq:cut}. In order to avoid overfitting with the proxy model, we require at least 5
valid data points in each arm, given that our 3rd order polynomial fit has 4 parameters. All 14 streams in Tab.~\ref{tab:selected_streams} have enough data points in all three observables and both arms after fitting the proxy model. On average, 80.3\% of the original length remains after enforcing the cut. (For comparison, the average length remaining is 84.8\% if a degree 4 polynomial is used and 71.9\% if a degree 2 polynomial is used.)  This indicates that the proxy model we introduced is a reasonable way to model smooth streams.

Fig.~\ref{fig:fit_proxy} illustrates the fitting procedure on all three observables for our fiducial simulation of the Jet stream. The red data points and error bars show the expected values $\langle X_i^\mathrm{smooth}\rangle$ and the per-bin uncertainty $\sigma_{X,i}$ of the observables from mock smooth streams. The orange line is the best-fit proxy model $X^{\mathrm{model}}(\phi_1;\hat{\bm \theta}_\mathrm{smooth})$. Below the fits, the dark blue lines are the residuals $\langle X_i^\mathrm{smooth}\rangle-X^{\mathrm{model}}_i(\hat{\bm \theta}_\mathrm{smooth})$ with error bar $\sigma_{X,i}$. The double vertical light blue lines show the longest valid region well described by the proxy model (satisfying Eq.~\ref{eq:cut}).
Fits for additional streams are shown in Fig.~\ref{fig:smooth_stream_modeling_all} in the appendix, where we see that our criteria Eq.~\ref{eq:cut} removes regions where epicyclic density fluctuations~\citep{2008MNRAS.387.1248K,2012MNRAS.420.2700K} are prominent.
For instance, the epicyclic density fluctuations for GD-1 around $-50^\circ<\phi_1<-30^\circ$ are excluded for the density observable $\rho$. However, this angular region is included for the $\phi_2$ and $v_r$ observables, which do not have such features.

\begin{figure}[t]
\centering
\includegraphics[width=\columnwidth]{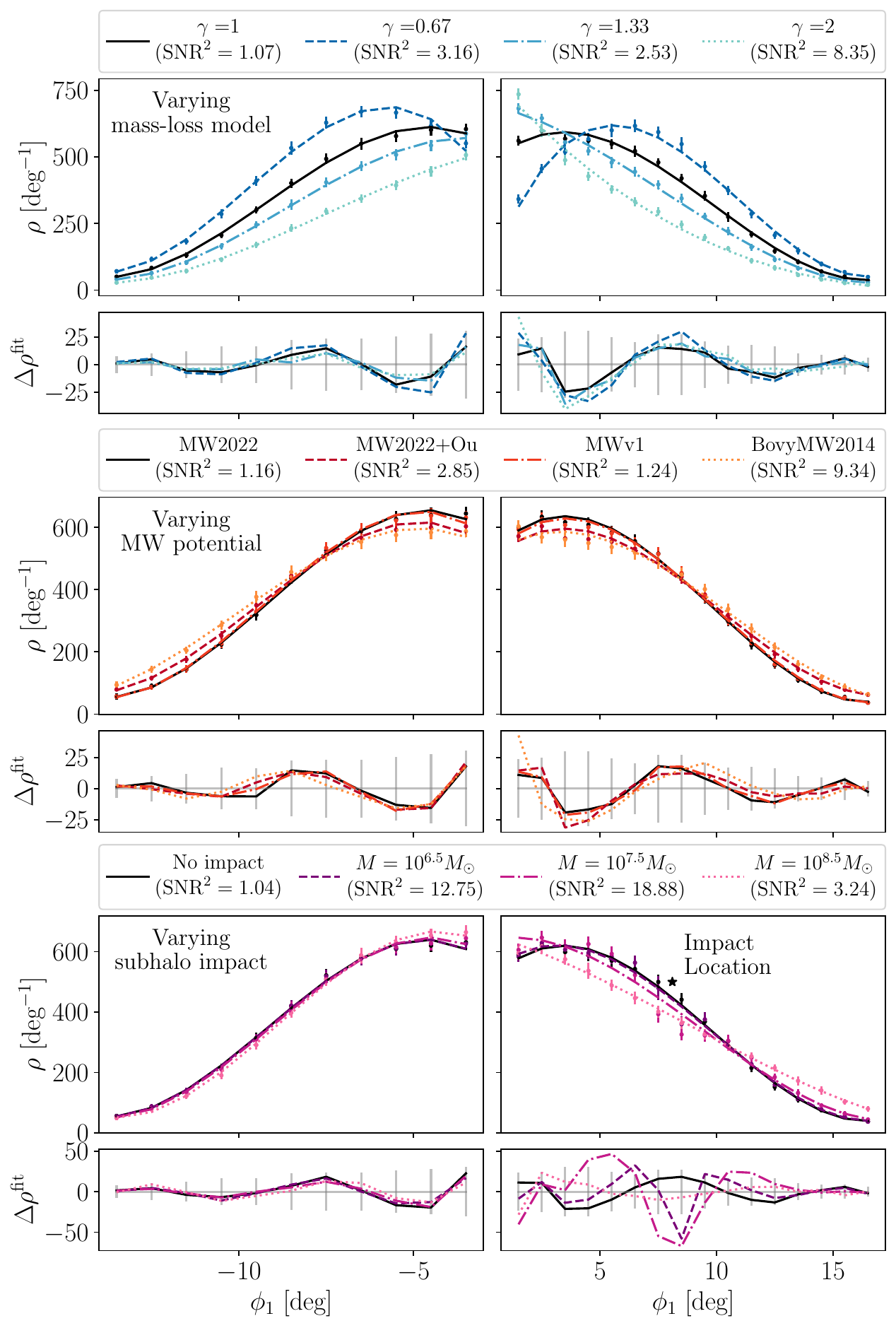}
\caption{Fitting the density observable to the proxy model for the different realizations of Jet shown in Fig.~\ref{fig:mass-loss-large-scale-sim}. Shown are different mass-loss models (top), different MW potentials (middle) and different subhalo impacts (bottom). The average value from 50 realizations of simulations (data points with error bars) and the best-fit proxy model (lines) are shown in the first, third and fifth rows. The residual to the fit is shown in the second, fourth and sixth rows, respectively. The gray error bar is the enlarged uncertainty $\tilde \sigma_{\rho,i}$ after the proxy-model fit. The quoted $\mathrm{SNR}^2$ is calculated from the impacted (right) arm only. Notice that the smaller-scale impacts with $M=10^{6.5}~M_\odot$ or $M=10^{7.5}~M_\odot$ are not entirely removed by the proxy model, and can be identified in the residual, while the larger scale impact with $M=10^{8.5}~M_\odot$ is consistent with the polynomial model. The impact parameter is chosen to be $b=1.05 \times (M/10^8 M_\odot)^{0.5}$ kpc.}
\label{fig:mass-loss-large-scale-fit}
\end{figure}

\subsection{Proxy-model fitting}
\label{sec:poly_model_fitting}

Given some stream data $X$, we now define a procedure to perform proxy-model subtraction and extract a residual. We set our fiducial stream model with $\gamma=1$ mass-loss model and MW2022 potential, and restrict to those $\phi_1$ region identified using Eq.~\ref{eq:cut}, where the smooth stream is well-described by the polynomial model.

To obtain the best-fit polynomial on $X$, we use a modified covariance matrix compared to the one obtained from the simulations, Eq.~\ref{eq:covariance_on_sims}. The covariance matrix from simulations $(C_X)_{ij}$ accounts for statistical and observational errors, but does not encode deviations from the polynomial model itself. We define an \emph{enlarged covariance matrix} accounting for the difference between the mean of simulations and the best-fit model:
\begin{align}
\label{eq:error_no_hartlap}
    (\tilde C_X)_{ij} &\equiv \Big \langle (X^{\rm smooth}_i-X^{\mathrm{model}}_i(\hat{\bm \theta}_\mathrm{smooth})) \nonumber \\
    &\qquad\qquad\;\;\times (X^{\rm smooth}_j-X^{\mathrm{model}}_j(\hat{\bm \theta}_\mathrm{smooth})) \Big \rangle \nonumber \\
    &=(C_X)_{ij}+ d_i d_j
\end{align}
where $\langle\cdot\rangle$ is the average over all realizations of the mock smooth stream and $d_i \equiv \langle X_i^\mathrm{smooth}\rangle-X^{\mathrm{model}}_i(\hat{\bm \theta}_\mathrm{smooth})$. Relative to the original covariance matrix $(C_X)_{ij}$, the enlarged covariance matrix $(\tilde C_X)_{ij}$ adds in the difference between the mean of simulations and the best-fit model. The corresponding enlarged per-bin uncertainty is defined by $\tilde{\sigma}_{X,i}\equiv\sqrt{(\tilde C_X)_{ii}}$.  For the inverse covariance matrix, we further apply a correction:
\begin{align}
\label{eq:error}
    (\hat C_X)^{-1}_{ij} \equiv \left(\frac{K-N^X_{\rm bin}-2}{K-1}\right) \times (\tilde C_X)^{-1}_{ij}
\end{align}
where $K=250$ is the number of smooth stream realizations, and $N_{\rm bin}^X$ is the number of bins used in the arm for that observable. This factor is a Hartlap correction~\citep{Hartlap_2007} accounting for the bias in the maximum-likelihood estimator for the inverse covariance matrix. %

The best-fit parameters for data $X_i$ are then given as
\begin{align}
    \hat {\bm \theta} =\argmin_{\bm \theta} \sum_{i,j} & \left( X_i - X^{\mathrm{model}}_i(\bm \theta)\right)\left(\hat C_X^{-1}\right)_{ij} \nonumber \\ & \times \left(X_j -X^{\mathrm{model}}_j(\bm \theta)\right).
    \label{eq:theta_fit}
\end{align}
We define the residual after polynomial model subtraction as:
\begin{align}
    \Delta X^{\rm fit}_i & \equiv X_i - X_i^{\rm model}( \hat {\bm \theta})
\end{align}
and the signal-to-noise (SNR) ratio as
\begin{equation}
    \label{eq:SNR_fit}
        \mathrm{SNR}^2=\sum_{X=\phi_2,v_r,\rho}\sum_{i,j}\Delta X_i^{\mathrm{fit}}\left(\hat C_X^{-1}\right)_{ij}\Delta X_j^{\mathrm{fit}}.
\end{equation}

Fig.~\ref{fig:mass-loss-large-scale-fit} shows the results of the polynomial fits for the different streams shown in Fig.~\ref{fig:mass-loss-large-scale-sim}. Streams with different mass-loss history are all well described by $\rho^{\rm model}$. Large-scale variations from varying the MW potential can also be accommodated by $\rho^{\rm model}$. This indicates profiling over the proxy model can account for these uncertainties when performing an impact analysis. Moreover, the bottom panel of Fig.~\ref{fig:mass-loss-large-scale-fit} suggests that the proxy model can absorb the effect of a distant impact of a high-mass subhalo $\sim 10^{8.5} M_\odot$, which affects the entire stream. However, crucially, it does not remove the more localized signal of an impact with $M\sim 10^{6.5} M_\odot$ or $M\sim 10^{7.5} M_\odot$. %

\section{Predicting subhalo impacts}
\label{sec:impacts}

\begin{figure}[t]
\centering

\includegraphics[width=0.8\columnwidth]{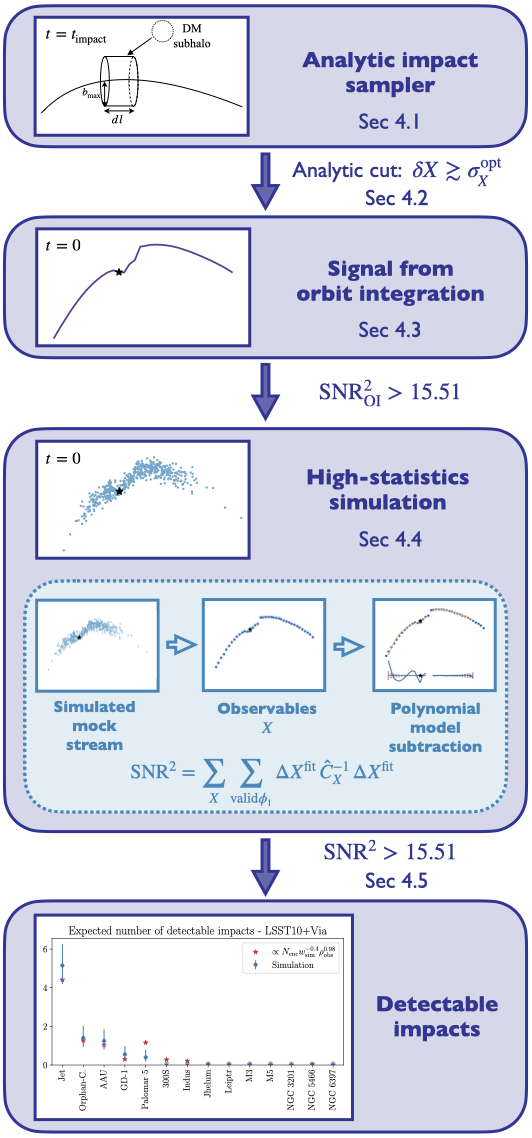}

\caption{Flow chart of our subhalo-impact analysis. The signal-to-noise ratio from orbit integration $\mathrm{SNR}^2_\mathrm{OI}$ is calculated analogously to the $\mathrm{SNR}^2$ from high-statistics simulation (light blue inset). The valid $\phi_1$ region and enlarged covariance matrix $\hat{C}$ are obtained by fitting smooth streams (see Sec.~\ref{sec:poly_model_fitting}).\label{fig:flowchart}}
\end{figure}

We generate samples of detectable impacts with a multi-step process, illustrated in Fig.~\ref{fig:flowchart}. The process begins with analytic sampling of encounters, which in general will include many encounters with low-mass subhalos that will not be individually detectable. To reduce the computational cost while maintaining good accuracy, we evaluate encounters for detection potential with multiple layers of cuts. The process ends with a high-statistics simulation of the impact to evaluate detectability. Each of the steps is detailed in the following subsections.

\subsection{Impact sampler}

We first sample subhalo encounters over the lifetime of the stream, largely following the analytic approach in~\cite{Erkal_2016}. In this first stage, we assume all stream stars are moving along the progenitor orbit and the stream grows linearly with time.

Over the lifetime of the stream, we consider snapshots at small time intervals $dt=20$ Myr.
We define a traversal time $t_\mathrm{trav}(t)$ as the time it would take the progenitor to traverse the stream at time $t$. %
We assume that this traversal time increases linearly with time, i.e.\ $t_\mathrm{trav}(t) = t_\mathrm{trav}(t=0) \, (T_\mathrm{form}-t)/T_\mathrm{form}$. The traversal time today $t_\mathrm{trav}(t=0)$ is fixed by the farthest valid $\phi_1$ bin from the progenitor across all observables and both arms.
At each time snapshot, we model the stream arms by traversing the progenitor $t_\mathrm{trav} / 2$ forward and $t_\mathrm{trav} / 2$ backward along its orbit.
This approximation for the stream length based on traversal time corresponds to an overall linearly growing stream while accounting for the fact that the stream gets stretched near pericenter and compressed near apocenter.
At each time snapshot, we split the stream into small chunks of length $dl$, which is decided by the distance traversed over 1 Myr.

For each stream segment $dl$, we consider the cylinder centered at the stream track with radius $b_\mathrm{max} = 10$ kpc. The expected number of subhalos entering this chunk through the cylinder in $dt$ is %
\begin{equation}
    \label{eq:dn_enc}
    dN_\mathrm{enc}=\sqrt{2\pi} n_\mathrm{sub} \sigma b_\mathrm{max} dl\,dt .
\end{equation}
Following \cite{Erkal_2016}, we assume the subhalo velocity dispersion for one component is $\sigma=180$ km/s and the CDM subhalo number density $n_\mathrm{sub}$ follows the profile:
\begin{equation}
\label{eq:nsub}
    \left(\frac{d n_\mathrm{sub}}{d M}\right)_\mathrm{CDM} \!\!\! =c_0 \left(\frac{M}{m_0}\right)^n \exp\left(-\frac{2}{\alpha_r}\left(\left(\frac{r_\mathrm{gc}}{r_{-2}}\right)^{\alpha_r}-1\right)\right)
\end{equation}
with $c_0=2.02 \times 10^{-13} M_\odot^{-1}~\mathrm{kpc}^{-3}$, $m_0=2.52 \times 10^7 M_\odot$, $n=-1.9$, $\alpha_r=0.678$ and $r_{-2}=162.4~\mathrm{kpc}$. This model was constructed by scaling a fit to the Aquarius~\citep{Springel2008} subhalo population to a fiducial MW mass of $10^{12}~M_\odot$.
Note that \cite{Erkal_2016}  includes an additional overall factor of $1/3$ in $n_{\rm sub}$ to approximate a reduction of subhalos due to the MW disk. We do not include this effect here, taking our fiducial as a scenario based on DM-only (DMO) simulations; we will consider alternate scenarios in Sec.~\ref{sec:subhalo_distribution}.
The model above does not include angle- or time-dependent terms, or correlations between the radius $r_\mathrm{gc}$ and subhalo mass $M$, in the subhalo model; this is an important area for future work.

Eq.~\ref{eq:nsub} is integrated over $dM$ with a lower bound subhalo mass of $10^5 M_\odot$ to get $n_\mathrm{sub}(r_\mathrm{gc})$.
Further integrating Eq.~\ref{eq:dn_enc}  over $dt$ and $dl$ gives the total number of encounters $N_{\rm enc}$. Assuming circular orbits of the stream, which have linear growth in stream length and constant $n_\mathrm{sub}$, we recover the $N_\mathrm{enc}^\mathrm{circ}$ quoted in Sec.~\ref{sec:intro}:
\begin{align}
\label{eq:n_enc}
    N^{\rm circ}_{\rm enc} = \sqrt{\frac{\pi}{2}}  n_\mathrm{sub} \sigma b_{\rm max} l T_{\rm form}.
\end{align}
More generally, we can obtain the expected number of subhalo encounters on a general orbit as:
\begin{equation}
    N_\mathrm{enc} =\int\sqrt{2\pi} n_\mathrm{sub}(l,t) \sigma b_\mathrm{max} dl\,dt ,
    \label{eq:Nenc_orbit}
\end{equation}
where $n_\mathrm{sub}(l,t)$ is evaluated across all stream segments $dl$ at each time.

Given the expected number of subhalo encounters for a given segment and time interval $dN_\mathrm{enc}$, we sample the number of encounters from a Poisson distribution with mean $dN_\mathrm{enc}$. For each encounter, we generate the following subhalo impact parameters:
\begin{itemize}
    \item Subhalo mass $M$ based on the distribution in Eq.~\ref{eq:nsub}, with lower bound of $10^5 M_\odot$.  Here $M$ represents present-day (tidally stripped) subhalo mass.
    \item Subhalo scale radius $r_s$, assuming a Hernquist potential and a mass-radius relation
    \begin{equation}
    \label{eq:mass-radius}
        r_s = c_{r_s}~\mathrm{kpc}\left(\frac{M}{10^8 M_\odot}\right)^{0.586}
    \end{equation}
    where $c_{r_s}$ is a random variable following a log-normal distribution whose natural log has a mean of 0.26 and a standard deviation of 0.38. This corresponds to a median value of 1.3 for $c_{r_s}$. Both the power-law exponent in Eq.~\ref{eq:mass-radius} and the parameters for $c_{r_s}$ distribution are obtained by fitting the $v_\mathrm{max}$-$M_\mathrm{tidal}$ relation from Via Lactea II\footnote{\url{https://www.ucolick.org/~diemand/vl/data.html}}~\citep{Diemand_2008} data.
    \item Time of impact $t_\mathrm{impact}$, where $t_\mathrm{impact} = 0$ corresponds to today. We fix it by the central value of the time segment being considered.
    \item An angle $\psi$ describing the entry point of the subhalo into the cylinder,
    sampled uniformly over $[0, 2\pi).$ %
    \item Subhalo velocity $w_r$, $w_\theta$, $w_\parallel$ sampled based on~\cite{Erkal_2016}. These are defined with respect to the cylinder, where $w_r$ and $w_\theta$ are the cylindrical radial and tangential velocities, and $w_\parallel$ is the relative velocity along the direction the stream is moving.
    \item Impact parameter $b$, determined by
    \begin{equation}
        b=b_\mathrm{max}\frac{w_\theta}{\sqrt{w_\theta^2+w_r^2}}
    \end{equation}
    \item Orbit position and velocity at the center of the cylindrical segment ($\bm r_*$, $\bm v_*$).
    This information, combined with the subhalo velocity $\vec w$ and angle $\psi$, determines the phase space coordinates of the subhalo at the time of impact $t_\mathrm{impact}$, and thus the orbit of the subhalo.
\end{itemize}
We will later explore deviations from the above assumptions for the subhalo profile and mass function.

Because we consider subhalo--stream encounters with $b_{\rm max} = 10$~kpc and subhalo masses down to $10^5 M_\odot$, this process will lead to many encounters that will not be individually detectable. We now consider detection prospects with multiple layers of cuts. The first layer uses an analytic estimate of signal size based on circular orbits, which is the least computationally expensive but most approximate calculation. Impacts that pass this cut are then promoted to a second layer based on the method of orbit integration, which requires a moderate computational cost and provides better accuracy than the analytic estimate, by accounting for elliptical orbits. The final layer is a high-statistics stream simulation, which is the most computationally expensive but also provides the most reliable signal prediction and accounts for the actual time-dependent stream length. We next introduce the three layers one by one.

\subsection{Analytic cut}
\label{sec:analytic_cut}

The analytic model for subhalo impacts  in~\cite{Erkal_2015, Erkal_2015_2}, which was derived for circular streams, provides a useful scaling for stream perturbations from subhalo impacts. We define the analytic perturbation scales
    \begin{equation}
        \delta \phi_2 \equiv \frac{GM}{\sqrt{b^2+r_s^2}w_\mathrm{rel}v_*}
        \label{eq:analytic_phi2}
    \end{equation}
    \begin{equation}
        \delta v_r \equiv \frac{GM}{\sqrt{b^2+r_s^2}w_\mathrm{rel}}
        \label{eq:analytic_vr}
    \end{equation}
    \begin{equation}
    \label{eq:analytic_density}
         \frac{\delta\rho}{\rho_0}\equiv \frac{GMt_{\rm impact}}{(b^2+r_s^2)w_\mathrm{rel}}
    \end{equation}
where $w_\mathrm{rel}=\sqrt{w_r^2+w_\theta^2+w_\parallel^2}$ is the relative subhalo velocity with respect to the stream star at the point of closest approach. Note from Eq.~\ref{eq:analytic_density} that the analytic model only predicts the density ratio rather than the density.

For each observable, we place a conservative cut comparing the analytic perturbation to an uncertainty estimate from our simulations. For this purpose we define an \emph{optimal uncertainty} $\sigma_X^{\mathrm{opt}}$.
For $X=\phi_2$ and $v_r$, this is given by the minimum enlarged uncertainty across bins,
\begin{equation}
\label{eq:sigma_opt_x}
\sigma_X^{\mathrm{opt}}\equiv \min_i \tilde{\sigma}_{X,i}.
\end{equation}
For density, we instead define the optimal fractional uncertainty,
\begin{equation}
\label{eq:sigma_opt_rho}
\sigma_{\rho/\rho_0}^{\mathrm{opt}}\equiv \min_i \frac{\tilde{\sigma}_{\rho,i}}{\left<\rho_i^\mathrm{smooth}\right>}.
\end{equation}
Using these optimal uncertainties means that we will pass many impacts that are not ultimately detectable, but it will not drop truly detectable impacts at this stage.

An impact will pass this stage if any of the following criteria are satisfied:
\begin{equation}
    \label{eq:analytic_cut_phi2}
        \delta \phi_2 >\sigma_{\phi_2}^{\mathrm{opt}}
    \end{equation}
    \begin{equation}
    \label{eq:analytic_cut_vr}
         \delta v_r >\frac{1}{2}\sigma_{v_r}^{\mathrm{opt}}
    \end{equation}
    \begin{equation}
    \label{eq:analytic_cut_rho}
         \frac{\delta\rho}{\rho_0} >\sigma_{\rho/\rho_0}^{\mathrm{opt}}
\end{equation}
The prefactors on the optimal uncertainties are chosen so that the detectable impacts have converged at the low-signal end, as shown in Fig.~\ref{fig:analytic_oi_cut}.

\subsection{Orbit integration}
\label{sec:OI}

For an impact which passes at least one of the analytic cuts in Eqs.~\ref{eq:analytic_cut_phi2}-\ref{eq:analytic_cut_rho}, we generate the signal profile using the method of orbit integration (OI). In this method, we consider stars perfectly aligned with the orbit of the progenitor, and perform
orbit integration on these stars with and without the subhalo potential to get per-bin data $X_{i}^\mathrm{impact,OI}$ and $X_{i}^\mathrm{smooth,OI}$, respectively. For $v_r$ and $\phi_2$, we extract observables by linearly interpolating at the center of each $\phi_1$ bin location. For the density observable, we simply count the number of stars per bin. The subhalo potential is turned on for a time interval of $\Delta t = \frac{4\sqrt{b^2+r_s^2}}{w_\mathrm{rel}}$, which is $4\times$ the time scale of duration of flyby~\citep{Erkal_2015_2}, centered at the time of impact $t_\mathrm{impact}$. This is essential to ensure that the stream is only affected by a single subhalo encounter. Since the stream is close to the progenitor orbit, this gives a good approximation for the effect of a subhalo impact. The advantage of OI compared to the analytic model is that it works for general stream orbits, realistic Milky Way potential, and actual observer location at the Sun, while being much less expensive than the full simulation.

We construct impacted stream data  for $\phi_2$ and $v_r$ via:
\begin{equation}
    \label{eq:impact_phi2_vr}
        X^\mathrm{OI}_i=X_{i}^\mathrm{impact, OI}-X_{i}^\mathrm{smooth,OI}+X_i^\mathrm{model}(\hat {\bm{\theta}}_\mathrm{smooth}).
\end{equation}
This ensures that the data reproduces the smooth-stream simulations in the limit of a weak impact where $X_{i}^\mathrm{impact,OI} - X_{i}^\mathrm{smooth,OI} \to 0$.
For the density ratio $\rho$, we similarly define
    \begin{equation}
    \label{eq:impact_rho}
    \rho_i^\mathrm{OI}=\frac{\rho_{i}^\mathrm{impact,OI}}{ \rho_{i}^\mathrm{smooth,OI}} \times \rho_i^\mathrm{model}(\hat {\bm{\theta}}_\mathrm{smooth}).
    \end{equation}

At this point, we could perform the same polynomial fit as described in Sec.~\ref{sec:poly_model_fitting}, and use the residual to compute an SNR. However, a notable limitation of the OI technique is that it might not accurately predict the location of the signal since it does not account for stream dispersion. This can significantly underestimate signal strengths if the impact signal moves  into the valid $\phi_1$ region or into a region with lower noise at time of observation. To avoid discarding such impacts, we allow for arbitrary shifts of the OI signal perturbation along $\phi_1$. We then repeat the process in Eq.~\ref{eq:impact_phi2_vr} and \ref{eq:impact_rho} to define a signal and perform polynomial subtraction to compute SNR. We define ${\rm SNR}_{\rm OI}^2$ as the maximum SNR$^2$ obtained over all possible $\phi_1$ shifts of the OI signal perturbation.
This procedure results in an overestimate of the signal strength, which serves as a loose filter on impacts.

In Appendix~\ref{app:extras}, Fig.~\ref{fig:detect_oi} shows examples where the unshifted signal predicted by OI would give overly low (left panel) or overly high (right panel) SNR compared to the SNR from the simulated impact.
Orbit integration does not capture the time-dependent growth of stream or the dispersion of stream stars, which may cause possible shifts in signal location as well as diminish the size of the signal. This is especially true for older impacts with $t_{\rm impact} \gtrsim 1$ Gyr.
In Fig.~\ref{fig:detect_oi}, the unshifted OI prediction is an impact centered on the black star, while the simulated impact has moved into the valid $\phi_1$ region (left panel) or become weaker (right panel) by the time of observation, thus causing the discrepancy between SNR from OI and SNR from simulation. The case of SNR underestimation leads us to calculate the maximal SNR from OI by sliding the signal profile over the valid $\phi_1$ range. The case of overestimation indicates the need to perform a more accurate SNR based on a high-statistics simulation in the next layer.

We place a cut of
\begin{equation}
\mathrm{SNR}_\mathrm{OI}^2 > 15.51
\label{eq:OI_cut}
\end{equation}
in order for an impact to be considered for the next layer for high-statistics simulation. This numerical threshold mimics the threshold we use below with simulated signals, but the much more optimistic procedure used to obtain $\mathrm{SNR}_\mathrm{OI}^2$ avoids cutting detectable signals. The last panel of Fig.~\ref{fig:analytic_oi_cut} in the Appendix shows that most of the impacts that are detectable in the final simulation step have SNR${}^2_\mathrm{OI}$ much higher than the 15.51 threshold.

\subsection{High-statistics simulation}
\label{sec:highstats_sim}

\begin{figure*}[t]
\centering
\includegraphics[width=0.495\textwidth]{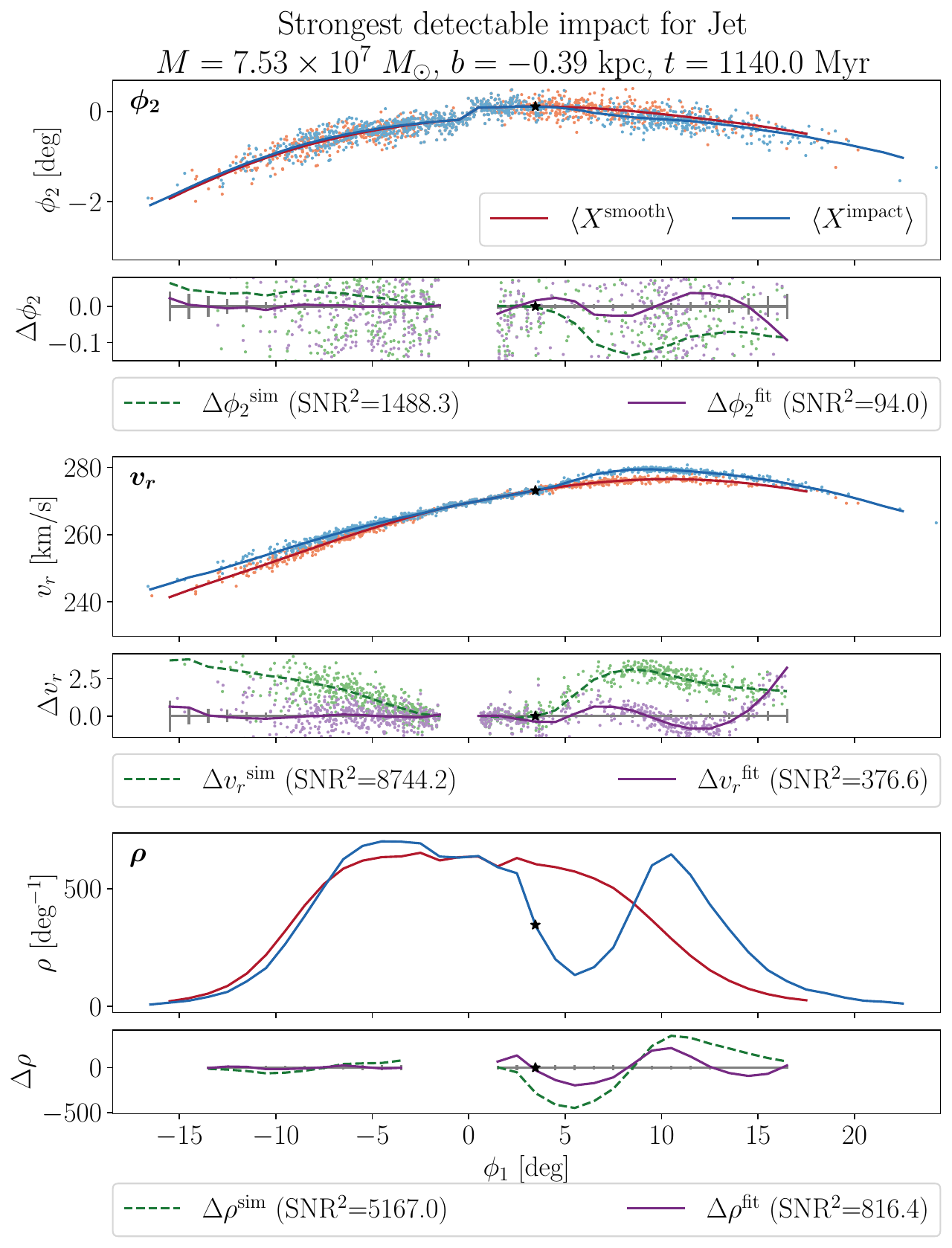}
\includegraphics[width=0.495\textwidth]{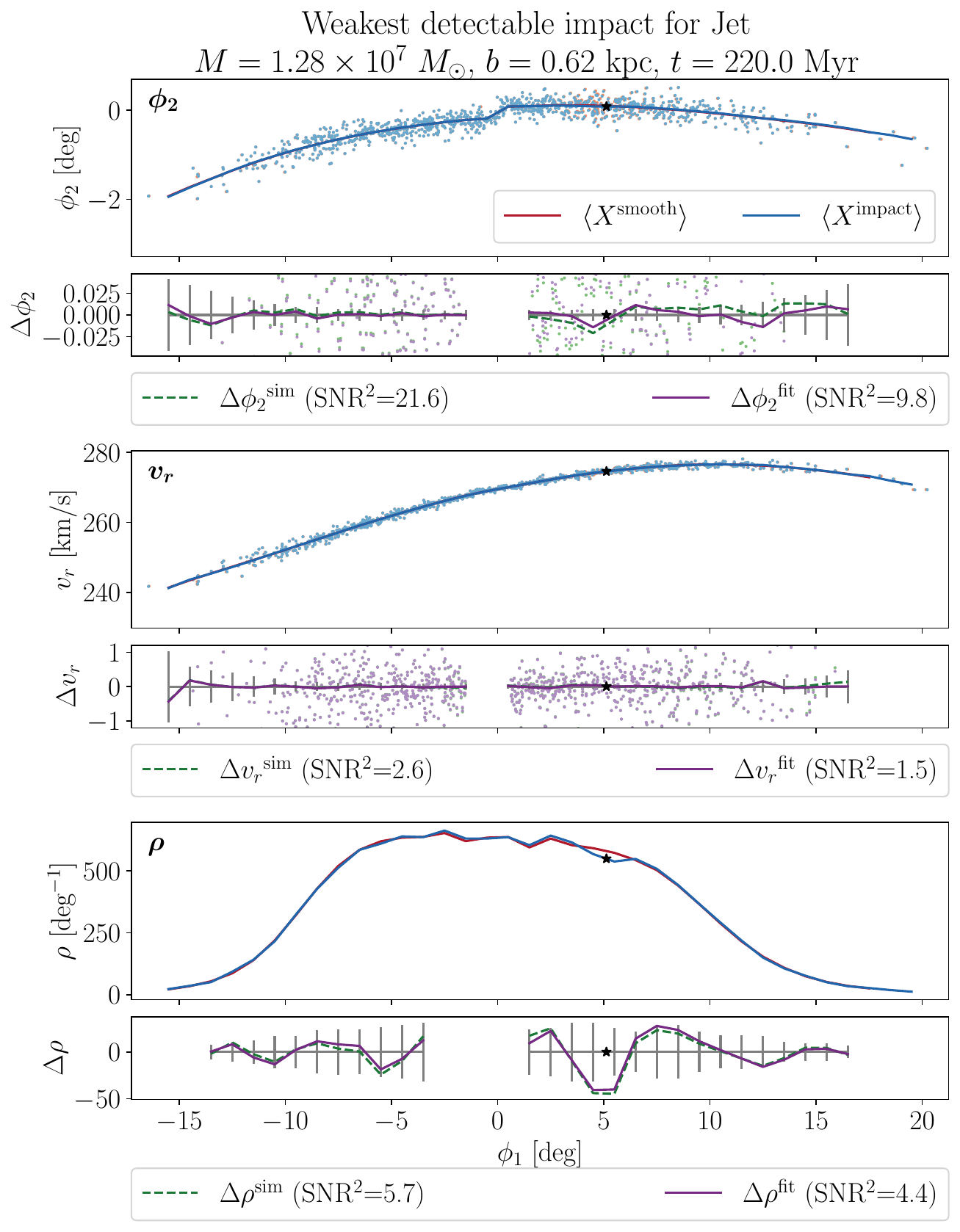}
\caption{Detectable impacts for Jet with the largest (left) and smallest (right) SNR across 20 runs of the impact history, and the resulting perturbations in $\phi_2$ (top), $v_r$ (center) and $\rho$ (bottom). Odd rows show per-star data from one realization of a simulation of the smooth stream (red) and the impacted stream (blue), along with the expected values of the observables over 250 realizations of smooth-stream simulations and $K$ (with $K=12$ for Jet) realizations of impacted stream simulations (solid lines). The black star marks the impact location. For visual clarity, we display 1000 stream stars rather than the expected number of stars for LSST10. Even rows show the respective residuals after subtracting the best-fit proxy model (purple solid) or the smooth stream (green dashed). The error bars (gray) correspond to the diagonal elements of the enlarged covariance matrix defined in Sec.~\ref{sec:poly_model_fitting}.}
\label{fig:detect}
\end{figure*}

For impacts passing the orbit integration layer, we perform a high-statistics simulation to get the most reliable signal profile. Similar to the procedure in Sec.~\ref{sec:modeling}, we use the particle spray model from~\cite{Fardal_2015} to generate the impacted stream. We adjust the orbit of the subhalo slightly relative to analytic impact sampler, such that the location of the star at the closest approach, $\bm r_*$, is on the simulated smooth stream rather than on the orbital track of the progenitor. This accounts for the small difference in stream vs. progenitor orbit and ensures that the impact parameter $b$ is accurate. To set the initial condition for the simulation, we first evolve the observed progenitor location today backwards under both the Milky Way and subhalo potential to determine the progenitor location at $t=T_\mathrm{form}$. Then the stream stars are released from the progenitor and evolved forward under both the Milky Way and subhalo potential. In both procedures, the subhalo potential is only turned on for a time interval of $\Delta t = \frac{4\sqrt{b^2+r_s^2}}{w_\mathrm{rel}}$ centered at $t=t_\mathrm{impact}$ to ensure a single encounter with the subhalo. We follow the same procedure as for smooth-stream simulations to incorporate observational errors to the impacted streams.

To reduce noise in the simulation, we generate $K$ realizations of each impacted stream simulation, and use the average over the $K$ realizations as the observables, $\langle X_{i}^\mathrm{impact}\rangle$.
$K$ is chosen to be of the same order as the number of $\phi_1$ bins over all observables and both arms, such that the expected simulation noise is at most $1/3$ of the final SNR$^2$ threshold value. The odd rows of Fig.~\ref{fig:detect} show example observables from simulations of impacted stream $\langle X^\mathrm{impact}_i\rangle$ (blue lines). As a comparison, the corresponding observables from smooth-stream simulations $\langle X^\mathrm{smooth}_i\rangle$ are shown as the red lines in the same panels.

\subsection{Detectability criterion}
\label{sec:test_statistic}

We evaluate the SNR$^2$ of Eq.~\ref{eq:SNR_fit} on the average of impacted stream simulations $\langle X^{\rm impact}_i \rangle$, following the polynomial fitting procedure of Sec.~\ref{sec:poly_model_fitting}.
As shown in App.~\ref{app:detectability_metric}, this SNR$^2$ is equivalent to the log-likelihood ratio under null and signal hypotheses, when evaluated on noise-free (Asimov) data and assuming a fixed signal template. We require that this noise-free SNR$^2$ for an impact lie above some threshold value, or equivalently that the median realization of that impact can be distinguished from the background at some confidence level.

As discussed above, since we rely on a finite number of realizations $K$ to obtain $\langle X^{\rm impact}_i \rangle$, simulation noise can add to the SNR$^2$. Assuming impacted stream simulations have similar covariance as the smooth-stream simulations, then $K$ realizations will contribute $ (N_{\rm bin}-24)/K$ to SNR$^2$, where $N_{\rm bin}$ is the total number of $\phi_1$ bins across all arms and observables and the subtraction by $24=2\times3\times4$ accounts for the free parameters from polynomial fitting. For some streams, $N_{\rm bin}$ might be O(100). To mitigate this effect while limiting computational cost, we select $K$ to be large enough such that $ (N_{\rm bin}-24)/K$ is less than 1/3 of the SNR$^2$ threshold for detection. We furthermore subtract off this expected contribution from simulation noise in our calculation of the SNR$^2$.

To determine whether an impact is detectable, we must compare this SNR$^2$ (equivalently Asimov value of the log-likelihood ratio) with the distribution of the log-likelihood ratio under the null hypothesis of no impact. In a full search, this statistic is profiled over the subhalo impact parameters.
In the asymptotic limit, if the parameters satisfy the regularity conditions of Wilks' theorem, the log-likelihood ratio test statistic is distributed as a chi-square variable with degrees of freedom equal to the number of parameters of interest~\citep{Wilks:1938dza,Cowan_2011}. For the subhalo-impact model considered here, there are eight relevant parameters: the subhalo mass $M$, subhalo scale radius $r_s$, impact parameter $b$, impact time $t_\mathrm{impact}$, the three components of the relative velocity vector $\vec{w}$, and the impact location along the stream.
We therefore set an approximate 95\% CL detectability threshold by comparing to a chi-square distribution with eight degrees of freedom:
\begin{equation}
    \mathrm{SNR}^2 > \chi^2_{8,\,0.95} = 15.51 .
    \label{eq:SNR_8dof}
\end{equation}
In practice, the null distribution might not follow the $\chi^2_{8}$ distribution and a more accurate threshold would be obtained by direct calibration with simulation data, which is computationally very expensive. We have performed this calibration for the analogous setup in circular streams in \cite{lu2025detectabilitydarkmattersubhalo}, using an analytic model to simplify the computational requirements. \cite{lu2025detectabilitydarkmattersubhalo} obtained values of 11-12 for the 95\% threshold on the log-likelihood ratio, similar to the corresponding value of $\chi^2_{6,\,0.95}=12.59$. In that work, we assumed fixed mass-radius relation, and fixed impact location along the stream, thus giving 6 degrees of freedom in total. %
We therefore adopt an analogous threshold based on the $\chi^2_8$ distribution corresponding to 8 degrees of freedom in this work.

An illustration of the polynomial fitting procedure and the residual $\Delta X^\mathrm{fit}$ is shown in the even rows of Fig.~\ref{fig:detect} for Jet.
More examples for other top-ranked streams can be found in Fig.~\ref{fig:detect_all} in Appendix~\ref{app:extras}. %
For comparison, we also show the residual if the polynomial model subtraction is not performed, and instead the signal is defined with respect to the fiducial smooth stream:
\begin{equation}
    \label{eq:deltaX_sim}
    \Delta X_i^\mathrm{sim}= \langle X_i^\mathrm{impact}\rangle-\langle X_{i}^\mathrm{smooth}\rangle.
\end{equation}
The ${\rm SNR}^2$ obtained with $\Delta X_i^\mathrm{sim}$ instead of $\Delta X_i^\mathrm{fit}$ is also shown, which amounts to an ideal scenario where one can simultaneously fit the impact and the smooth-stream model.
The polynomial fitting removes much of the larger-scale perturbations in $\phi_2$ and $v_r$ induced by changes in the orbit, while preserving localized perturbations.

\subsection{Convergence of detectable impacts with $b$}
\label{sec:convergence_with_b}

We now quantify the convergence problem introduced in Sec.~\ref{sec:proxy_model}, where number of detectable subhalo encounters grows with impact parameter $b$.
We use the analytic scaling for stream perturbations, Eqs.~\ref{eq:analytic_phi2}-\ref{eq:analytic_density}, combined with an estimate of the angular width of the signal:
\begin{align}
    \Delta \phi_1 \sim \frac{\sqrt{b^2 + r_s^2}}{r_h}.
    \label{eq:analytic_delta_phi1}
\end{align}
Without performing subtraction with a third-order polynomial and assuming a diagonal covariance matrix, we can estimate the SNR for each of the observables as ${\rm SNR}_X^2 = \delta X^2 \Delta \phi_1 / \sigma_X^2$. Focusing just on the $b$ and $M$ dependence, this gives the scaling
\begin{align}
    {\rm SNR}_{\phi_2}^2 & \propto \frac{M^2}{(b^2 + r_s^2)^{1/2}} \sim \frac{M^2}{b} \\
    {\rm SNR}_{v_r}^2 & \propto \frac{M^2}{(b^2 + r_s^2)^{1/2}} \sim \frac{M^2}{b} \\
    {\rm SNR}_\rho^2 & \propto \frac{M^2}{(b^2 + r_s^2)^{3/2}}\sim \frac{M^2}{b^3}.
\end{align}
For large $b$, the SNR$^2$ only drops as a factor of $1/b$ for the $\phi_2$ and $v_r$ observables. Holding our SNR threshold fixed, this means as we consider larger $b$, then we must consider more massive subhalos with a scaling of $M \propto \sqrt{b}$ to achieve the same SNR.

Now consider the rate of encounters, which scales as $N_{\rm enc} \propto b \, n_{\rm sub}(>M)$ (Eq.~\ref{eq:n_enc}).  The subhalo density above some mass threshold $n_{\rm sub}(>M) \propto M^{-0.9}$. Combining this with the scaling $M \propto \sqrt{b}$ (for $\phi_2$ and $v_r$) and $M \propto b^{1.5}$ (for density) gives
\begin{align}
    N^{\phi_2, v_r}_{\rm enc} &\propto b^{0.55} \\
    N^{\rho}_{\rm enc} &\propto b^{-0.35}.
\end{align}
The number of impacts grows with $b$ for the $\phi_2$ and $v_r$ observables, and converges weakly with larger $b$ for the density observable. Physically, the $\phi_2$ and $v_r$ observables are sensitive to changes in the stream orbit, while the density observable is primarily sensitive to differential changes in the orbit and thus has better convergence properties.

There are a few effects that can regulate the growth of impacts with $b$. First, eventually the number of more massive subhalos does cut off. However, at this high end there are fewer and fewer subhalos and their effects will depend on the specific orbits; one must explicitly model the effect of the actual massive subhalos of the Milky Way rather than using our sampling procedure. Second, the angular width we used above, Eq.~\ref{eq:analytic_delta_phi1}, is eventually cut off by the angular length of the stream itself, $\ell$. This modifies the above scaling to
\begin{align}
    {\rm SNR}_{\phi_2}^2 & \propto \frac{M^2}{(b^2 + r_s^2)} \sim \frac{M^2}{b^2} \\
    {\rm SNR}_{v_r}^2 & \propto \frac{M^2}{(b^2 + r_s^2)} \sim \frac{M^2}{b^2} \\
    {\rm SNR}_\rho^2 & \propto \frac{M^2}{(b^2 + r_s^2)^2}\sim \frac{M^2}{b^4}.
\end{align}
This gives a scaling of $N_{\rm enc} \propto  b^{0.1}$ for $\phi_2, v_r$ which still grows with $b$. Even if the subhalo density has a slightly steeper dependence on $M$ rather than $n_{\rm sub}(>M) \propto M^{-0.9}$, this would still be just barely convergent.

\begin{figure*}[t]
\centering
\includegraphics[width=0.49\textwidth]{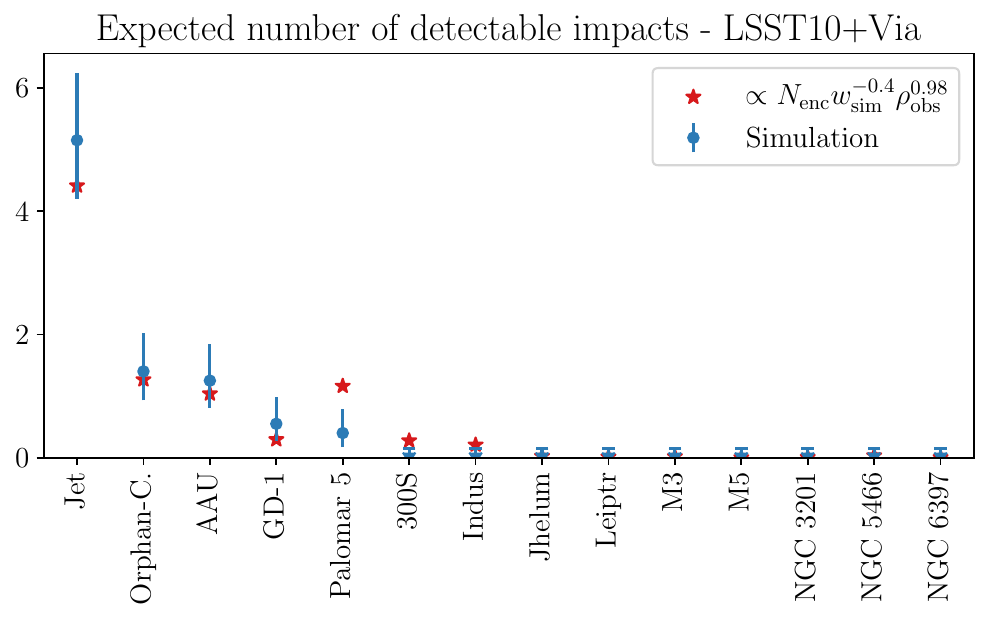}
\includegraphics[width=0.5\textwidth]{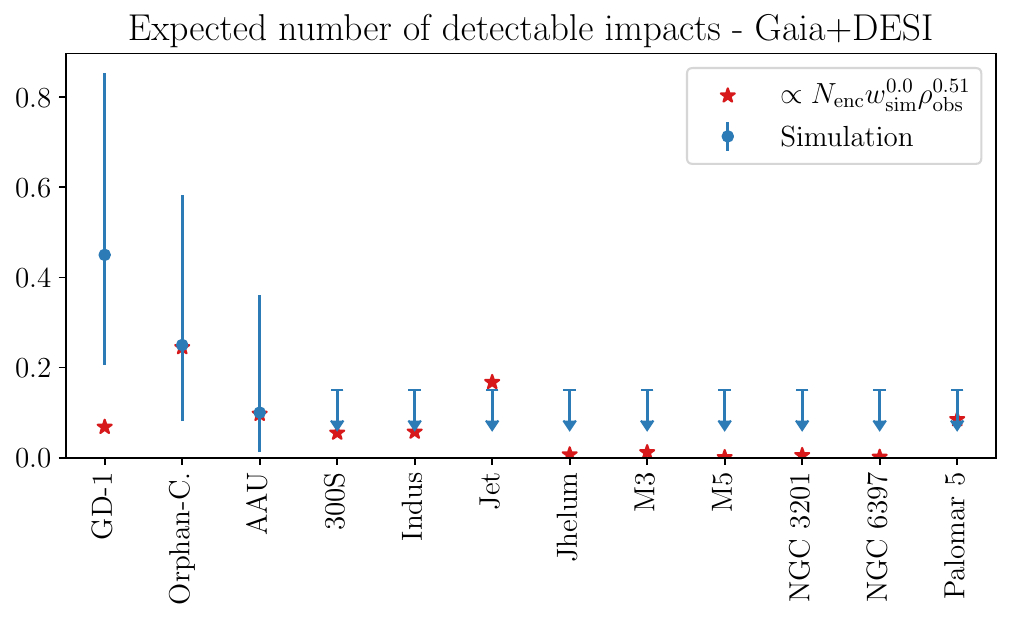}
\caption{Expected number of detectable impacts (blue) with near-future LSST10+Via (left) and present Gaia+DESI (right) data. Detectable impacts are required to satisfy SNR$^2 > 15.51$, corresponding to the 95\% CL threshold for a $\chi_8^2$ distribution (see Eq.~\ref{eq:SNR_8dof}).
The error bars show 95\% confidence interval on the Poisson mean, while for cases with no detectable impacts across 20 realizations we present a one-sided 95\% upper bound. %
In addition, we show the number of detectable impacts predicted by the fitting function defined in Eq.~\ref{eq:Ndet_scaling}.}
\label{fig:results}
\end{figure*}

We have chosen the polynomial subtraction method to regulate the growth of impacts at large $b$.  Large $b$ impacts modify the orbital track of the stream, which shifts $\phi_2$ and $v_r$ along the whole stream without being identifiable as an impact.  By subtracting a third-order polynomial in our definition of the SNR, we have a robust way of removing these impacts, without relying on a hard cutoff on impacts or expensive modeling of the MW potential. We will evaluate the impact of this subtraction technique in the next section.

\section{Results}
\label{sec:results}

\begin{figure*}[t]
\centering
\includegraphics[width=0.495\textwidth]{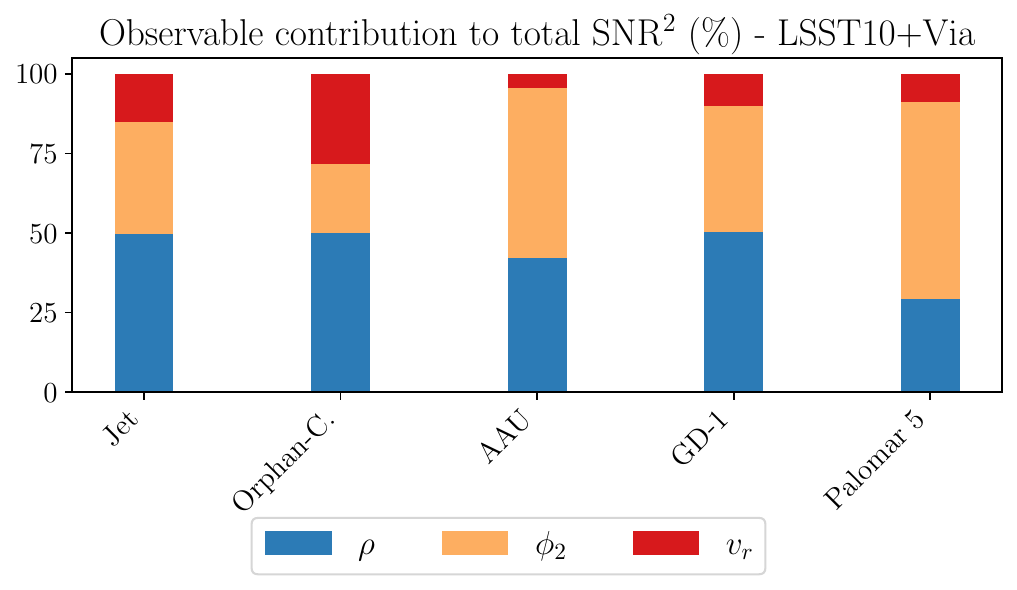}
\includegraphics[width=0.49\textwidth]{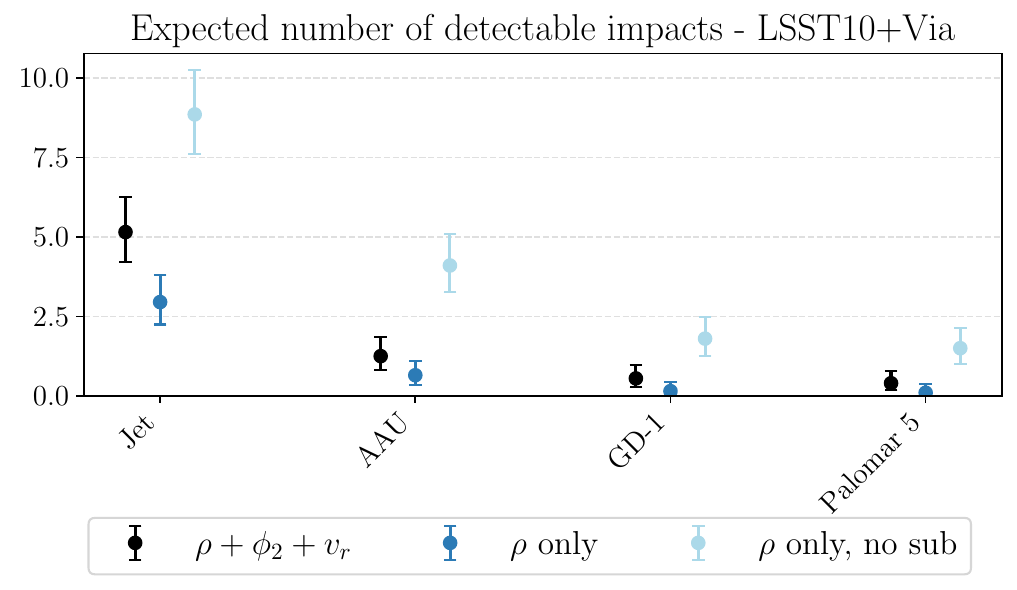}
\caption{\textbf{Left}: Contributions of different observables to the SNR${}^2$ for detectable impacts. \textbf{Right}: Number of detectable impacts for all three observables (black), density only (dark blue), and density only without proxy-model subtraction (light blue) for top-ranked streams. Orphan-Chenab is omitted because the additional simulations required when no proxy model is used are too computationally expensive. }
\label{fig:observable_contribution}
\end{figure*}

Using the methods described in Sec.~\ref{sec:modeling} and Sec.~\ref{sec:impacts}, we estimate the expected number of detectable impacts for all the streams selected in Sec.~\ref{sec:stream_catalog} and listed in Tab.~\ref{tab:selected_streams}. For each stream, we generate 20 realizations of encounters over the stream lifetime, in order to obtain sufficient statistics on the number and distribution of impacts.

Under the best-case observational scenario (LSST10+Via), the expected number of detectable impacts $N_\mathrm{det}$ for each stream is shown as the blue data points in the left panel of Fig.~\ref{fig:results}, as well as Tab.~\ref{tab:stream-entries}. Note that error bars on $N_{\rm det}$ give the 95\% containment on the Poisson mean from the 20 realizations of the stream history, while an individual realization has Poisson fluctuations about the mean.

As we can see, only a small subset of streams are expected to experience non-zero detectable impacts. Jet is the clear standout with $N_\mathrm{det}=5.15$, followed by Orphan-Chenab, ATLAS-Aliqa Uma, GD-1, and Palomar 5 with expected values in the range 0.4-1.4, while $N_\mathrm{det}$ for most other streams remains close to zero. The top ranked streams all have the properties of being especially old, thin, and dense, among which Jet has a particularly old age of $T_\mathrm{form}=4.8$ Gyr. Furthermore, it has the largest physical length $l$ among the streams considered, and its orbit with $r_{\rm peri} = 13.2$ kpc enables it to encounter high subhalo density over its lifetime.

As a comparison, the results for a present-era observational scenario (Gaia+DESI) are shown in the right panel of Fig.~\ref{fig:results}. Most streams have an expected number of detectable impacts consistent with zero. GD-1, Orphan-Chenab, and ATLAS-Aliqa Uma are the only three streams with detectable impacts over 20 runs, with expected values in the range of 0.1-0.45. This means that we can barely detect any single strong impact from present-era data. Interestingly, the improvement for Jet from Gaia to LSST10 is much larger than that for ATLAS-Aliqa Uma and other top-ranked streams: the expected number increases from 0 to 5.15 for Jet, compared with an improvement of 0.1 to 1.25 for ATLAS-Aliqa Uma. This difference is likely driven by the large heliocentric distance of Jet of 30.7 kpc, where the deeper photometry expected from LSST10 provides a substantially larger gain in the number of observable stream stars.

The pattern of encounters across different streams can be understood in terms of a simple scaling behavior. We find that for the LSST10+Via observational scenario, the number of expected detectable impacts, $N_{\rm det}$, is approximately described by the fitting function:
\begin{equation}
 N_\mathrm{det}\simeq 0.00088~N_\mathrm{enc} \left(\frac{w_\mathrm{sim}}{0.2~\mathrm{deg}}\right)^{-0.40}\left(\frac{\rho_\mathrm{obs}}{200~\mathrm{deg}^{-1}}\right)^{0.98} ,
 \label{eq:Ndet_scaling}
\end{equation}
where $N_\mathrm{enc}$ is the prediction obtained using Eq.~\ref{eq:Nenc_orbit}, with $b_\mathrm{max}=10~\mathrm{kpc}$ and $M_\mathrm{min}=10^5 M_\sun$ in the subhalo mass function. For the present-day observational scenario Gaia+DESI, the best-fit fitting function gives a scaling of $N_\mathrm{det}\simeq 0.00044~N_\mathrm{enc} \left(\dfrac{\rho_\mathrm{obs}}{200~\mathrm{deg}^{-1}}\right)^{0.51}$. On top of $N_\mathrm{enc}$, the stream width and the observed density further modulate the detectability of a single impact, as we showed in our previous work~\citep{lu2025detectabilitydarkmattersubhalo}.
The best-fit scaling law is shown as the red stars in the same plot.
Broadly speaking, streams with larger $N_{\rm enc}$, narrower width, and higher observed densities are the most promising targets for subhalo searches, although additional geometric effects can still matter at the order-unity level. A table of all stream properties entering into Eq.~\ref{eq:Ndet_scaling}, as well as values for $N_{\rm det}$, can be found in Tab.~\ref{tab:stream-entries}. For an analytic estimate of $N_\mathrm{enc}$ based on circular orbits, we also give all the properties that enter $N_\mathrm{enc}^\mathrm{circ}$ as in Eq.~\ref{eq:n_enc}. The physical length is estimated by $l=r_h \ell^\mathrm{90-10}$, and the subhalo density $n_\mathrm{sub}$ is evaluated at the time-averaged distance to Galactic Center $\left<r_\mathrm{gc}\right>=\frac{r_\mathrm{apo}+r_\mathrm{peri}}{2}(1+\frac{ecc^2}{2})$ for a Keplerian elliptical orbit.

\begin{figure}[t]
\centering
\includegraphics[width=\columnwidth]{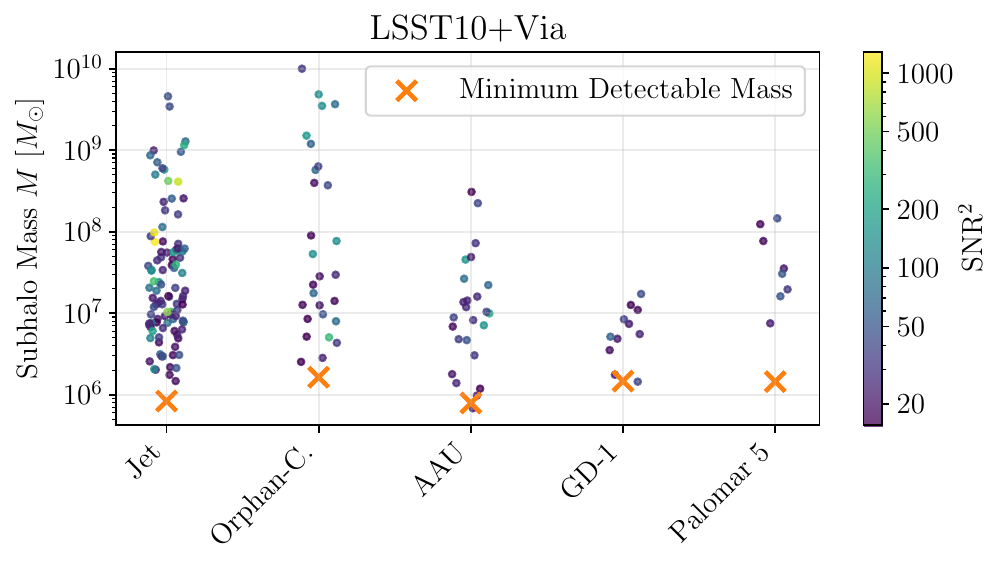}
\caption{For detectable impacts in the top-ranked streams, distribution of subhalo masses from 20 realizations of each stream. The orange cross shows the minimum detectable subhalo mass predicted in~\cite{lu2025detectabilitydarkmattersubhalo} based on a circular stream model and simple scaling of stream properties. Each data point is given a small random horizontal offset for visual clarity.}
\label{fig:min_mass}
\end{figure}

The detectability of impacts is not driven by density perturbations alone. Rather, the total SNR${}^2$ is shared across multiple observables, with the decomposition in Fig.~\ref{fig:observable_contribution} (left panel) indicating contributions of $48.8\%$ from $\rho$, $35.5\%$ from $\phi_2$, and $15.7\%$ from $v_r$ on average across all detectable impacts in all streams. The angular deflection in the vertical direction, $\phi_2$, is as important as density fluctuation. The radial velocity is a noisier channel for impacts, given the lower statistics and larger observational errors, but it still contributes a non-negligible $\sim15\%$ to the total SNR${}^2$.

For the top 5 streams, we show in Fig.~\ref{fig:min_mass} the individual subhalo masses for detectable impacts (data points, with color indicating SNR$^2$) over 20 realizations. The orange cross shows the minimum detectable subhalo mass obtained in~\cite{lu2025detectabilitydarkmattersubhalo} based on a circular stream model and simple scaling of stream properties. The close agreement indicates that the present detailed analysis is broadly consistent with that earlier estimate.

\begin{figure}[t]
\centering
\includegraphics[width=\columnwidth]{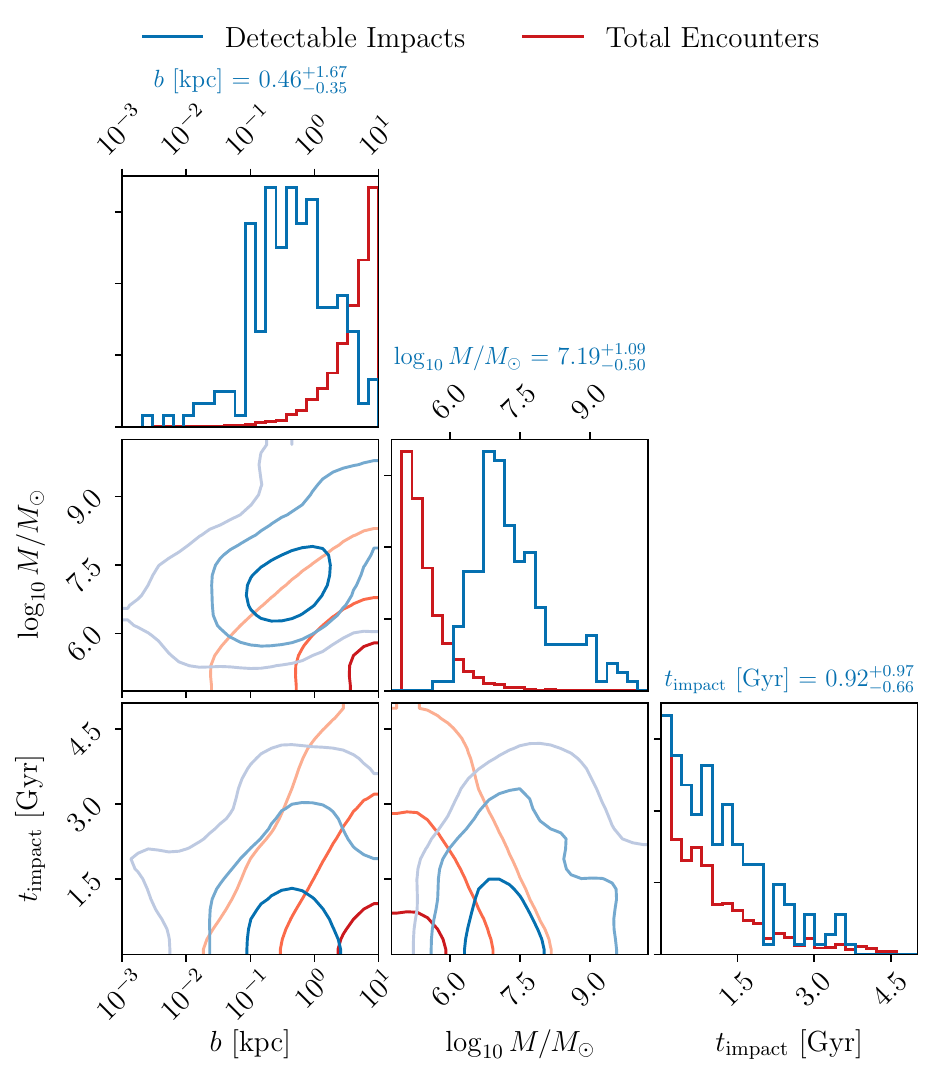}
\caption{Distribution of detectable impacts combined for 20 realizations under LSST10+Via (blue) vs distribution of total encounters for a single realization (red) for all streams. Contours from dark to light indicate the $1\sigma$, $2\sigma$, and $3\sigma$ confidence regions, respectively. The histograms are normalized such that the peak bins for both setups are at the same levels.}
\label{fig:corner_det_vs_enc}
\end{figure}

Fig.~\ref{fig:corner_det_vs_enc} shows the distribution over subhalo impact parameters $M$, $b$ and $t_{\rm impact}$, including all streams. The blue contours and histograms include all detectable impacts from 20 realizations under the LSST10+Via observational scenario. The red contours and histograms include all subhalo encounters, detectable or not, from a single realization. The distribution in $t_{\rm impact}$ of detectable impacts follows similar scaling as the total encounters - the number decreases as $t_{\rm impact}$ gets larger. However, the distribution is less peaked at $t_{\rm impact} = 0 $ for detectable impacts because recent impacts do not have enough time to form gap features. %

\subsection{Effect of polynomial model subtraction}

\begin{figure*}[t]
\centering
\includegraphics[width=\textwidth]{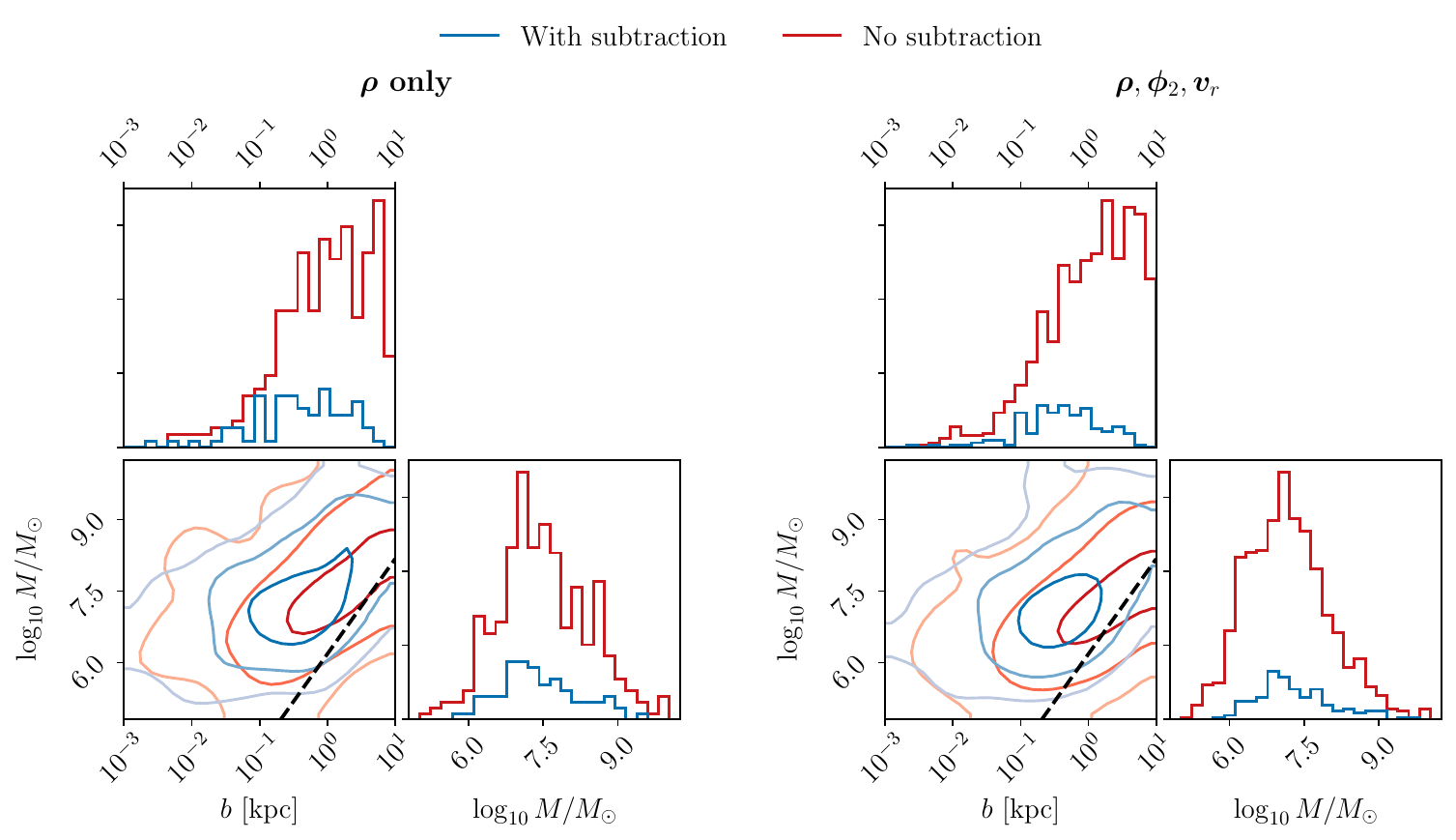}
\caption{ Distribution in subhalo mass $M$ and impact parameter $b$ for detectable impacts on all streams (except Orphan-Chenab) over 20 runs. Orphan-Chenab is omitted because the additional simulations required when no proxy model is used are too computationally expensive. Contours from dark to light indicate the $1\sigma$, $2\sigma$, and $3\sigma$ confidence regions, respectively. We show detectable impacts for analyses considering $\rho$ only (\textbf{left}) and all three observables (\textbf{right}), with proxy-model subtraction (\textbf{blue}) and without proxy-model subtraction (\textbf{red}). The proxy-model subtraction effectively makes the number of detectable impacts converge for both sets of observables. Black dashed lines correspond to the cut $b<5 r_s$ used in previous studies~\cite{Erkal_2016}, where $r_s$ is the scale radius of the Plummer subhalo profile.}
\label{fig:corner_mb}
\end{figure*}

A core aspect of our impact analysis is polynomial model fitting, which  ensures convergence of impact number with impact parameter $b$. We assess the effects of this procedure by comparing with scenarios where we do not perform the polynomial model subtraction. In these cases, we modify the impact sampler pipeline by evaluating the ${\rm SNR}^2$ directly on the signal profile without subtraction. This is done both at the level of orbit integration and at the level of simulated streams, where we directly use Eq.~\ref{eq:deltaX_sim} for the signal perturbation.  We keep the same valid $\phi_1$ region and covariance matrix as in the subtracted case.

In Fig.~\ref{fig:observable_contribution} (right panel), we show the number of detectable impacts if we consider only the density observable with and without proxy-model subtraction. For this observable, the results can converge even without subtraction. The expected number of detectable impacts for Jet changes from $\sim 8.85$ without subtraction (light blue) to $\sim 2.95$ after subtraction (dark blue), while ATLAS-Aliqa Uma drops from $\sim 4.10$ to $\sim 0.65$. Incorporating the $\phi_2$ and $v_r$ observables with subtraction (black) raises the number of impacts back up, although still not to the level for the density observable without subtraction.

The impacts removed by polynomial subtraction are preferentially at large $(b, M)$, as can be seen in Fig.~\ref{fig:corner_mb}. The left panel compares the distribution of detectable impacts in the $(b,M)$ plane with and without proxy-model subtraction, assuming the density observable only. With proxy-model subtraction, the population of detectable impacts is peaked at relatively small $b \lesssim$ 0.5 kpc. Without subtraction, the detectable population extends to larger $b$ and $M$,  and the population has not converged within the wide range we have set. %
Even for the density observable only, where we expect better $b$ scaling, we see that the distribution peaks at $b\gtrsim 1$ kpc and has not converged at $b_\mathrm{max} \sim 10$ kpc. As a reference, we also plot the cut $b<5 r_s$ (black dashed line) typically used for generating subhalo impacts~\cite{Erkal_2016}. It can be seen that the detectable impacts mostly fall within this bound only after applying the proxy-model subtraction, while without subtraction, many more impacts lie beyond it.
When considering all observables (right panel of  Fig.~\ref{fig:corner_mb}) without proxy-model subtraction, the growth of detectable impacts with $b, M$ is even more severe. Here the distribution is much more similar to that of all subhalo encounters, Fig.~\ref{fig:corner_det_vs_enc}.
Hence, polynomial model subtraction preferentially preserves lower $(b,M)$ impacts producing localized stream perturbations, while ensuring convergence on the high $M, b$ ends.

\begin{table*}
\centering
\begingroup
\setlength{\tabcolsep}{2pt}
\renewcommand{\arraystretch}{1.25}
\begin{tabular*}{\textwidth}{@{}c@{\extracolsep{\fill}}rrrrrrrrrrrcc@{}}
\toprule
Name & \makecell{$T_\mathrm{form}$ \\ $[\mathrm{Gyr}]$} & \makecell{$\ell^{90-10}$ \\ $[^\circ]$} & \makecell{$r_\mathrm{h}$ \\ $[\mathrm{kpc}]$} & \makecell{$r_\mathrm{peri}$ \\ $[\mathrm{kpc}]$} & \makecell{$r_\mathrm{apo}$ \\ $[\mathrm{kpc}]$} & \makecell{$\langle r_\mathrm{gc}\rangle$ \\ $[\mathrm{kpc}]$} & \makecell{$10^3 n_\mathrm{sub}$ \\ $[\mathrm{kpc}^{-3}]$} & $N_\mathrm{enc}^\mathrm{circ}$ & $N_\mathrm{enc}$ & \makecell{$w^\mathrm{sim}$ \\ $[^\circ]$} & \makecell{$\rho_\mathrm{obs}$ \\ $[\mathrm{deg}^{-1}]$} & \makecell{$N_\mathrm{det}$ \\ $[\mathrm{LSST10}]$} & \makecell{$N_\mathrm{det}$ \\ $[\mathrm{Gaia}]$} \\
\midrule
Jet & 4.8 & 22 & 30.7 & 13.2 & 34.7 & 26.4 & 6.6 & 855 & 2181 & 0.11 & 371.2 & $5.15^{+1.10}_{-0.95}$ & $<0.15$ \\
Orphan-C. & 1.9 & 76 & 20.7 & 16.0 & 54.6 & 40.6 & 4.9 & 587 & 1260 & 0.32 & 276.7 & $1.40^{+0.62}_{-0.47}$ & $0.25^{+0.33}_{-0.17}$ \\
AAU & 3.0 & 25 & 21.4 & 12.4 & 38.1 & 28.5 & 6.3 & 418 & 687 & 0.11 & 269.4 & $1.25^{+0.60}_{-0.44}$ & $0.10^{+0.26}_{-0.09}$ \\
GD-1 & 2.0 & 60 & 8.0 & 14.3 & 25.5 & 20.7 & 7.5 & 287 & 416 & 0.18 & 154.2 & $0.55^{+0.43}_{-0.28}$ & $0.45^{+0.40}_{-0.24}$ \\
Palomar 5 & 2.0 & 18 & 21.3 & 12.6 & 19.1 & 16.2 & 8.4 & 266 & 533 & 0.07 & 330.0 & $0.40^{+0.39}_{-0.23}$ & $<0.15$ \\
300S & 1.1 & 15 & 15.9 & 4.0 & 45.4 & 33.3 & 5.7 & 57 & 180 & 0.19 & 347.9 & $<0.15$ & $<0.15$ \\
Indus & 0.6 & 72 & 16.6 & 12.3 & 21.4 & 17.5 & 8.2 & 252 & 294 & 0.69 & 265.2 & $<0.15$ & $<0.15$ \\
Jhelum & 0.1 & 23 & 13.0 & 9.4 & 30.5 & 22.7 & 7.2 & 12 & 27 & 0.95 & 369.2 & $<0.15$ & $<0.15$ \\
Leiptr & 0.9 & 54 & 7.1 & 12.2 & 51.9 & 38.2 & 5.2 & 72 & 93 & 0.32 & 33.2 & $<0.15$ & -- \\
M3 & 0.7 & 39 & 9.8 & 5.5 & 16.4 & 12.3 & 9.4 & 94 & 141 & 0.50 & 41.6 & $<0.15$ & $<0.15$ \\
M5 & 0.2 & 22 & 14.2 & 2.7 & 28.5 & 20.9 & 7.5 & 18 & 9 & 0.24 & 31.0 & $<0.15$ & $<0.15$ \\
NGC 3201 & 0.6 & 81 & 4.9 & 8.5 & 33.1 & 24.4 & 6.9 & 66 & 65 & 0.44 & 24.2 & $<0.15$ & $<0.15$ \\
NGC 5466 & 1.0 & 19 & 17.4 & 6.3 & 54.9 & 40.2 & 5.0 & 65 & 102 & 0.09 & 50.6 & $<0.15$ & -- \\
NGC 6397 & 0.1 & 20 & 2.5 & 3.1 & 6.6 & 5.2 & 11.8 & 3 & 11 & 0.74 & 73.8 & $<0.15$ & $<0.15$ \\
\bottomrule
\end{tabular*}
\endgroup
\caption{Stream properties and number of detectable impacts. $N_\mathrm{enc}^{\rm circ}$ is obtained from the analytic estimate for circular orbits, Eq.~\ref{eq:n_enc}. $N_\mathrm{enc}$ is the number of encounters obtained using Eq.~\ref{eq:Nenc_orbit}. We assume $\sigma=180~\mathrm{km/s}$, $b_\mathrm{max}=10~\mathrm{kpc}$, $M_\mathrm{min}=10^5 M_\sun$ for both $N_\mathrm{enc}^\mathrm{circ}$ and $N_\mathrm{enc}$. The density $\rho_\mathrm{obs}$ is the average density in the LSST10 observational scenario in the valid $\phi_1$ region.
The uncertainty quoted for $N_\mathrm{det}$ corresponds to the 95\% confidence interval of the Poisson mean, while for cases with no detectable impacts we quote a one-sided 95\% upper bound. %
Some streams are missing $N_\mathrm{det}$ results (dashed entry) in the Gaia observational scenario because they do not have a sufficient valid $\phi_1$ region due to the low observable density.}
\label{tab:stream-entries}
\end{table*}

\subsection{Dependence on modeling assumptions}

\subsubsection{Mass-loss model}

Our main results assume a constant mass-loss scenario for streams ($\gamma=1$ in Eq.~\ref{eq:mass-loss}), while varying $\gamma$ leads to different inferred stream age (see  Fig.~\ref{fig:stream_age_gamma}) and density profile along $\phi_1$ (see Fig.~\ref{fig:mass-loss-large-scale-sim}). As can be seen from Eq.~\ref{eq:n_enc}, the age of the stream is a particularly important factor that affects the number of subhalo encounters and detectable impacts.

To explore the impact of this systematic uncertainty on detectable impacts, we consider the case $\gamma=2$ in Eq.~\ref{eq:mass-loss} and rerun the entire pipeline.  We choose this as it induces the greatest difference from our fiducial $\gamma = 1$ in both stream age and density profile, while still being in the range of values considered for real systems.
According to~\cite{Gieles_2023}, this is an extreme case with an increasing mass-loss rate, corresponding to the case when the progenitor has a high black-hole fraction.

\begin{figure}[t]
\centering
\includegraphics[width=\columnwidth]{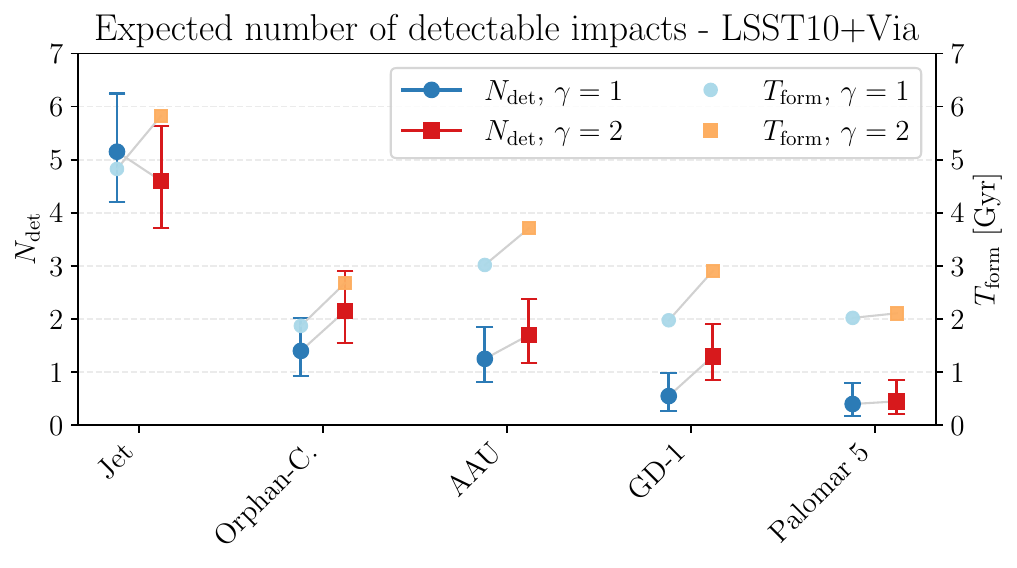}
\caption{Best-fit stream age (light points) and number of detectable impacts (dark error bars) for $\gamma=1$ (blue) and $\gamma=2$ (red) mass-loss models.}
\label{fig:mass_loss_and_age}
\end{figure}

The comparison of the $\gamma=1$ (fiducial) and $\gamma=2$ mass-loss models with respect to both stream age and expected number of detectable impacts under LSST10+Via can be found in Fig.~\ref{fig:mass_loss_and_age}.
The number of detectable impacts increases when we consider $\gamma =2$, consistent with the increase in stream age. The increase in detectable impacts is at the $O(1)$ level, shifting results at a level comparable to our  error bars.

\subsubsection{Subhalo density profile}

We assume a Hernquist subhalo profile for our main results because it shares the cuspy profile of a truncated NFW profile expected for stripped subhalos~\citep{2017MNRAS.466.4974M}, while it is simpler to sample the subhalo profile parameters (scale radius).

To explore the dependence of our results on the subhalo profile, we also consider a Plummer potential representing a ``core''-like density profile, in contrast to the ``cuspy'' profile of the Hernquist potential. We again obtain the best-fit mass-radius relation for Plummer potentials from Via Lactea II\footnote{\url{https://www.ucolick.org/~diemand/vl/data.html}}~\citep{Diemand_2008} data. The scale radius $r_s$ also follows Eq.~\ref{eq:mass-radius}, but here the coefficient $c_{r_s}$ follows a log-normal distribution whose natural log has a mean of 0.69 and a standard deviation of 0.38. This corresponds to a median value of 2.0 for $c_{r_s}$.
As shown in the left panel of Fig.~\ref{fig:vs_dm_models}, assuming a Plummer subhalo profile gives results very similar to the fiducial Hernquist case. %

\subsubsection{Subhalo radial distribution and velocity dispersion}
\label{sec:subhalo_distribution}

\begin{figure*}[t]
\centering
\includegraphics[width=0.485\textwidth]{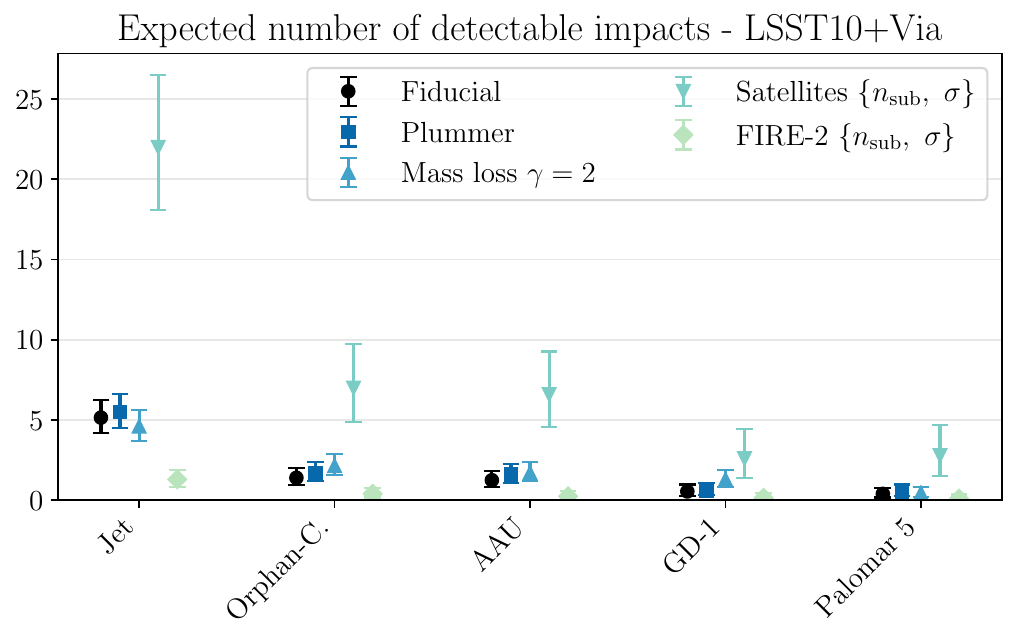}
\includegraphics[width=0.495\textwidth]{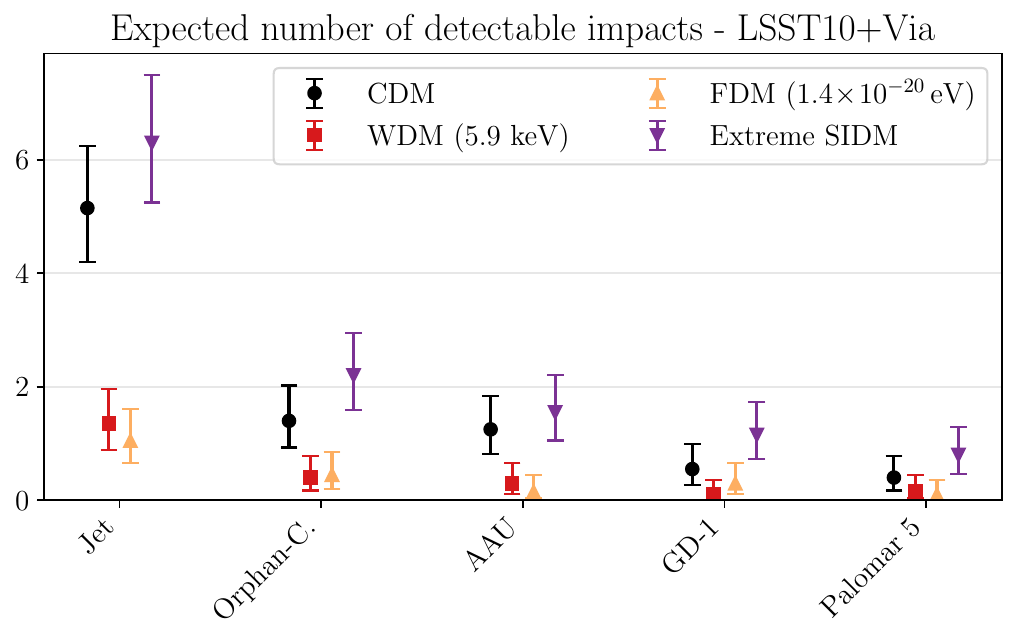}
\caption{Number of detectable impacts for our fiducial scenario (CDM SHMF, Hernquist subhalo density profile, $\gamma=1$ mass-loss model, DMO subhalo population and velocity dispersion, shown as black in both panels) compared to different CDM variations (left) and DM models (right). On the left we show CDM with a Plummer subhalo density profile, CDM with a $\gamma=2$ mass-loss model, CDM with subhalo distributions from MW satellites or FIRE-2. On the right we show WDM, FDM, and an extreme SIDM scenario with core-collapsed halos (Burkert profile with $r_s$ given by Eq.~\ref{eq:rs_SIDM} for $M<10^8 M_\odot$ subhalos).}
\label{fig:vs_dm_models}
\end{figure*}

The subhalo radial dependence in Eq.~\ref{eq:nsub} plays a role in the  subhalo density encountered, $n_{\rm sub}$, and uncertainties in the overall subhalo density directly translate into uncertainties in $N_{\rm det}$. For a fixed host potential, the concentration of the radial profile will also be correlated with the velocity dispersion of the subhalos, with more concentrated profiles having lower velocity dispersions~\citep{10.1111/j.1365-2966.2004.07940.x}. %
The velocity dispersion $\sigma$ affects the overall encounter rate (Eq.~\ref{eq:dn_enc}) as well as the distribution used to sample the subhalo velocities. Our fiducial scenario uses an Einasto subhalo profile, Eq.~\ref{eq:nsub}, and velocity dispersion $\sigma = 180$ km/s based on DMO simulations. However, at the low subhalo masses $\sim 10^6 M_\odot$ and small $r_\mathrm{gc} \sim 10-30$ kpc relevant for subhalo encounters using the currently-known population of stellar streams, there are large uncertainties on the subhalo density. Here we consider two cases beyond our fiducial model to illustrate the impact of this uncertainty.

First, we consider a case where the radial dependence is modified to a cored-NFW profile with $r_s = 21$ kpc, according to measurements of the MW satellite population~\citep{Tan_2026}. The total integrated mass within the virial radius $r_\mathrm{200}$ is fixed to the same value as in the Einasto profile. Since satellite galaxies are thought to occupy subhalos selected by quantities such as peak circular velocity that are not affected by tidal stripping, unlike present-day properties~\citep{2020ApJ...893...48N}, this provides a plausible centrally-concentrated radial distribution closer to a case of no tidal disruption of subhalo.
This is analogous to what was found in \cite{2021MNRAS.503.4075G} if all subhalos are tracked by their peak mass (rather than tidally stripped present-day mass), in which case the subhalo radial distribution follows the underlying host profile. Accordingly, we set the velocity dispersion to $\sigma=120$ km/s in this case, based on measurements of a collection of stars, globular clusters, and satellite galaxies~\citep{Battaglia_2005}. The results are shown as ``Satellites'' in Fig.~\ref{fig:vs_dm_models}. We find an enhancement by a factor of $\sim 4$ on the number of detectable impacts. This is driven primarily by the relatively larger subhalo density in the range $r_\mathrm{gc} = 10-30$ kpc for the more concentrated cored-NFW profile, while the reduced velocity dispersion works in the other direction to mildly reduce the encounter rate. This enhancement would potentially be smaller if we considered the correlation between $r_\mathrm{gc}$ and $M$ in the SHMF, where more massive subhalos are expected to be more concentrated~\citep{Nadler_2023}, since rare high-mass subhalos contribute less to the stream perturber signal than more common low-mass subhalos.

Another scenario we consider is based on the subhalo population from the FIRE-2 simulations for MW-mass halos~\citep{Barry:2023ksd}. In this simulation, baryonic effects lead to increased tidal stripping of subhalos, with a 2-10$\times$ suppression in the subhalo population $n_\mathrm{sub}$ at distances $r_\mathrm{gc}\lesssim50$ kpc. We adopt both the time-dependent SHMF and radius-dependent velocity dispersion in~\cite{Barry:2023ksd}. $n_\mathrm{sub}$ is scaled down by 1.4$\times$ to match the same total MW halo mass of $10^{12}~M_\odot$ as our fiducial Einasto model. The velocity dispersion $\sigma$ ranges from $185-260$ km/s for subhalos at $r_\mathrm{gc}=10-30$ kpc, larger than our fiducial DMO case of $\sigma=180$ km/s. We do not apply any LMC boost factor for this scenario (see discussion of LMC effects in Sec.~\ref{sec:add-uncertainties}). The results based on this FIRE-2 setup are shown in Fig.~\ref{fig:vs_dm_models}. The number of detectable impacts is $\sim4\times$ lower than the fiducial DMO case. Again, the difference is mainly driven by the much lower subhalo number density, but is mitigated by the higher velocity dispersion. We caution that this estimate uses the extrapolated mass function below $10^7 M_\odot$, where simulation results remain uncertain due to a combination of limited resolution (i.e., stripping of subhalos below the mass resolution limit and artificial disruption) and the effects of the Galactic disk. The results may also be sensitive to the algorithm used for subhalo finding and tracking, which may spuriously lose halos as they undergo stripping~\citep{2024ApJ...970..178M,2025ApJ...986..147W}.

To properly account for subhalo mass loss and tidal tracks, it would be valuable to integrate our framework with high resolution simulations or semi-analytic models, similar to what was done in \cite{Adams:2024zhi,menker2024}. Note that despite the different assumptions, our results seem roughly consistent with these works for GD-1 and Palomar 5 (see Sec.~\ref{sec:comparison_w_previous}). However, we see from the discussion above that there remain potentially large systematic uncertainties in the number of detectable impacts from assumptions about how low-mass subhalos undergo tidal stripping or disruption.

\subsubsection{Non-CDM models }

For WDM and FDM, we adopt the SHMF suppression model from~\cite{Nadler_2025}, given by
\begin{align}
    \frac{\left(dn_\mathrm{sub}/dM\right)_\mathrm{XDM} }{ \left(dn_\mathrm{sub}/dM\right)_\mathrm{CDM} } =
    \left(1+\left(\frac{\alpha_\mathrm{s} M_\mathrm{hm}}{M}\right)^{\beta_\mathrm{s}}\right)^{-\gamma_\mathrm{s}}
\end{align}
where the CDM SHMF comes from Eq.~\ref{eq:nsub}. We use their best-fit $\alpha_\mathrm{s}$, $\beta_\mathrm{s}$ and $\gamma_\mathrm{s}$ for WDM and FDM as well as their relations for $M_\mathrm{hm}(m_\mathrm{WDM})$ and $M_\mathrm{hm}(m_\mathrm{FDM,22})$. While \cite{Nadler_2025} primarily presented the suppression in the SHMF for peak (unstripped) halo masses, they found that suppression for present-day halo mass is very similar, and in the WDM case found results consistent with the earlier study \cite{2014MNRAS.439..300L}. We assume that the subhalo radial distribution in WDM and FDM is unchanged relative to CDM, which is consistent with the \cite{Nadler_2025} simulation results for subhalos with peak masses down to $\approx 10^8~M_{\mathrm{\odot}}$; we do not model potential changes to the WDM or FDM radial distribution at lower subhalo masses (e.g., \citealt{2021MNRAS.507.4826L}). We also do not account for the possibility of prompt cusps in the inner region of these halos~\citep{2019PhRvD.100b3523D,2023MNRAS.522L..78D}; the median impact parameter for our detectable impacts is $b= 0.46$ kpc, beyond the region where such features would be present for WDM of mass 6 keV.

Setting the DM mass to be at the current 95\% CL bound from MW satellite abundances in each model, $m_\mathrm{WDM} = 5.9$ keV and $m_\mathrm{FDM} = 1.4 \times 10^{-20}$ eV~\citep{Nadler_2025}, we obtain the results shown in Fig.~\ref{fig:vs_dm_models}. The suppression scales at these masses are $\alpha_s M_{\rm hm}=9.7\times10^{7}\,M_\odot$ for WDM and $2.3\times10^{8}\,M_\odot$ for FDM. Both models result in a factor of $\sim 4$ suppression in the number of detectable impacts, which is significant in the context of the error on the Poisson mean. However, any single realization of the stream history will have Poisson fluctuations. Given the low expected $N_{\rm det}$ in our fiducial scenario, the Poisson fluctuations will make it challenging to distinguish between the CDM and beyond-CDM models. Many more streams with detectable impacts, or other distinguishing features beyond the number of impacts, will be needed for statistical power.

To mimic an extreme SIDM scenario, we assume that all subhalos with present-day mass below $10^8 M_\odot$ are in the core-collapsed phase.  We choose this extreme case to highlight the sensitivity to core collapse; note that, in recently-studied SIDM models, the core-collapsed fraction peaks at $\approx 50\%$~\citep{2025JCAP...02..053A,2025ApJ...991...69N}, although the details are simulation and model-dependent~\citep{2025arXiv251005258G,2026arXiv260602566S}. Meanwhile, while many subhalos with masses above $\approx 10^8~M_{\mathrm{\odot}}$ are expected to be cored in SIDM models, these systems do not appreciably affect our results.

\begin{figure}[t]
\centering
\includegraphics[width=\columnwidth]{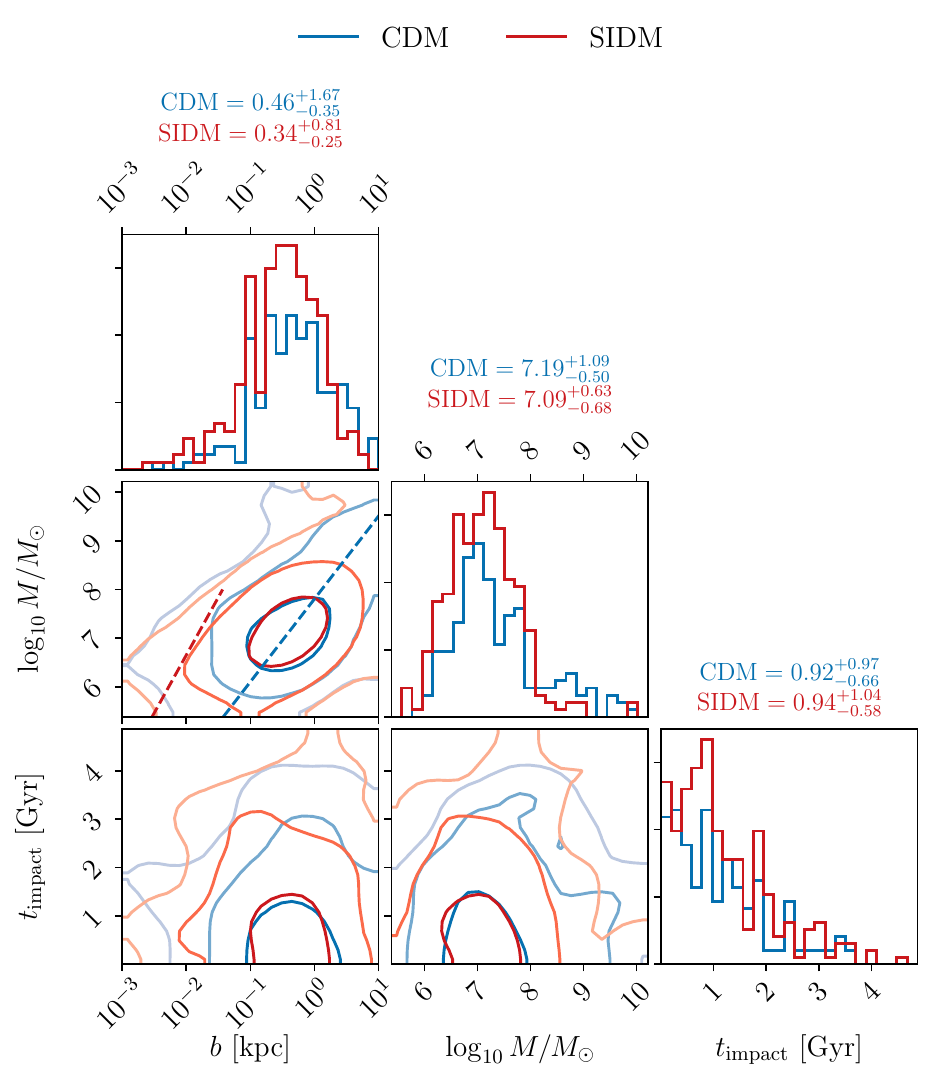}
\caption{Distribution of detectable impacts combined for 20 realizations under LSST10+Via for CDM (blue) vs extreme SIDM (red) for all streams. Contours from dark to light indicate the $1\sigma$, $2\sigma$, and $3\sigma$ confidence regions, respectively. The scale radius of the Hernquist potential for CDM is shown as the blue dashed line. The scale radius of the Burkert potential for SIDM below $10^8~M_\odot$ is shown as the red dashed line.}
\label{fig:corner_cdm_vs_sidm}
\end{figure}

For core-collapsed SIDM subhalos, we use a Burkert density profile with a scale radius much smaller than the CDM case:
\begin{equation}
    r_s = 37~\mathrm{pc}\left(\frac{M}{10^8 M_\odot}\right)^{0.42}
    \label{eq:rs_SIDM}
\end{equation}
Our choice of the Burkert potential comes from fitting the SIDM simulation results in~\cite{zhang2024gd1stellarstreamperturber} for cross section per mass of 50 cm${}^2$ g${^{-1}}$, which gives a scale radius of 12 pc for $M = 7 \times 10^6 M_\odot$. The scaling in mass comes from fitting the bound mass and core radius relation for core-collapsed subhalos using the package \texttt{sashimi-si}\footnote{\url{https://github.com/shinichiroando/sashimi-si}} ~\citep{Yang_2024,Ando_2025}. The result of this SIDM setup is shown in Fig.~\ref{fig:vs_dm_models}. The core-collapsed SIDM increases the number of detectable impacts compared to the fiducial CDM case,  due to the highly concentrated density profile of core-collapsed subhalo. From the comparison of corner plot for CDM and SIDM in Fig.~\ref{fig:corner_cdm_vs_sidm}, we see that SIDM has more detectable impacts for subhalos below $10^8~M_\odot$. Also, the effect from more concentrated subhalo is more important on impacts with smaller impact parameter $b$, shifting the median impact parameter of detectable impacts from 0.46 kpc for CDM to 0.34 kpc for SIDM. %

\subsubsection{Additional uncertainties}
\label{sec:add-uncertainties}

We have considered different mass-loss parameters $\gamma$ as a proxy for stream modeling uncertainty, but our stream simulations are still based on the particle spray algorithm, which does not model the full dynamical history of the stream. Stream dispersion and observations are sensitive to the progenitor black hole population, an effect we have approximated by varying $\gamma$~\citep{roberts2025stellarstreamsblackholerich}. Accreted streams are also sensitive to the DM density profile of their parent satellite, which could lead to perturbations~\citep{2021MNRAS.501..179M}.

We have assumed here a static Milky Way. However, Milky Way dynamics over time can influence subhalo encounter rates and stream evolution.
As a prominent example, the merger of the Large Magellanic Cloud (LMC) leads to increased subhalo density and modified subhalo encounter rates~\citep{Barry:2023ksd,Arora2024ApJ...974..286A}. The size of the effect depends on the individual stream orbit, with a roughly 50\% boost in encounters in some cases. Orphan-Chenab in particular has features that have been attributed to a close encounter with the LMC~\citep{Erkal_2019}. The boost expected for other top-ranked streams is expected to be more mild, however, with \cite{Arora2024ApJ...974..286A} noting no enhancement expected for GD-1. Similarly, the effect on Jet is also not expected to be significant~\citep{Ferguson_2021}.

More generally, stream evolution is affected by time dependence in the Milky Way potential, which is an additional source of perturbations in streams. Perturbations can arise from a time-dependent smooth halo~\citep{Panithanpaisal2025,arora2026streamleftunscathedimprint}, mergers~\citep{Guillaume_asymmetry}, the spiral arms and Galactic bar~\citep{pearson2017gapslengthasymmetrystellar,2017MNRAS.470...60E,Banik_2019_Pal5}, and other localized baryonic perturbers such as globular clusters and molecular clouds~\citep{2016MNRAS.463L..17A,DokeHattori2022}. In general, the effects above raise the threshold for detection. Addressing all of these effects represents important research directions for subhalo detection in streams, but requires dedicated analysis for individual streams as well as significant computational resources to address them in a cosmological context. Our work thus represents an idealized analysis that helps highlight promising streams for further study.

\begin{table}[t]
\centering
\renewcommand{\arraystretch}{1.25}
\begin{tabular}{cccc}
\toprule
 &  & GD-1 & Palomar 5 \\
\midrule
\multirow{2}{*}{ERKAL16}
& $f < 0.75$ & 0.15 & 0.21 \\
& $f < 0.9$  &      & 0.35 \\
\midrule
MENKER24
& $f < 0.9$  &      & 0.44 \\
\midrule
ADAMS24
& $f < 0.9$  & 0.36 &      \\
\midrule
\multirow{3}{*}{This work}
& $\rho$, no sub          & $0.91^{+0.35}_{-0.27}$ & $0.74^{+0.32}_{-0.24}$ \\
& $\rho$                  & $0.08^{+0.15}_{-0.06}$ & $0.05^{+0.13}_{-0.04}$ \\
& $\rho$, $\phi_2$, $v_r$ & $0.28^{+0.22}_{-0.14}$ & $0.20^{+0.19}_{-0.11}$ \\
\bottomrule
\end{tabular}
\caption{Comparison with previous work on subhalo impact rates, in units of
$\mathrm{Gyr}^{-1}$. The labels \textnormal{ERKAL16},
\textnormal{MENKER24}, and \textnormal{ADAMS24} denote results from
\cite{Erkal_2016}, \cite{menker2024}, and \cite{Adams:2024zhi},
respectively. The density ratio $f=\rho/\rho_0$ denotes the gap depth assumed to be required for an impact detection.}
\label{tab:previous}
\end{table}

\subsection{Comparison with previous results}
\label{sec:comparison_w_previous}

A comparison of our results with previous work is shown in Tab.~\ref{tab:previous} for GD-1 and Palomar 5. To account for differences in the assumed stream age, we have divided the total number of impacts by the estimated stream age assumed in each work to calculate the encounter rate.

Most previous works identify a detectable impact as a gap with density ratio $f=\rho/\rho_0$, and require it to be less than some threshold, either $f<0.9$ or $f<0.75$.
\cite{Erkal_2016} only provides an explicit result for $f<0.75$, but a result for $f<0.9$ for Palomar 5 can be obtained from their figures. Note also that \cite{Erkal_2016} assumes a factor of 3 lower subhalo density than us to approximate the effects of the MW disk.
Compared to our analysis, both~\cite{menker2024} and~\cite{Adams:2024zhi} use a subhalo population profile based on semi-analytic modeling, with   \cite{menker2024} considering a subhalo population based on a DMO treatment and \cite{Adams:2024zhi} including the MW disk.
While~\cite{menker2024} uses an analytic model for gap depth, the gap depths in \cite{Adams:2024zhi} are obtained from simulated streams.

Our results without proxy-model subtraction and considering the density observable only are most comparable to those previous works.
Our expected rates are somewhat higher even in this setup because our detection threshold for LSST10 statistics can be lower than requiring $f<0.9$. Overall, our results for detectable impacts are roughly consistent with previous studies, given the different statistical measures and (in some cases) different subhalo populations assumed. %

\section{Conclusions}
\label{sec:conclude}

Close encounters of DM subhalos with stellar streams produce localized features in stream morphology and kinematics. In the near future, stellar stream observations offer a way to detect DM subhalos with present-day mass as low as $\sim 10^6$~$M_\odot$, potentially probing the mass scale at which DM subhalos start to be completely dark, as well as constrain DM theories beyond CDM.

In this work, we have made systematic predictions for the number of individually detectable subhalo impacts in Milky Way streams. These predictions represent the expectation for an idealized realization of the Milky Way, with a static potential hosting a CDM subhalo population described by analytic mass function and radial distribution. Streams are evolved with a particle spray model. Within this framework, we forecast the possibility of detecting subhalo impacts with three promising observables: the density of stars $\rho$, their transverse angle $\phi_2$, and their radial velocity $v_r$, considering both present-day (Gaia+DESI) and future (LSST10+Via) observational scenarios.

Starting from the catalog of 131 Milky Way stellar streams in \cite{BONACA2025101713}, we derived best-fit estimates of stream ages and determined a subset of 14 streams which are sufficiently well-modeled with a particle-spray algorithm to be included in our subsequent analysis of DM subhalo impacts. %
For each stream, we sampled subhalo-encounter histories over the stream lifetime and simulated their effect on observables to assess detectability.

In our framework, an impact is considered detectable if it produces a statistically significant deviation (SNR$^2 > 15.51$)  relative to the best-fit third-order polynomial of the stream. This polynomial serves as a proxy model of the smooth stream. Defining deviations relative to the polynomial proxy model serves two purposes. First, it absorbs large-angular-scale uncertainties in smooth-stream modeling; we show that it can capture variations in mass-loss history and Milky Way potential, which would otherwise appear statistically significant. Second, it regulates the effect of distant encounters, which grow in number with impact parameter $b$ and perturb the stream as a whole. Without proxy-model subtraction, the number of detectable impacts would diverge for the observables $\phi_2$ and $v_r$ for large impact parameter $b$ and subhalo mass $M$ (corresponding to impacts affecting the whole stream at once), while the convergence behavior for $\rho$ is marginal. This method gives predictions for detectable impacts that are convergent with $b$, and preferentially selects for localized impacts.

Our main results are summarized as follows:

\begin{itemize}
    \item With near-future data (LSST10+Via), only 5 of the 14 streams have expected number of detectable impacts $N_\mathrm{det} > 0.2$: Jet ($5.15^{+1.10}_{-0.95}$), Orphan-Chenab ($1.40^{+0.62}_{-0.47}$), ATLAS-Aliqa Uma ($1.25^{+0.60}_{-0.44}$), GD-1 ($0.55^{+0.43}_{-0.28}$), and Palomar 5 ($0.40^{+0.39}_{-0.23}$). With present-day data (Gaia+DESI), $N_\mathrm{det}$ is in the range 0.1-0.45 for 3 streams (Orphan-Chenab, AAU, and GD-1). For 12 streams included in our analysis, our predictions of detectable impact rates are, to our knowledge, the first in the literature.
    \item We have identified Jet as a promising target for further study, owing to its large best-fit age (4.8~Gyr), high physical length, and orbit that samples high subhalo density. Moreover, observations of Jet will particularly profit from the deeper photometry of LSST10. Being on a retrograde orbit and located far from the Galactic Center, Jet is unlikely to be significantly impacted by baryonic structures such as the bar or by the LMC~\citep{Ferguson_2021}.
    \item Whereas previous studies of impact rates have focused on the stream density $\rho$, we have considered the additional observables $\phi_2$ and $v_r$, which contribute between 50\% and 70\% of the total SNR$^2$ for the 5 highest-ranked streams in our analysis. A density-only analysis with proxy-model subtraction yields lower rates of observable impacts in GD-1 and Palomar 5 than previously predicted in the literature. However, inclusion of $\phi_2$ and $v_r$ largely recovers this loss and yields comparable rates as previously predicted.
    \item Without proxy-model subtraction, the number of detectable impacts grows with $b$ and $M$. Previous studies imposed cuts of $b < 5 r_s$ for scale radius $r_s$ of a subhalo of mass $M$. With proxy-model subtraction, the detectable population converges with $b$ and is concentrated within the region $b < 5 r_s$. The distribution of detectable subhalo impacts is peaked at (present-day) subhalo mass of $10^7 M_\odot$ and impact parameter $\sim 0.46$ kpc, with minimum detectable subhalo mass reaching $10^6 M_\odot$.
    \item
    The dominant source of systematic uncertainty in our CDM predictions comes from the subhalo population for $r_\mathrm{gc} \lesssim 30$ kpc in the Milky Way. Compared to our fiducial result, we find that it could be a factor of few larger if we assume a scenario based on measurements of MW satellites, or a factor of few smaller if we assume a scenario based on FIRE-2 simulations.
    \item Changing the underlying DM model leads to a factor of $\sim 4$ decrease of impacts for WDM and FDM and an $O(1)$ increase for an extreme SIDM-like scenario. Distinguishing DM models through the number of impacts alone will require multiple streams with detectable impacts, as well as a reduction in the systematic uncertainty on the MW subhalo population. Constraining the mass and impact properties of individual impacts could help in disentangling the effects of different DM models and that of baryonic physics on the subhalo population~\citep{2021ApJ...920L..11N}.
\end{itemize}

Our predictions should be read as best-case expectations for the number of localized subhalo impacts. We have assumed that impacts do not overlap with each other, which may affect detectability when $N_{\rm det} \gg 1$, such as for Jet or in the scenario with high MW subhalo density.
There are a number of confounding effects that can also induce perturbations in streams, including Milky Way dynamics~\citep{Panithanpaisal2025,Guillaume_asymmetry,arora2026streamleftunscathedimprint} and other baryonic perturbers~\citep{DokeHattori2022,Banik_2019_Pal5,pearson2017gapslengthasymmetrystellar}, which raise the threshold for a detection. At the same time, for a sufficiently strong encounter, it is possible to constrain the subhalo mass and radius, orbit, and impact time~\citep{Erkal_2015_2,hilmi2024inferringdarkmattersubhalo,2025arXiv251207960N}, which can help distinguish the subhalo origin of a perturbation. Assessing the importance of perturbations induced by baryonic effects and disentangling them from subhalo impacts requires dedicated modeling and observation for individual streams. Our results provide additional context for where to invest effort.

In addition, we have focused on 14 streams that are relatively well-characterized with a particle spray model of an unperturbed stream. Other streams beyond these 14 may still be promising: among the additional 35 streams where we performed fits to the age, there are a number that are narrow and at least a few Gyr old, leading to a large estimated number of subhalo encounters (see Table~\ref{tab:fit_all_streams} in App.~\ref{app:extras}). However, the estimated stellar mass is generally lower than for our top-ranked streams. Kwando is an exception to this, and thus a potentially promising object of study. The stream C-19 may also have a particularly high length,  age, and stellar mass~\citep{2025A&A...698A..82Y}. Furthermore, there are many streams missing a stellar mass estimate. Improved characterization and modeling of streams will be key in leveraging many Milky Way streams for DM studies. %

\section*{Acknowledgments}

JL and TL were supported by the US Department of Energy Office of Science under Award No. DE-SC0022104 and a Harold and Suzy Ticho Endowed Fellowship. EB was supported, in part, by the US National Science Foundation under Grant PHY-2210177. VL was supported by the Network for Neutrinos, Nuclear Astrophysics and Symmetries (N3AS) through the National Science Foundation Physics Frontier Center, Grant No. PHY-2020275.

{\emph{Software:}}
Python \citep{python},
numpy \citep{numpy:2020}, scipy \citep{scipy:2020},
astropy \citep{astropy_2013, astropy_2018, The_Astropy_Collaboration_2022},
jupyter \citep{jupyter}, matplotlib \citep{matplotlib}, gala \citep{gala}.

The authors used Claude and ChatGPT to assist with code development and debugging, and to proofread portions of the manuscript. AI-assisted code and suggestions were reviewed and validated by the authors, and the authors take full responsibility for the accuracy and content of the work.

\bibliographystyle{mnras}
\bibliography{refs}

\newpage

\appendix

\section{Additional Tables and Figures}
\label{app:extras}

This appendix contains additional tables and figures supplementing those in the main text.

\begin{table*}[h]
	\begin{center}
		\begin{tabular}{cccccccccc}
            \toprule
            Name & progenitor type & $M_\mathrm{prog}$ & $M_{\mathrm{prog},0}$ & $M_\mathrm{stellar}$ & $\ell^{90-10}$ & $T_\mathrm{form}$ & $w^\mathrm{sim}$ & $w^\mathrm{data}$ & $N_\mathrm{enc}^\mathrm{circ}$\\
            \addlinespace[3pt]
             &  & [$M_\odot$] & [$M_\odot$] & [$M_\odot$] & [$^\circ$] & [Gyr] & [$^\circ$] & [$^\circ$] & \\
            \midrule
            300S & GC & $5.0 \times 10^4$ & $0$ & $5.0 \times 10^4$ & 15 & 1.1 & 0.19 & 0.24 & 57 \\
            ATLAS-Aliqa Uma & GC & $2.0 \times 10^4$ & $0$ & $2.0 \times 10^4$ & 25 & 3 & 0.11 & 0.18 & 418 \\
            C-11 & unknown & $1.1 \times 10^3$ & $0$ & $1.1 \times 10^3$ & 26 & 2.1 & 0.056 & 0.57 & 127 \\
            C-12 & unknown & $1.4 \times 10^4$ & $0$ & $1.4 \times 10^4$ & 10 & 0.99 & 0.16 & 0.46 & 41 \\
            C-13 & unknown & $7.5 \times 10^2$ & $0$ & $7.5 \times 10^2$ & 8 & 0.86 & 0.065 & 0.41 & 20 \\
            C-23 & unknown & $3.1 \times 10^2$ & $0$ & $3.1 \times 10^2$ & 17 & 3.1 & 0.029 & 0.11 & 158 \\
            C-24 & unknown & $2.7 \times 10^3$ & $0$ & $2.7 \times 10^3$ & 26 & 3.2 & 0.049 & 0.36 & 350 \\
            C-25 & unknown & $1.3 \times 10^3$ & $0$ & $1.3 \times 10^3$ & 16 & 1.6 & 0.039 & 0.98 & 61 \\
            C-7 & unknown & $1.5 \times 10^3$ & $0$ & $1.5 \times 10^3$ & 21 & 0.88 & 0.031 & 0.39 & 22 \\
            C-9 & unknown & $6.3 \times 10^2$ & $0$ & $6.3 \times 10^2$ & 17 & 1.8 & 0.15 & 1.3 & 66 \\
            GD-1 & GC & $2.0 \times 10^4$ & $0$ & $2.0 \times 10^4$ & 60 & 2 & 0.18 & 0.23 & 287 \\
            Gaia-1 & unknown & $1.1 \times 10^3$ & $0$ & $1.1 \times 10^3$ & 29 & 1 & 0.088 & 0.28 & 27 \\
            Gaia-11 & unknown & $1.1 \times 10^3$ & $0$ & $1.1 \times 10^3$ & 30 & 3.5 & 0.027 & 0.26 & 215 \\
            Gaia-12 & unknown & $1.2 \times 10^3$ & $0$ & $1.2 \times 10^3$ & 13 & 1.9 & 0.047 & 0.22 & 52 \\
            Gaia-6 & unknown & $1.8 \times 10^3$ & $0$ & $1.8 \times 10^3$ & 12 & 0.91 & 0.068 & 0.24 & 20 \\
            Gaia-8 & unknown & $3.6 \times 10^3$ & $0$ & $3.6 \times 10^3$ & 24 & 1.2 & 0.1 & 0.64 & 77 \\
            Gaia-9 & unknown & $9.5 \times 10^2$ & $0$ & $9.5 \times 10^2$ & 24 & 1.1 & 0.079 & 1.2 & 38 \\
            Hrid & unknown & $2.0 \times 10^3$ & $0$ & $2.0 \times 10^3$ & 47 & 1.1 & 0.13 & 1.1 & 38 \\
            Indus & dwarf & $6.5 \times 10^6$ & $0$ & $3.4 \times 10^4$ & 72 & 0.64 & 0.69 & 0.75 & 252 \\
            Jet & unknown & $2.5 \times 10^4$ & $0$ & $2.5 \times 10^4$ & 22 & 4.8 & 0.11 & 0.15 & 855 \\
            Jhelum & dwarf & $1.3 \times 10^7$ & $0$ & $1.7 \times 10^4$ & 23 & 0.14 & 0.95 & 0.58 & 12 \\
            Kshir & unknown & $2.2 \times 10^3$ & $0$ & $2.2 \times 10^3$ & 21 & 2 & 0.068 & 0.19 & 157 \\
            Kwando & unknown & $4.0 \times 10^4$ & $0$ & $4.0 \times 10^4$ & 42 & 6.3 & 0.19 & 1.2 & 547 \\
            Leiptr & dwarf & $1.0 \times 10^5$ & $0$ & $3.0 \times 10^3$ & 54 & 0.89 & 0.32 & 0.36 & 72 \\
            M2 & GC & $6.2 \times 10^5$ & $6.2 \times 10^5$ & $3.0 \times 10^2$ & 49 & 1 & 0.66 & 0.29 & 197 \\
            M3 & GC & $4.1 \times 10^5$ & $4.1 \times 10^5$ & $2.0 \times 10^3$ & 39 & 0.66 & 0.5 & 0.43 & 94 \\
            M5 & GC & $3.9 \times 10^5$ & $3.9 \times 10^5$ & $7.1 \times 10^2$ & 22 & 0.19 & 0.24 & 0.28 & 18 \\
            M68 & GC & $1.3 \times 10^5$ & $1.3 \times 10^5$ & $1.7 \times 10^3$ & 74 & 1.4 & 0.34 & 0.74 & 157 \\
            NGC 288 & GC & $9.8 \times 10^4$ & $9.6 \times 10^4$ & $2.2 \times 10^3$ & 10 & 1.3 & 1.3 & 0.47 & 60 \\
            NGC 3201 & GC & $2.0 \times 10^5$ & $1.9 \times 10^5$ & $2.1 \times 10^3$ & 81 & 0.6 & 0.44 & 0.39 & 66 \\
            NGC 5466 & GC & $5.8 \times 10^4$ & $5.6 \times 10^4$ & $1.9 \times 10^3$ & 19 & 0.97 & 0.087 & 0.12 & 65 \\
            NGC 6397 & GC & $8.5 \times 10^4$ & $8.2 \times 10^4$ & $2.5 \times 10^3$ & 20 & 0.13 & 0.74 & 0.57 & 3 \\
            New-1 & unknown & $4.2 \times 10^2$ & $0$ & $4.2 \times 10^2$ & 25 & 8.6 & 0.012 & 0.052 & 822 \\
            New-10 & unknown & $1.0 \times 10^2$ & $0$ & $1.0 \times 10^2$ & 9.2 & 0.33 & 0.11 & 0.83 & 3 \\
            New-13 & unknown & $2.5 \times 10^2$ & $0$ & $2.5 \times 10^2$ & 20 & 2.2 & 0.04 & 0.37 & 70 \\
            New-16 & unknown & $1.7 \times 10^2$ & $0$ & $1.7 \times 10^2$ & 11 & 0.94 & 0.058 & 0.29 & 20 \\
            New-2 & unknown & $7.3 \times 10^1$ & $0$ & $7.3 \times 10^1$ & 34 & 0.56 & 0.21 & 2.2 & 15 \\
            New-23 & unknown & $1.2 \times 10^2$ & $0$ & $1.2 \times 10^2$ & 18 & 1.2 & 0.0073 & 0.48 & 15 \\
            New-3 & unknown & $1.2 \times 10^2$ & $0$ & $1.2 \times 10^2$ & 27 & 0.77 & 0.12 & 1.6 & 10 \\
            New-5 & unknown & $1.6 \times 10^1$ & $0$ & $1.6 \times 10^1$ & 70 & 2.2 & 0.013 & 1.1 & 57 \\
            New-6 & unknown & $3.4 \times 10^2$ & $0$ & $3.4 \times 10^2$ & 54 & 2.5 & 0.091 & 2.2 & 127 \\
            New-7 & unknown & $1.1 \times 10^2$ & $0$ & $1.1 \times 10^2$ & 17 & 0.96 & 0.095 & 0.47 & 17 \\
            Orphan-Chenab & dwarf & $1.1 \times 10^6$ & $0$ & $1.3 \times 10^5$ & 76 & 1.9 & 0.32 & 0.39 & 587 \\
            Palomar 5 & GC & $3.0 \times 10^4$ & $1.3 \times 10^4$ & $1.7 \times 10^4$ & 18 & 2 & 0.07 & 0.14 & 266 \\
            Phlegethon & unknown & $1.7 \times 10^3$ & $0$ & $1.7 \times 10^3$ & 45 & 1.1 & 0.14 & 0.86 & 53 \\
            SGP-S & unknown & $5.3 \times 10^2$ & $0$ & $5.3 \times 10^2$ & 23 & 3.3 & 0.034 & 0.2 & 151 \\
            Slidr & unknown & $8.4 \times 10^2$ & $0$ & $8.4 \times 10^2$ & 19 & 0.6 & 0.078 & 1.4 & 9 \\
            Sylgr & unknown & $7.0 \times 10^2$ & $0$ & $7.0 \times 10^2$ & 15 & 0.89 & 0.14 & 0.8 & 12 \\
            Ylgr & unknown & $1.1 \times 10^4$ & $0$ & $1.1 \times 10^4$ & 18 & 0.71 & 0.13 & 0.5 & 46 \\
            \bottomrule
		\end{tabular}
	\end{center}
	\caption{Streams from Ref.~\cite{BONACA2025101713} considered in the fit described in Sec.~\ref{sec:stream_catalog}. These streams satisfy our requirement of having a minimum catalog length of 20$^\circ$ and a reported $M_{\rm stellar}$. Shown are the progenitor type, the estimated original progenitor mass $M_\mathrm{prog}$, the mass of the progenitor today $M_{\mathrm{prog}, 0}$,  the stellar mass $M_\mathrm{stellar}$, the 90$^\mathrm{th}$-to-10$^\mathrm{th}$-percentile length $\ell^{90-10}$, the best-fit stream age $T_\mathrm{form}$, the width of the resulting simulated stream $w_\mathrm{sim}$, the width of the observed stream $w_\mathrm{data}$, and the estimated number of subhalo encounters in the circular approximation $N_\mathrm{enc}^\mathrm{circ}$ assuming $\sigma=180~\mathrm{km/s}$, $b_\mathrm{max}=10~\mathrm{kpc}$, $M_\mathrm{min}=10^5 M_\sun$ (see Sec.~\ref{sec:stream_catalog} for details). \label{tab:fit_all_streams}}
\end{table*}

\begin{figure*}[t]
\centering
\includegraphics[width=0.49\textwidth]{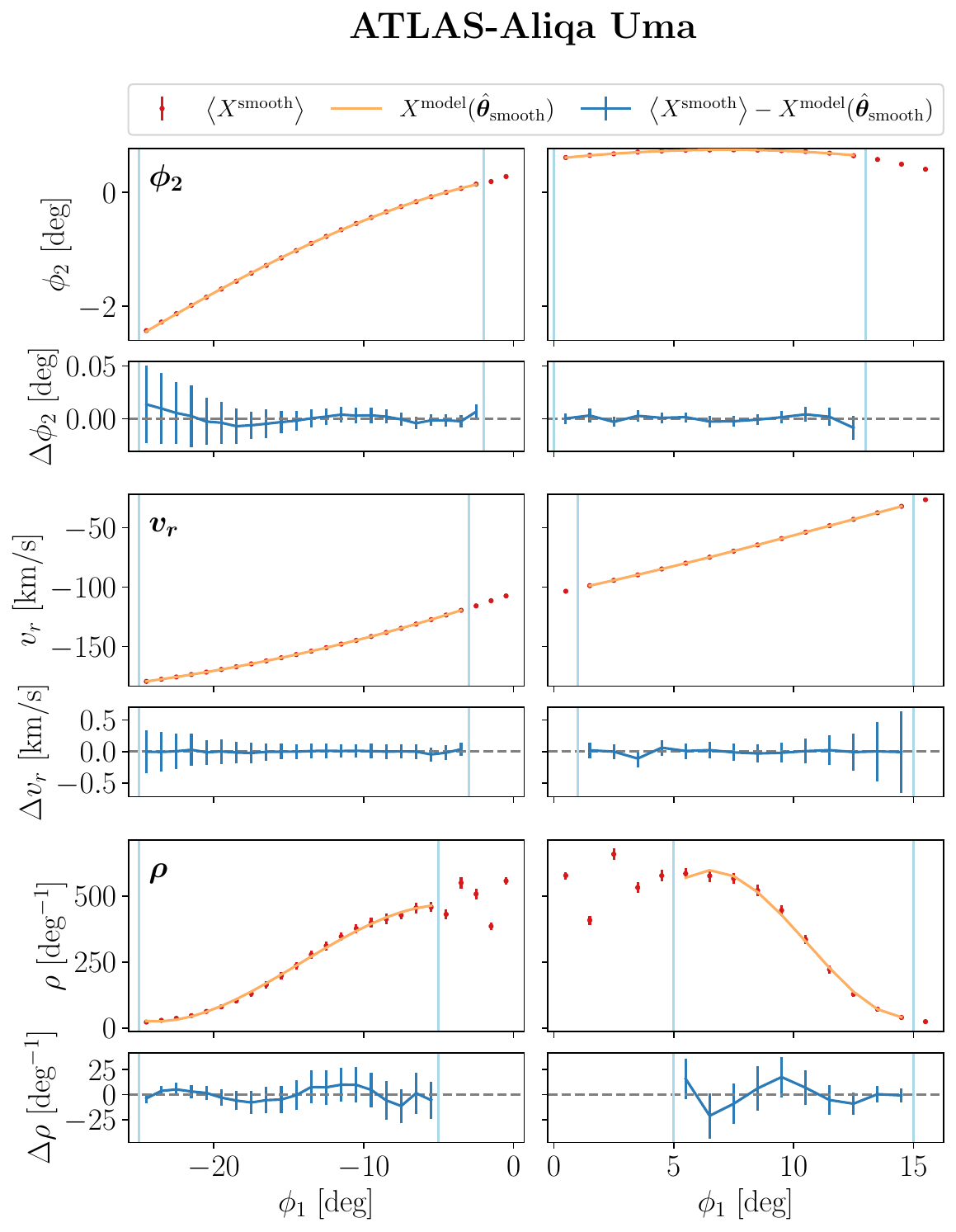}
\includegraphics[width=0.49\textwidth]{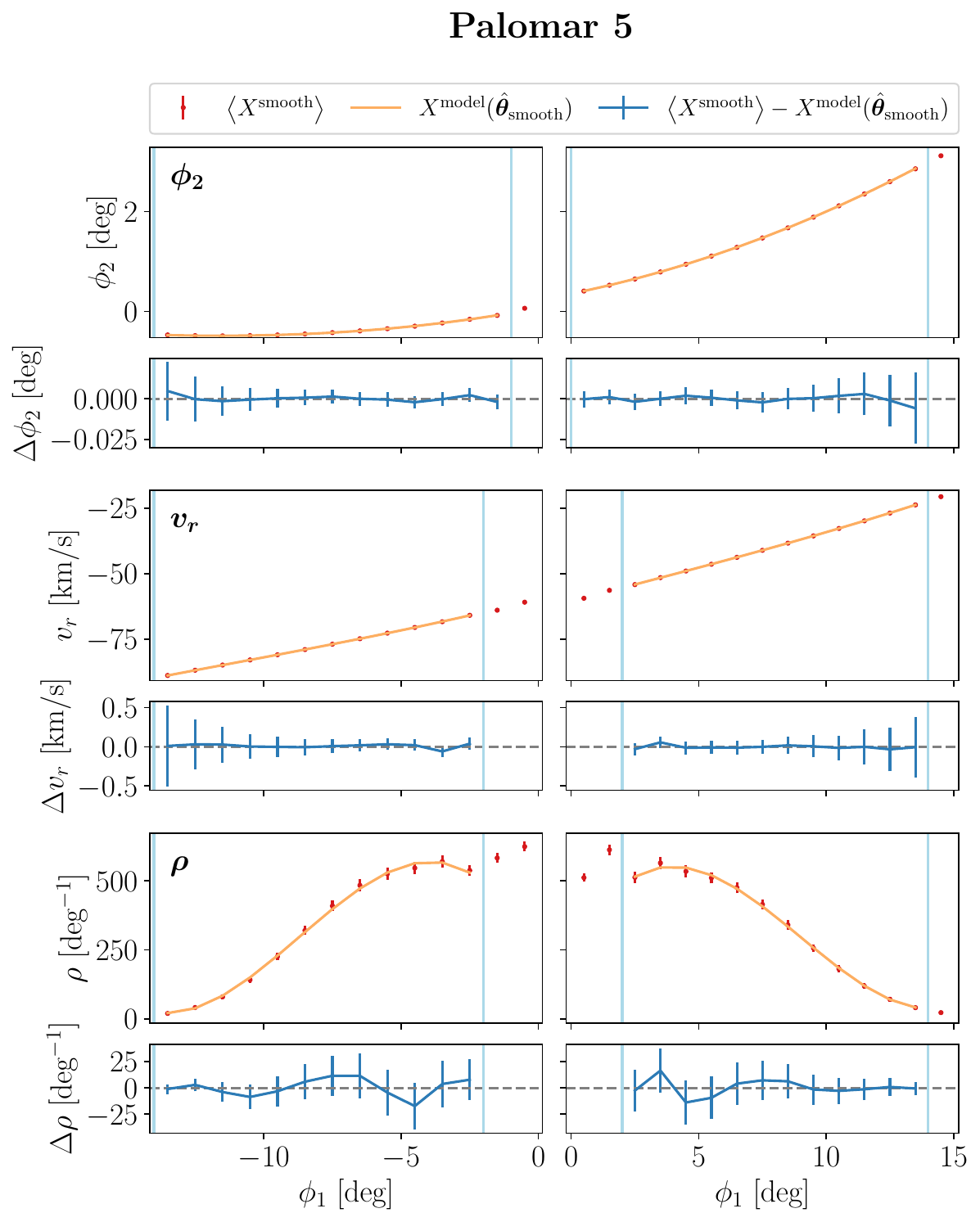}
\includegraphics[width=0.49\textwidth]{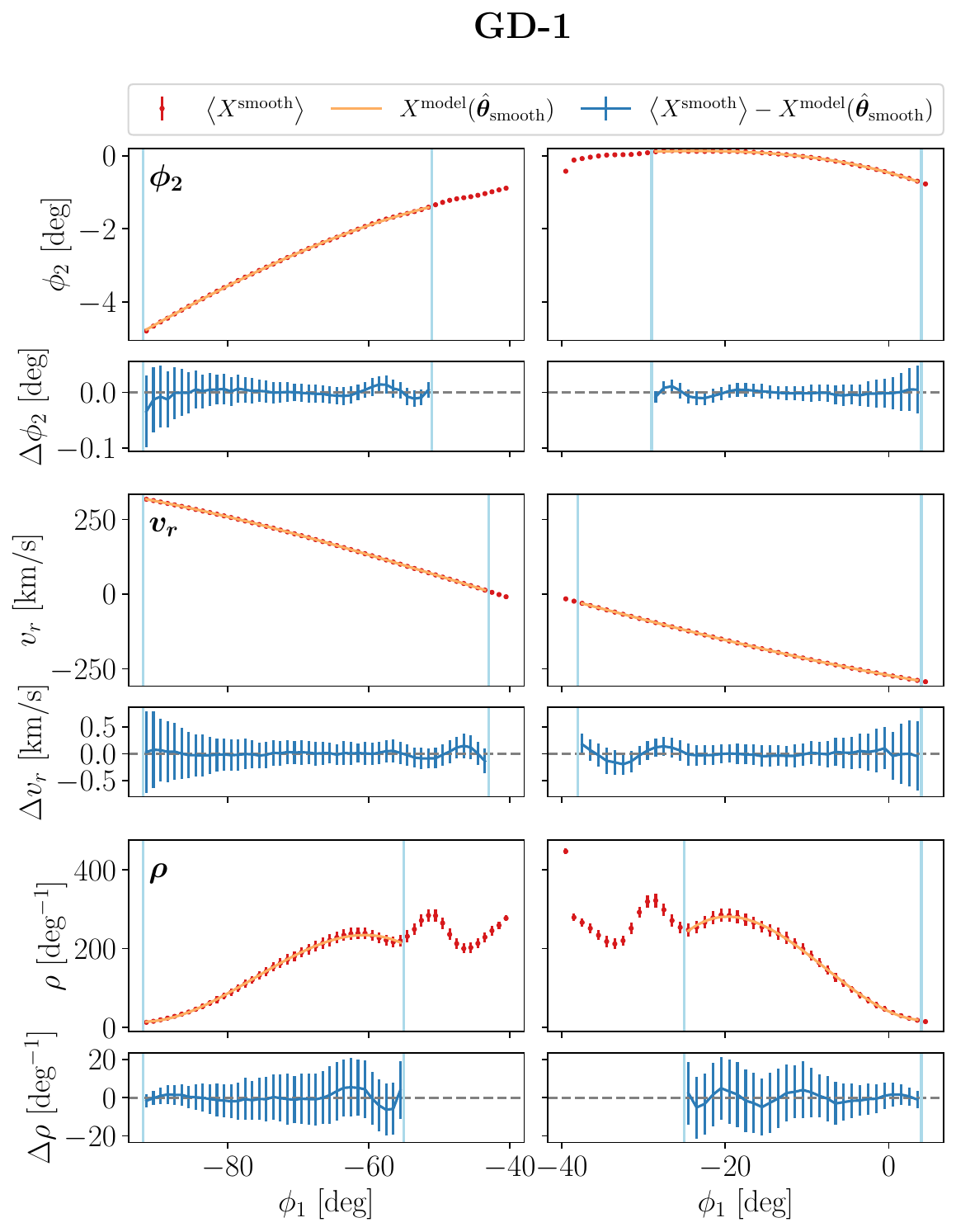}
\includegraphics[width=0.49\textwidth]{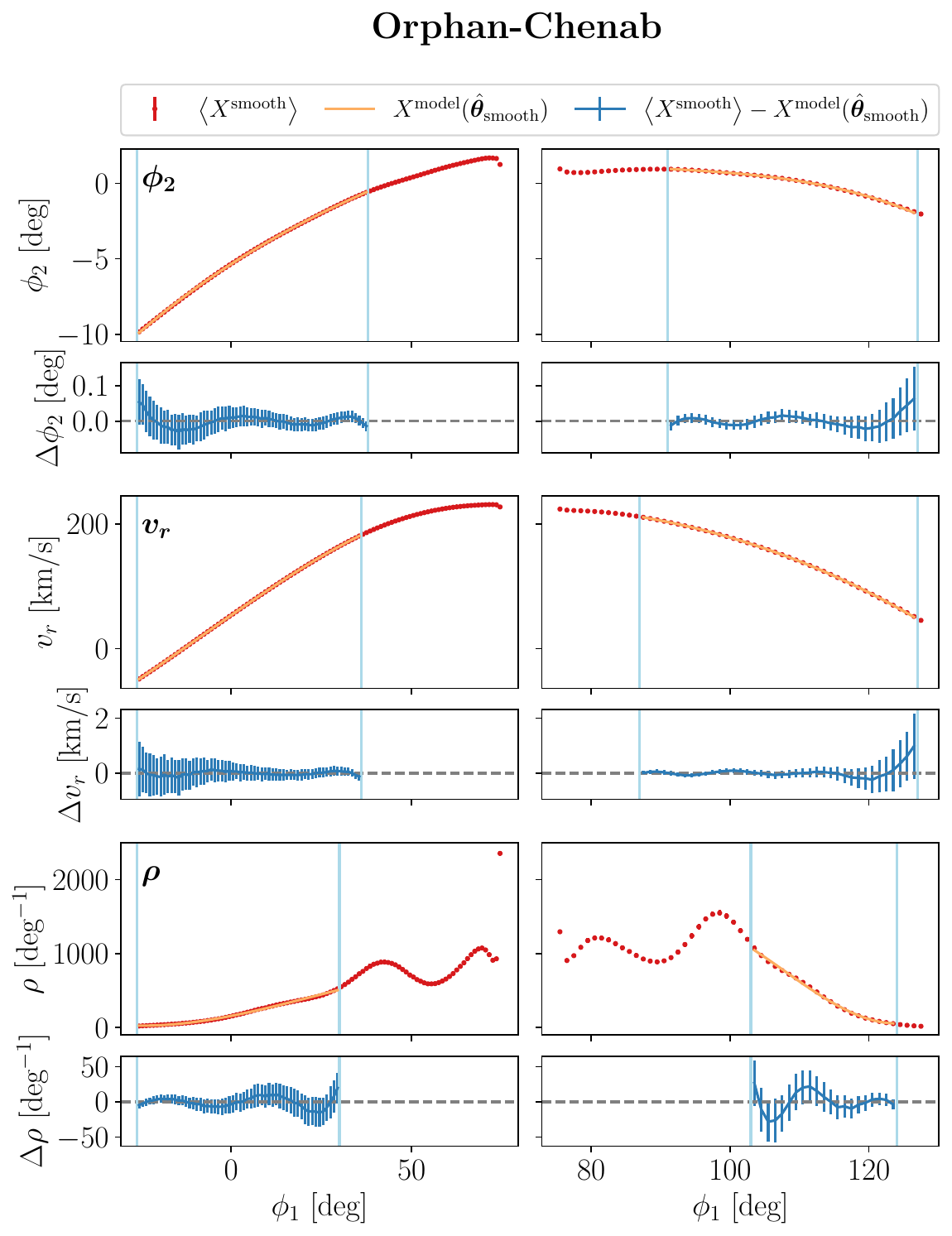}
\caption{Same as Fig.~\ref{fig:fit_proxy} for the next four highest-ranked streams behind Jet.}
\label{fig:smooth_stream_modeling_all}
\end{figure*}

\begin{figure*}[t]
\centering
\includegraphics[width=\textwidth]{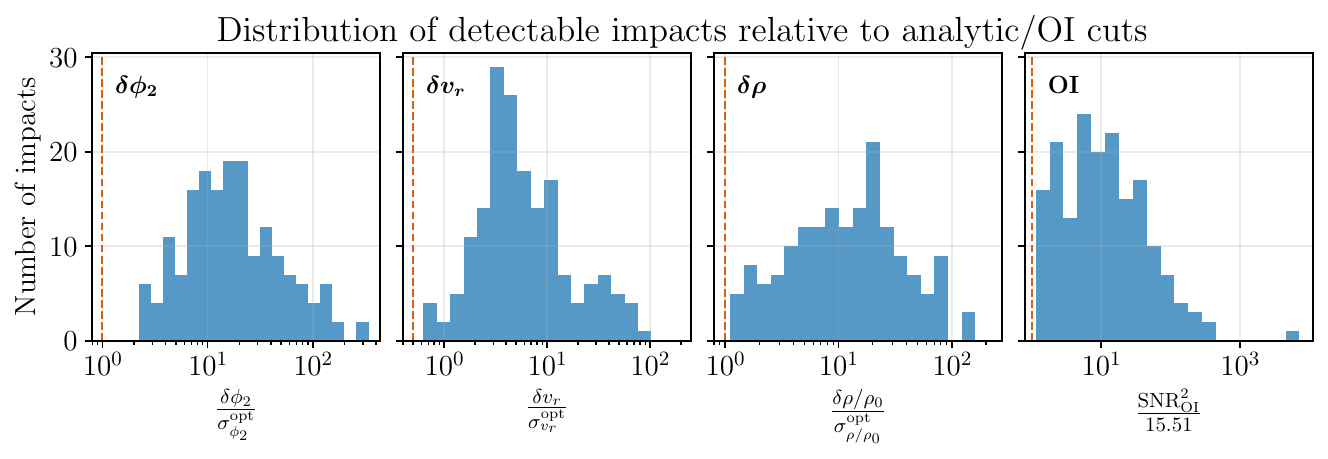}
\caption{For an impact to be considered detectable, we require it to pass at least one of the analytic cuts given in Eqs.~\ref{eq:analytic_cut_phi2}-\ref{eq:analytic_cut_rho}, as well as the OI cut in Eq.~\ref{eq:OI_cut}. %
Here we show the distribution of these analytic estimates for signal strength (first three panels) and SNR${}^2_\mathrm{OI}$ (last panel) for
impacts that are classified as detectable after all steps of the impact analysis. The dashed lines correspond to the cut thresholds given in Eqs.~\ref{eq:analytic_cut_phi2}-\ref{eq:analytic_cut_rho} and Eq.~\ref{eq:OI_cut}. We find that the vast majority of detectable impacts are concentrated at values much higher than these thresholds, which justifies using these cuts as loose filters in our impact analysis.}
\label{fig:analytic_oi_cut}
\end{figure*}

\begin{figure*}[t]
\centering
\includegraphics[width=0.495\textwidth]{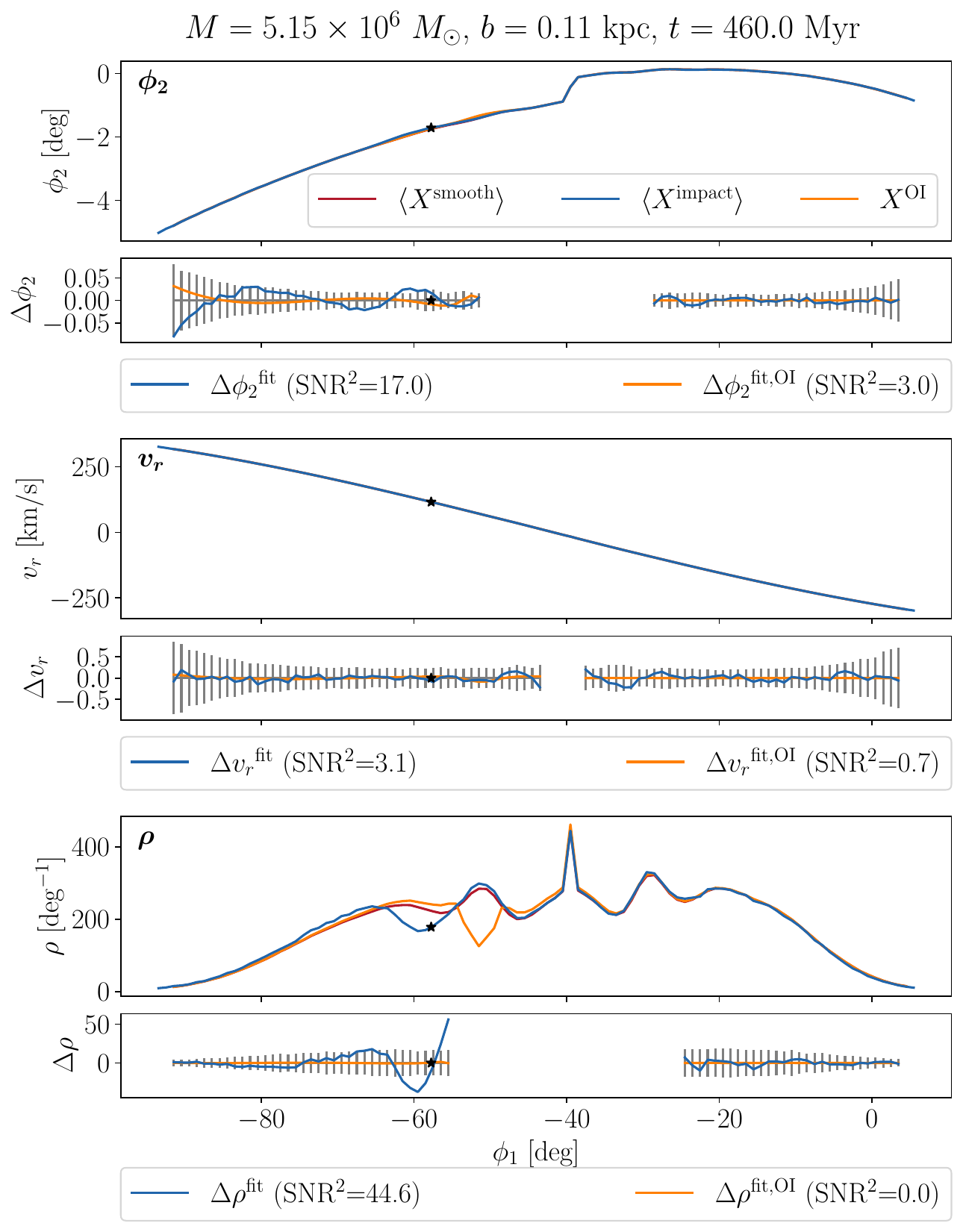}
\includegraphics[width=0.495\textwidth]{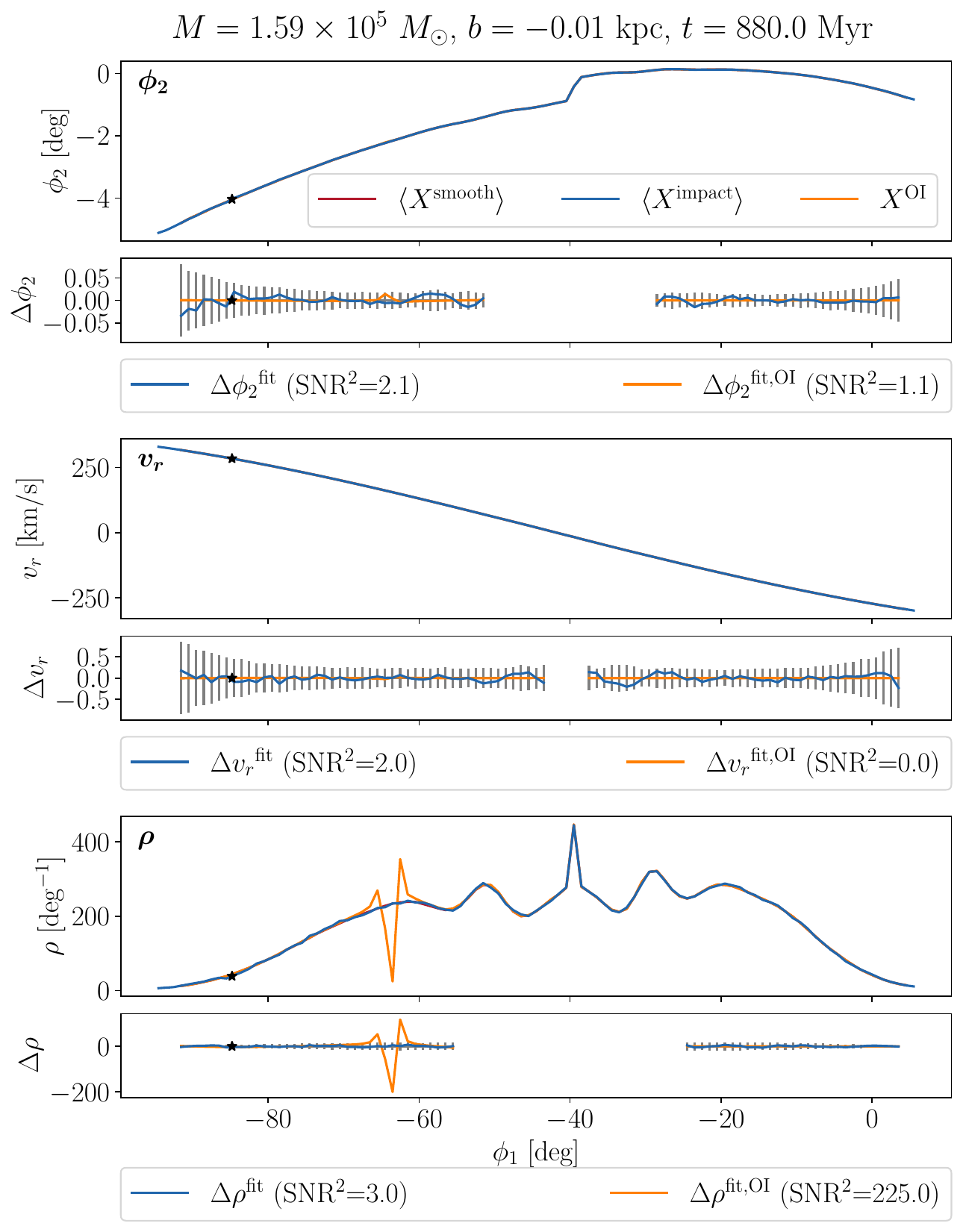}
\caption{Example subhalo encounters for GD-1 where calculating a naive SNR based on orbit integration does not give reliable results. \textbf{Left}: example where the naive SNR from OI is much lower than the SNR from simulation, indicating the necessity of using an impact-location-independent SNR from OI, as defined in Sec.~\ref{sec:OI}. \textbf{Right}: example where the naive SNR from OI is much higher than the SNR from simulation, indicating the necessity to further perform high-statistics simulations for a more accurate SNR. Shown are the observables $\phi_2$ (top), $v_r$ (center) and $\rho$ (bottom). Odd rows show the observables from 250 realizations of smooth-stream simulations (red), $K$ ($K=34$ for GD-1) realizations of impacted stream simulations (blue), and from orbit integration of the impacted stream (orange). The black star marks the impact location at the time of impact. Even rows show the residuals after subtracting the respective best-fit proxy model for the impacted stream from simulation (blue) and orbit integration (orange). The error bars (gray) correspond to the diagonal elements of the enlarged covariance matrix defined in Sec.~\ref{sec:poly_model_fitting}.}
\label{fig:detect_oi}
\end{figure*}

\begin{figure*}[t]
\centering
\includegraphics[width=0.49\textwidth]{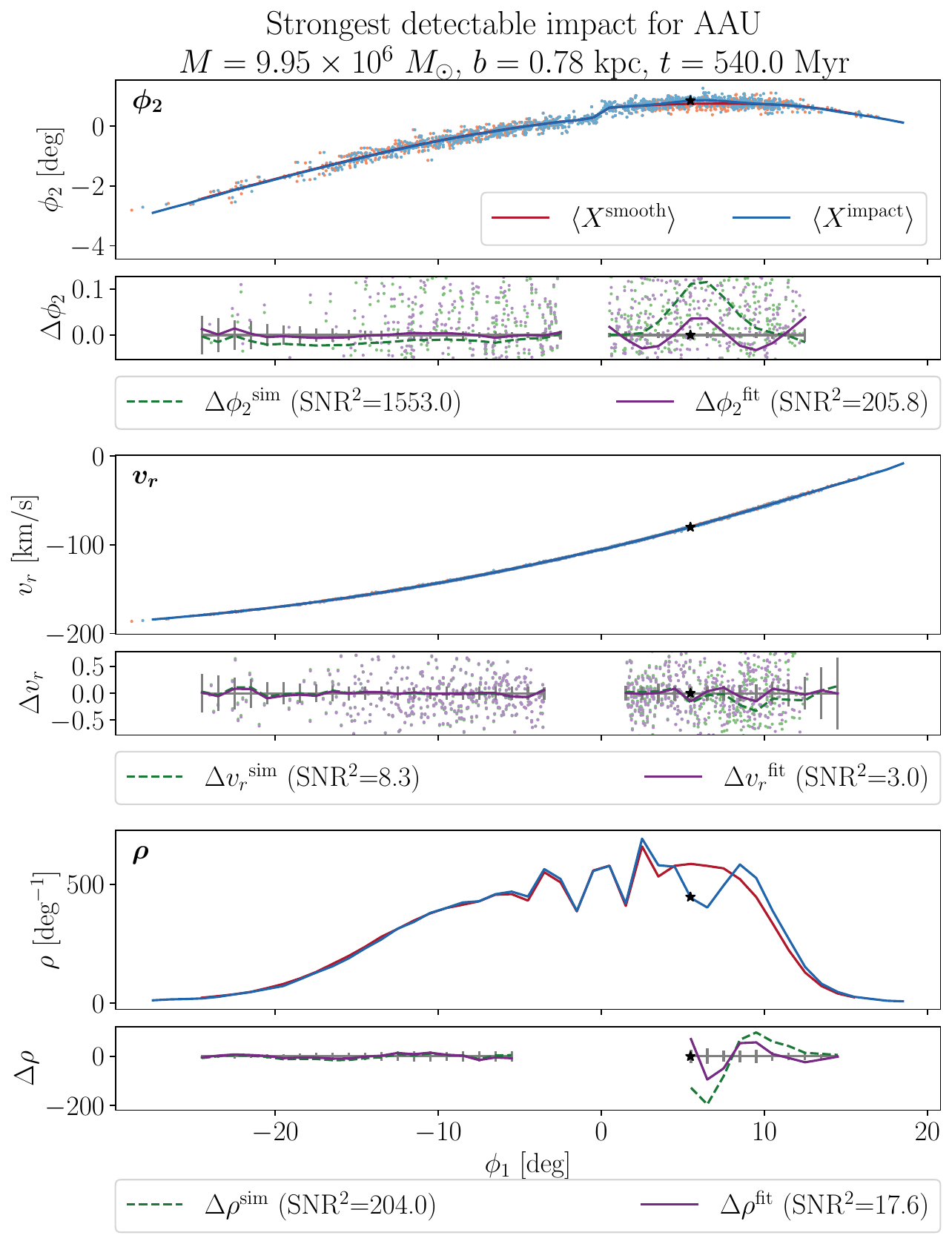}
\includegraphics[width=0.49\textwidth]{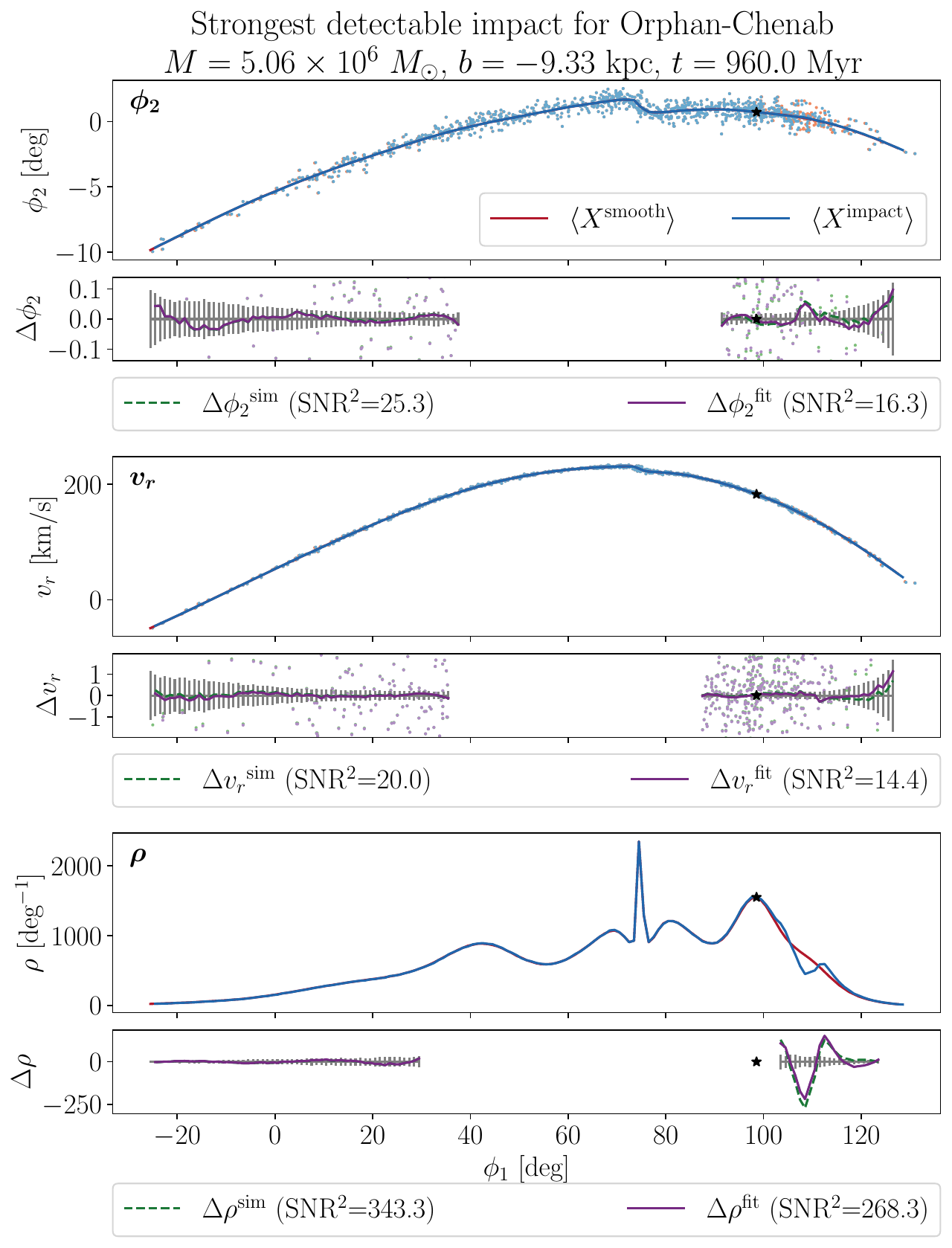}
\includegraphics[width=0.49\textwidth]{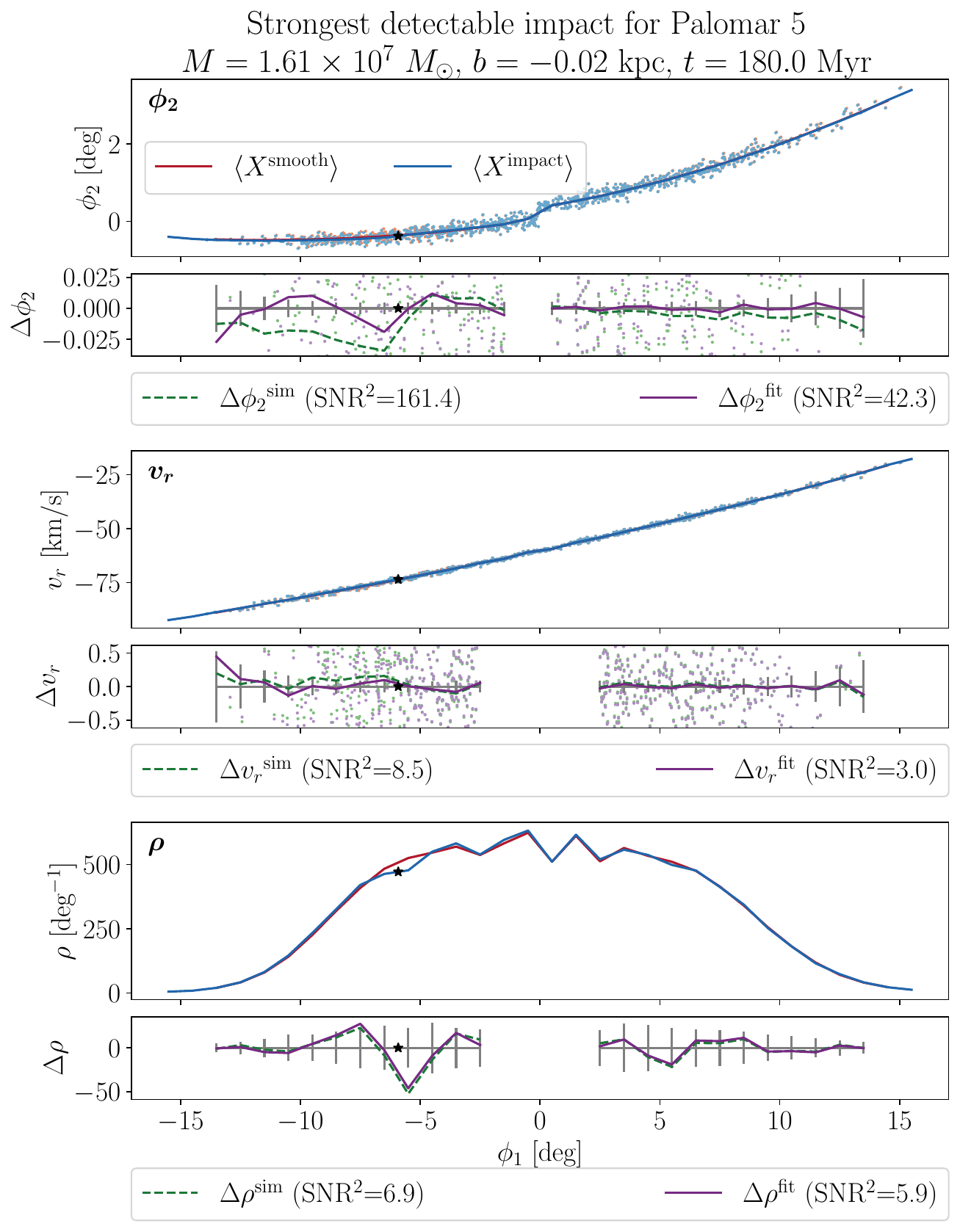}
\includegraphics[width=0.49\textwidth]{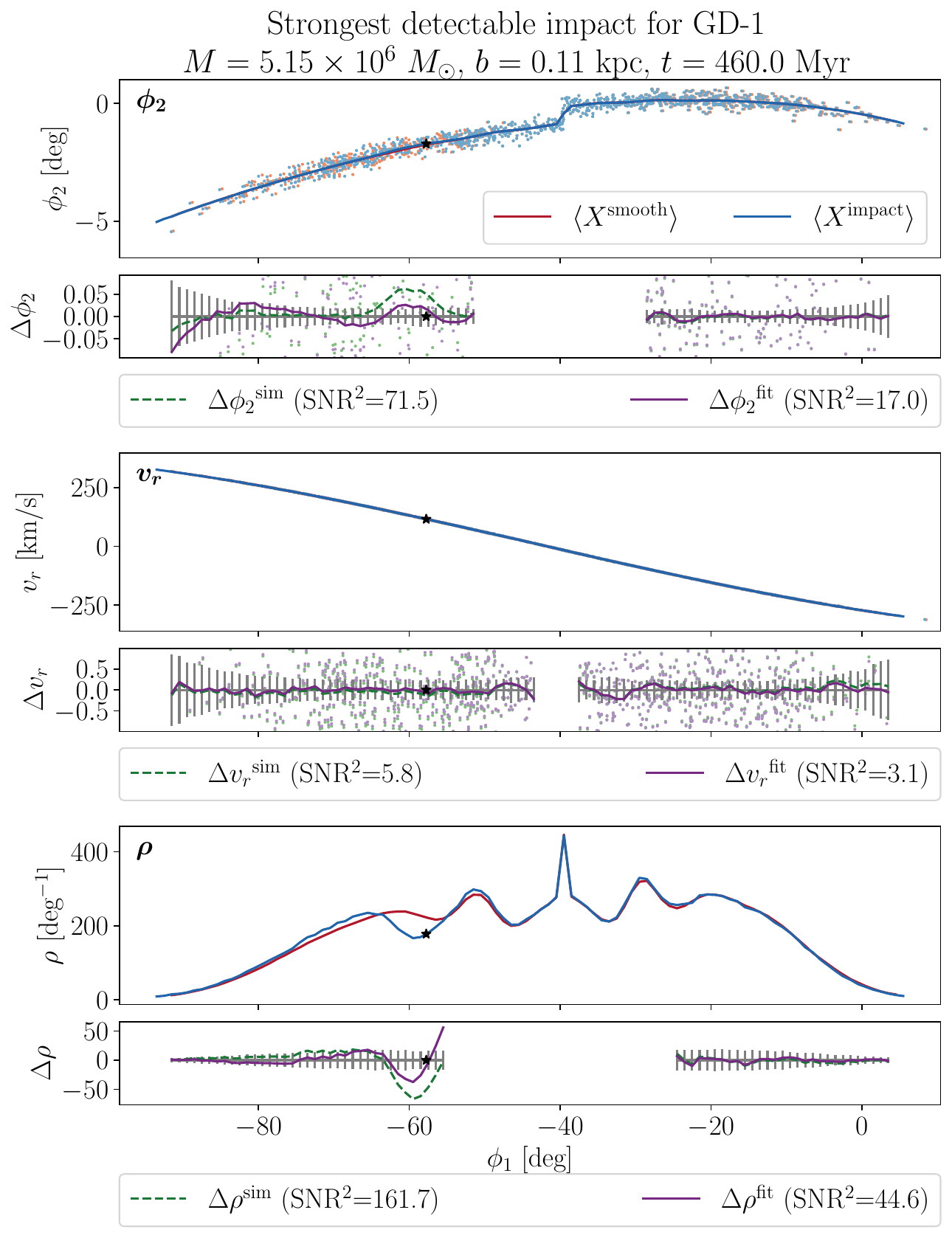}
\caption{Same as Fig.~\ref{fig:detect} for the four highest-ranked streams behind Jet, except that here only the impacts with the largest SNR across 20 runs of the impact history are shown.}
\label{fig:detect_all}
\end{figure*}

\section{Detectability metric}\label{app:detectability_metric}

Here we show that the noise-free SNR$^2$ used in Sec.~\ref{sec:test_statistic} to assess detectability is equivalent to the log-likelihood ratio evaluated on noise-free data for a fixed subhalo impact. We work with the binned observables $X_i$ for $X\in\{\phi_2, v_r, \rho\}$ defined over the valid $\phi_1$ bins of a given stream arm, where it is well-described by the polynomial proxy model. %

In the absence of any subhalo impact, we model the stream data in each bin as $X_i = X^{\mathrm{model}}_i(\bm\theta) + n_{i}$, where $X^{\mathrm{model}}_i(\bm\theta)$ is the polynomial proxy model with parameters $\bm\theta$, and $n_i$ is zero-mean Gaussian noise with covariance $\langle n_{i} n_{j}\rangle = (\tilde C_X)_{ij}$ (given in Eq.~\ref{eq:error_no_hartlap}). In computing likelihoods, we use our finite-sample estimator of the inverse covariance matrix $\hat C_X^{-1}$, Eq.~\ref{eq:error}.

Given stream data $X_i$,  consider two hypotheses:
\begin{enumerate}
    \item $H_0$: no impact is present, so $X_i = X^{\mathrm{model}}_i(\bm\theta) + n_i$.
    \item $H_1$: an impact with signal perturbation $\delta X_i$ is present, so $X_i = X^{\mathrm{model}}_i(\bm\theta) + \delta X_i + n_i$.
\end{enumerate}
Here $\delta X_i$ is the perturbation obtained from high-statistics simulations for a given set of subhalo impact parameters, $
\delta X_i = \langle X^{\rm impact}_i \rangle - \langle X^{\rm smooth}_i \rangle$. We have assumed the perturbed stream is well-modeled by the polynomial model plus $\delta X_i$ and noise, which is valid since we restrict to $\phi_1$ regions where $\langle X^{\rm smooth}_i \rangle$ is well-modeled by the polynomial model.

The log-likelihood under $H_1$, in matrix and vector notation, is
\begin{align}\label{eq:logL1}
    &\ln\mathcal{L}_1(\bm\theta) = \nonumber \\
    -&\frac{1}{2}\left(\vec{X} - \vec{X}^{\mathrm{model}}(\bm\theta) - \delta \vec{X}\right)^T  \!\hat{\bm{C}}_{X}^{-1} \! \left(\vec{X} - \vec{X}^{\mathrm{model}}(\bm\theta) - \delta \vec{X}\right) \, ,
\end{align}
and $\ln\mathcal{L}_0(\bm\theta)$ is the same expression with $\delta \vec{X} = 0$. Under both hypotheses, the parameters $\bm\theta$ are nuisance parameters. Maximizing over $\bm\theta$ in each case gives the profile log-likelihoods $\ln\mathcal{L}_1(\hat{\bm\theta}_1)$ and $\ln\mathcal{L}_0(\hat{\bm\theta}_0)$.

Since the polynomial proxy model is linear in $\bm\theta$, the best-fit parameters can be obtained in closed form. Define the design matrix $U_{i\alpha}\equiv \partial X^{\mathrm{model}}_i/\partial\theta_\alpha$, where $\vec{X}^{\mathrm{model}}(\bm\theta) = \vec{U} \bm\theta$. The best-fit parameters, obtained by maximizing the likelihood under $H_0$ and $H_1$, are respectively found to be
\begin{align}\label{eqn:theta_hat}
    \begin{split}
    \hat{\bm\theta}_{0} &= (\vec{U}^T \hat{\bm{C}}_{X}^{-1} \vec{U})^{-1}\vec{U}^T \hat{\bm{C}}_{X}^{-1} \vec{X}  \\
    \hat{\bm\theta}_{1} &= (\vec{U}^T \hat{\bm{C}}_{X}^{-1} \vec{U})^{-1}\vec{U}^T \hat{\bm{C}}_{X}^{-1} (\vec{X}-\delta \vec{X})\, .
    \end{split}
\end{align}
The log-likelihood ratio,  $\Lambda\equiv -2\ln(\mathcal{L}_0(\hat{\bm\theta}_0)/\mathcal{L}_1(\hat{\bm\theta}_1))$, then takes the form
\begin{equation}\label{eqn:Lambda}
    \Lambda = \vec{r}_0^T \hat{\bm{C}}_{X}^{-1} \vec{r}_0 - \vec{r}_1^T \hat{\bm{C}}_{X}^{-1} \vec{r}_1 \, ,
\end{equation}
where $\vec{r}_{0} \equiv \vec{X} - \vec{U}\hat{\bm\theta}_0$ and $\vec{r}_{1} \equiv \vec{X} - \vec{U}\hat{\bm\theta}_1 - \delta\vec{X}$ are the residuals under $H_0$ and $H_1$, respectively.

The residuals above can be written as
\begin{align}
    \vec{r}_{0} &= \vec{P}  \vec{X}  \\
    \vec{r}_1 &= \vec{r}_0 - \vec{P} \delta\vec{X} \label{eqn:r1_r0}
\end{align}
where the projection matrix $ \vec{P} $ is given by
\begin{align}
     \vec{P} = \bm{1} - \vec{U} (\vec{U}^T \hat{\bm{C}}_{X}^{-1} \vec{U})^{-1}\vec{U}^T \hat{\bm{C}}_{X}^{-1}.
\end{align}
The projection matrix removes the component of the data lying in the polynomial model subspace and satisfies $\vec{P} \vec{U} \bm{\theta} = 0$. Hence, the component of the impact template that cannot be absorbed by refitting the polynomial model is
\begin{align}
    \delta \vec{X}_\perp \equiv \vec{P} \delta \vec X .
\end{align}
Note that $\vec{U}^T \hat{\bm{C}}_{X}^{-1} \delta\vec{X}_\perp = 0$ by construction so that $\delta\vec{X}_\perp$ is orthogonal to the polynomial model subspace, under the inner product defined by $\hat{\bm{C}}_{X}^{-1}$.

We can now obtain the expected log-likelihood ratio for a given impact by considering the Asimov dataset, defined as the data realization in which all observables take their expected values under a given hypothesis, {\emph{i.e.}},\ the noise realization is set to zero~\citep{Cowan_2011}.
Taking the Asimov dataset under the signal hypothesis, $\vec{X} = \vec{X}^{\mathrm{model}}(\bm\theta_\mathrm{true}) + \delta\vec{X}$, the residuals are $\vec{r}_{0} = \delta \vec{X}_\perp$ and $\vec{r}_1 = 0$, giving
\begin{align}
    \Lambda_\mathrm{Asimov} &=
    \delta\vec{X}_\perp^T\, \hat{\bm{C}}_{X}^{-1} \,\delta\vec{X}_\perp = \mathrm{SNR}^2 .
\end{align}
This is precisely the $\mathrm{SNR}^2$ evaluated on noise-free data, used in Sec.~\ref{sec:test_statistic}. $\Delta X^{\rm fit}$ is the residual after polynomial model subtraction, equivalent to $\delta\vec{X}_\perp $.
In practice, we evaluate the $\mathrm{SNR}^2$ by directly performing polynomial subtraction on  $\langle X_i^{\rm impact} \rangle $, rather than first defining $\delta X_i = \langle X_i^{\rm impact} \rangle  - \langle X_i^{\rm smooth} \rangle $. These two differ by a  small amount since we selected on regions where the smooth stream is well described by the polynomial model, Eq.~\ref{eq:cut}.

The log-likelihood ratio above has been defined for a fixed template (fixed impact parameters). Detecting an impact, however, requires searching over all possible impact parameters $\bm \psi$ and the relevant test statistic is a log-likelihood ratio that profiles over these parameters, $q_0 = 2 \log [ {\cal L}( \hat {\bm \psi}, \hat \theta)/{\cal L}_0(\hat \theta) ] $.
To determine whether an impact is detectable, we must compare the expected log-likelihood ratio with the distribution of the test statistic $q_0$ under the null hypothesis of no impact. This distribution was explicitly obtained in \cite{lu2025detectabilitydarkmattersubhalo} for impacts on circular streams and used to set a 95\% CL threshold. In this work, the computational requirements to determine the relevant threshold are far greater and so we use an approximate threshold as discussed in Sec.~\ref{sec:test_statistic}.

The SNR above is also equivalent to the matched-filter SNR for the projected signal template $\delta X_\perp$. To see this, define
the linear statistic
\begin{equation}
T \equiv \delta \vec{X}_\perp^T \, \hat{\vec{C}}_X^{-1} \, \vec{X} \, .
\label{eq:matched_filter_T}
\end{equation}
Under $H_0$ and $H_1$, $T$ is Gaussian with
\begin{equation}
\langle T \rangle_{H_0} = 0 \, , \quad
\langle T \rangle_{H_1} = \mathrm{SNR}^2 \, , \quad
\mathrm{Var}(T) = \mathrm{SNR}^2
\end{equation}
where $\mathrm{Var}(T)$ is the same under both hypotheses. Here we have used that $\delta \vec{X}_\perp^T \, \hat {\vec{C}}_X^{-1} \vec{U} = 0$.  The standard matched-filter signal-to-noise
ratio is then
\begin{equation}
\frac{\langle T \rangle_{H_1} - \langle T
\rangle_{H_0}}{\sqrt{\mathrm{Var}(T)}} = \mathrm{SNR} \, .
\end{equation}
Thus the same quantity can be viewed either as the square root of the Asimov profile log-likelihood ratio or as the matched-filter SNR for the projected impact template. As in the profile likelihood case, this is only for a single fixed template and in a full search one must consider templates over the entire subhalo parameter space. This introduces look-elsewhere effects that raise the threshold for a detectable impact, which must be separately calibrated.

\begin{figure*}[t]
\centering
\includegraphics[width=\textwidth]{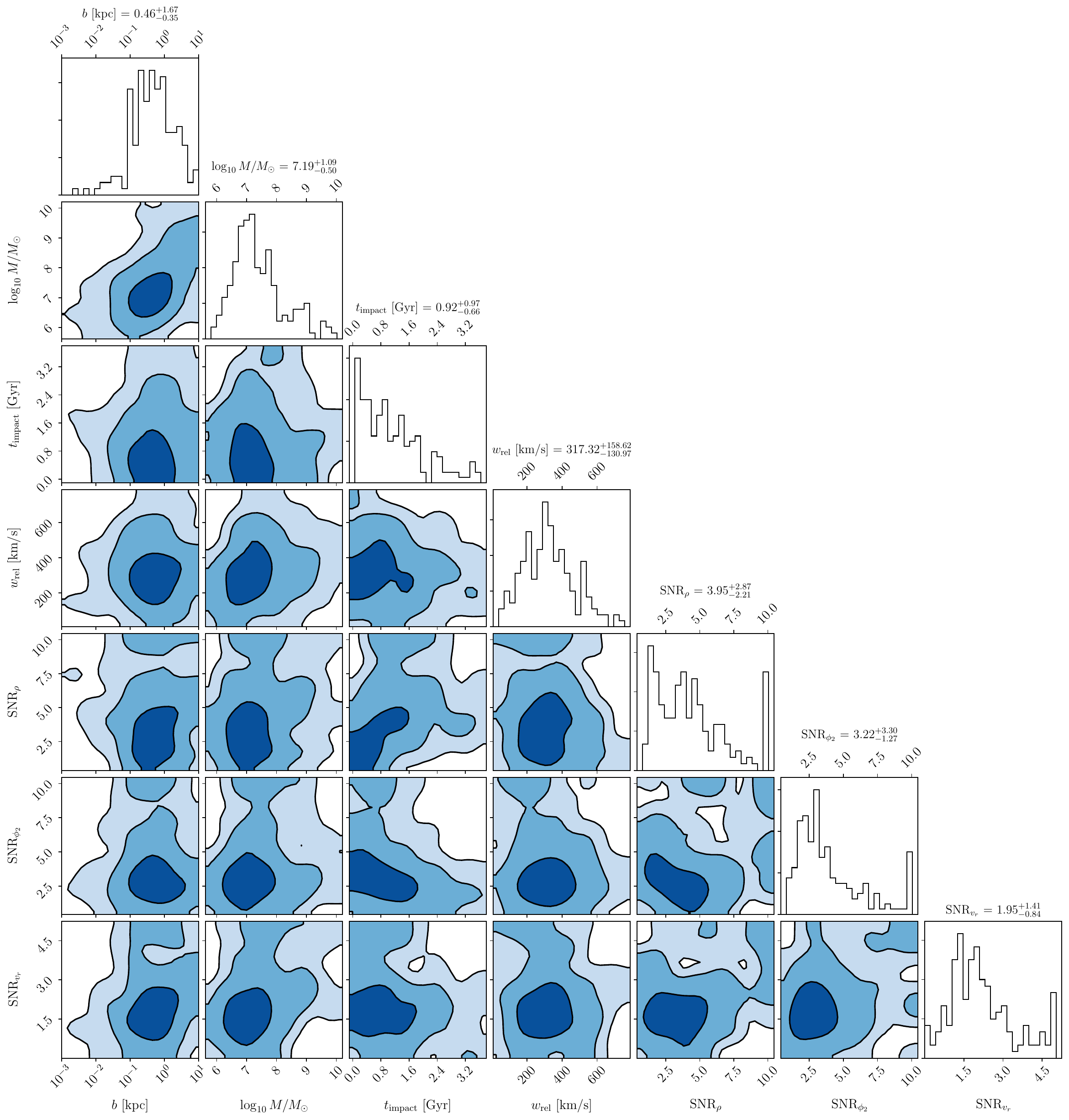}
\caption{Distribution of detectable impacts with respect to subhalo mass $M$, time of impact $t_{\rm impact}$, relative velocity $w_\mathrm{rel}$, and SNR contributions from $\rho$, $\phi_2$, and $v_r$. Contours from dark to light indicate the $1\sigma$, $2\sigma$, and $3\sigma$ confidence regions, respectively. The distributions are compiled from 20 realizations of all streams, under the LSST10+Via observational scenario. %
}
\label{fig:corner}
\end{figure*}

\begin{figure*}[t]
\centering
\includegraphics[width=\textwidth]{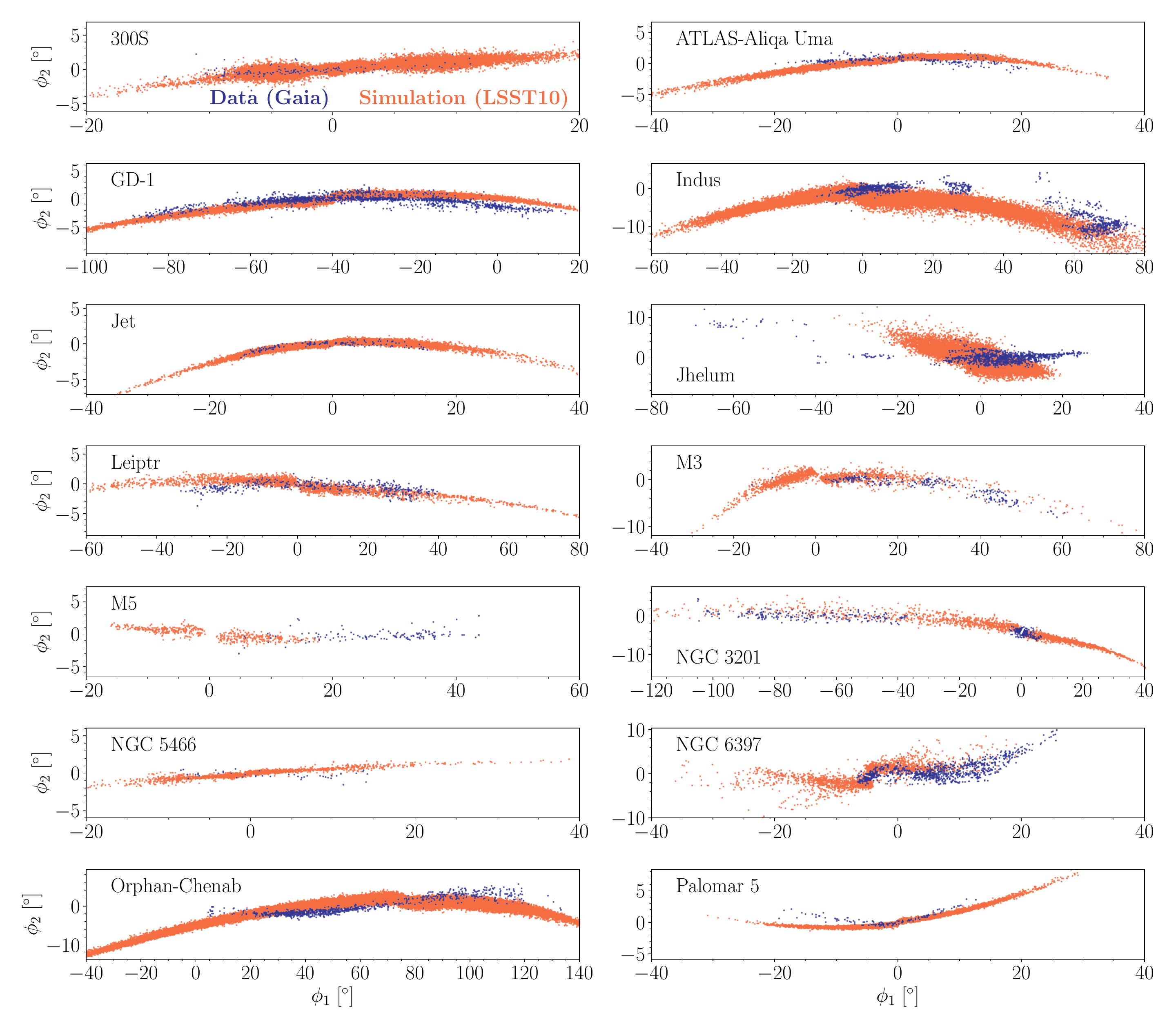}
\caption{Same as Fig.~\ref{fig:sky_plots}, but with expected LSST10 densities for the simulated streams.}
\label{fig:sky_plots_lsst10}
\end{figure*}

\begin{figure*}[t]
\centering
\includegraphics[width=0.9\textwidth]{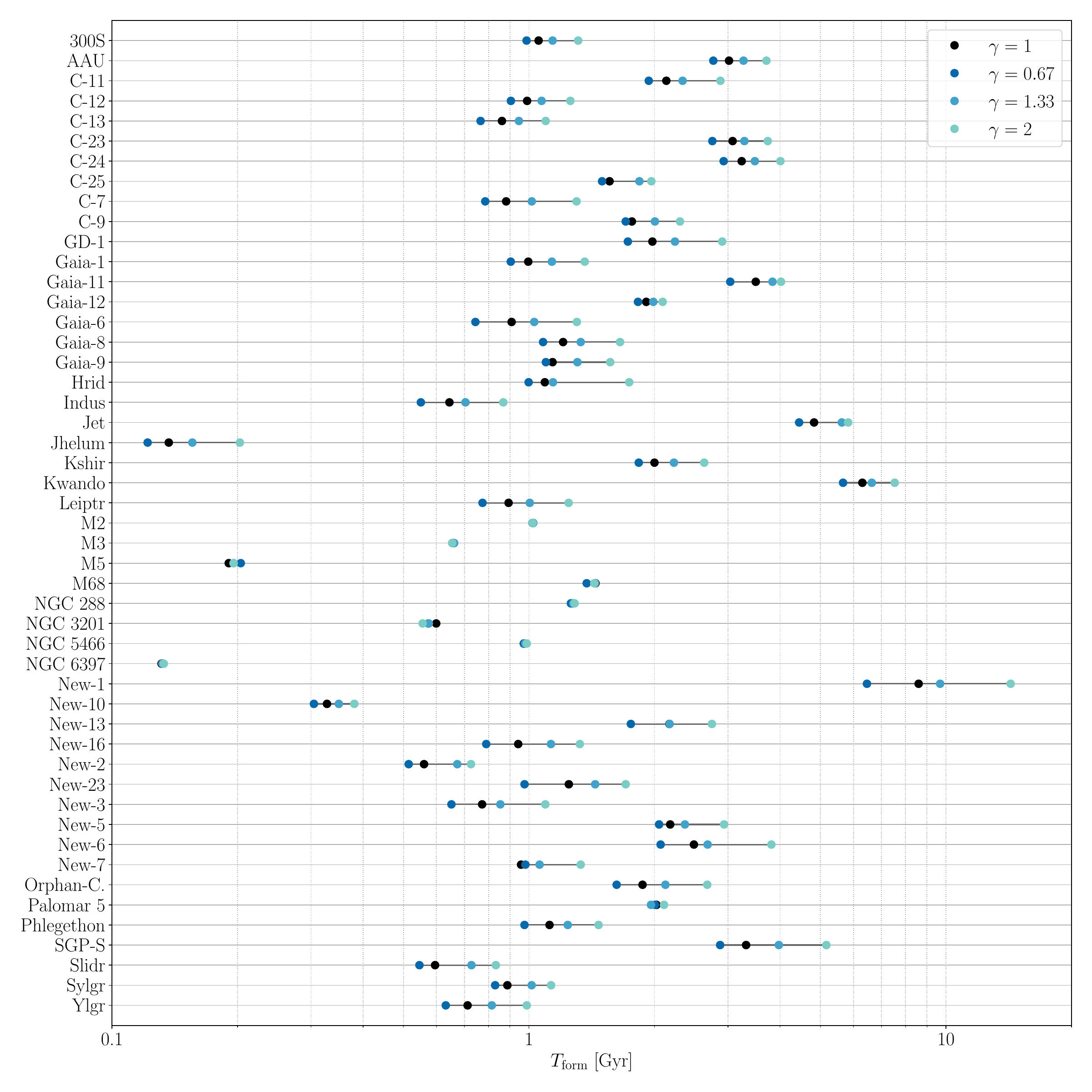}
\caption{Stream ages $T_\mathrm{form}$ resulting from the fit described in Sec.~\ref{sec:stream_catalog} for different mass-loss models, parameterized by $\gamma$ as in Eq.~\ref{eq:mass-loss}. The fiducial value ($\gamma=1$) used for the impact analysis is shown in black.}
\label{fig:stream_age_gamma}
\end{figure*}

\end{document}